\documentclass[reqno]{amsart}
\usepackage[utf8]{inputenc}
\usepackage[margin=1in]{geometry}
\usepackage{amsmath, amssymb}
\usepackage{xcolor}
\usepackage{enumitem}
\usepackage{varwidth}
\usepackage{siunitx}
\usepackage{graphicx}
\graphicspath{{images/}}
\usepackage{tikz}
\usetikzlibrary{patterns,arrows.meta}

\newcommand\bxi{\boldsymbol{\xi}}
\newcommand\bx{\boldsymbol{x}}
\newcommand\bv{\boldsymbol{v}}

\title[R13 solutions for flows between parallel plates]{Flows Between Parallel Plates: Analytical Solutions of Regularized 13-Moment Equations for Inverse-Power-Law Models}

\author{Zhicheng Hu}
\address[Zhicheng Hu]{Department of Mathematics, College of Science,
    Nanjing University of Aeronautics and Astronautics, Nanjing
    210016, China}
\email{huzhicheng@nuaa.edu.cn}

\author{Siyao Yang}
\address[Siyao Yang]{Department of Mathematics, National University of Singapore,
  Level 4, Block S17, 10 Lower Kent Ridge Road, Singapore 119076}
\email{matsiya@nus.edu.sg}

\author{Zhenning Cai}
\address[Zhenning Cai]{Department of Mathematics, National University of Singapore,
  Level 4, Block S17, 10 Lower Kent Ridge Road, Singapore 119076}
\thanks{Zhicheng Hu was supported by the National Natural Science Foundation of China under grant No. 11601229 and 11702135. Zhenning Cai and Siyao Yang were supported by the Academic Research Fund of the Ministry of Education of Singapore under grant No. R-146-000-305-114.}
\email{matcz@nus.edu.sg}

\DeclareMathOperator\re{Re}
\DeclareMathOperator\im{Im}
\newcommand\dd{\mathrm{d}}
\newcommand\Kn{\mathrm{Kn}}
\newcommand\ee{\mathrm{e}}
\newcommand\C{\textbf{C}}

\begin{document}

\begin{abstract}
We study the structure of stationary channel flows predicted by the regularized 13-moment equations. Compared with the previous work \cite{Taheri2009}, we focus on gases whose molecules satisfy the general inverse power law. The analytical solutions are obtained for the semi-linear equations, and the structures of Couette, Fourier, and Poiseuille flows are solved by coupling the general solutions with newly derived boundary conditions. The results show excellent agreement with the reference solution in the slip-flow regime. Our results also show that the R13 equations derived from inverse power law models can have better accuracy than the R13 equations of Maxwell molecules with altered viscosity.
\end{abstract}


\maketitle

\section{Introduction}
Microflows play important roles in the simulations of microelectromechanical systems. To accurately describe the fluid states inside the tiny devices, the gases have to be treated as rarefied gases in modelling and simulations, since the mean free path for the gases may appear to be comparative to the size of the devices. For early slip flows, the gas can still be considered to be relatively dense, so that the classical Navier-Stokes equations, one of the classical models in continuum mechanics, are still applicable in the simulations \cite{Karniadakis2005}. The rarefaction effects due to the gas-wall interaction, such as velocity slip and temperature jump, can be captured by high-order boundary conditions \cite{Hadjiconstantinou2003}. However, some other phenomena, such as heat transfer from cold to hot \cite{Akhlaghi2018}, cannot be described by Navier-Stokes equations, and thus finer models are required in the simulations.

Compared with continuum mechanics, gas kinetic theory is able to provide much more accurate description for the states of rarefied gases. One of the fundamental models in the gas kinetic theory is the Boltzmann equation \cite{Boltzmann1872}. However, the simulation of the Boltzmann equation is highly resource-demanding due to its high dimensionality \cite{John2011, Dimarco2018}. In particular, for slip flows and early transitional flows, instead of using the Boltzmann equation, one may expect that applying some extension of Navier-Stokes equations is already sufficient to provide solutions with adequate accuracy. Approaches to deriving such models include Chapman-Enskog expansion \cite{Chapman1916,Enskog1921} and the moment method \cite{Grad1949}. In the recent years, the classical Chapman-Enskog expansion has been extended to more involved cases including reacting flows, mixtures and granular gases\cite{Nagnibeda2009, Brey2015, Baranger2018, Garzo2018}. Meanwhile, significant progress on moment methods has been made by the researchers \cite{Singh2016, Bobylev2018, Fox2018, Alldredge2019, Patel2019, Bohmer2020}, and applications to more complicated gases such as polyatomic and relativistic gases are also rising\cite{Arima2018,Tinti2019,Arima2020}. In this work, we focus on one of the moment methods initially derived in \cite{Struchtrup2003}, and is now known as regularized 13-moment (R13) equations.

The R13 equations were initially derived for Maxwell molecules, which are a special type of molecules whose mutual force is always repulsive with magnitude inversely proportional to the fifth power of their distance. For this case, the model, theory and numerical methods of R13 equations have been well developed: the derivation of the equations is simplified in \cite{Struchtrup}; the boundary conditions are formulated in \cite{Torrilhon2008, Rana2016, Struchtrup2017}; the H-theorems for linear and nonlinear equations are established in \cite{Struchtrup2007, Torrilhon2012}; and its numerical solver has been studied in \cite{Rana2013, Claydon2017, Theisen2020}. Meanwhile, the analytical solutions of linear/semi-linear R13 equations have been found for problems with simple geometry \cite{Taheri2009, Taheri2010, Torrilhon2010, Rana2018}, and it has also been applied to a number of benchmark problems such as shock structure \cite{Torrilhon2004, Timokhin2017} and cavity flows \cite{Rana2013}. The method has also been extended to polyatomic gases and gas mixtures\cite{Rahimi2014,Gupta2016}. These works have shown the reliability of the R13 equations in describing slip and early transitional flows. Recently, the R13 equations have been generalized to non-Maxwell gases. In \cite{Struchtrup2013}, the linear R13 equations for the hard-sphere model is formulated. In \cite{Cai2020}, the authors derived nonlinear R13 equations based on the Boltzmann equation with linearized collision operator for all inverse power law models, and the models have been validated in \cite{Cai2020} using the shock structure problem which does not require boundary conditions.

The nonlinear R13 equations for general inverse-power-law models have a highly complicated collision term containing more than one hundred terms, which is inconvenient in practical applications. One possible scenario for simpler applications is the linear regime such as slow microflows, so that we can apply linear/semi-linear R13 equations, whose expressions are much neater. For microflow applications, boundary conditions have to be formulated. In this work, we will provide the general procedure to derive boundary conditions and simplify both R13 equations and the boundary conditions by semi-linearization about a global equilibrium state, so that for one-dimensional channel flows, we may find the analytical solutions. These solutions will then be compared with the DSMC (direct simulation of Monte Carlo) solutions \cite{Bird1994} to show the validity of R13 equations and the boundary conditions. The difference between different gas models can also be observed by these analytical results. Besides, we will also compare our results with the results of R13 equations for Maxwell molecules with a modified viscosity coefficient, which have been used in some literature to approximate the dynamics for non-Maxwell gases \cite{Torrilhon2006, Torrilhon2008, Taheri2010, Timokhin2015, Timokhin2019}.

The rest of this paper is arranged as follows. In Section \ref{sec:R13}, we state the problem settings and formulate the semi-linear R13 equations together with the boundary conditions. The analytical solutions to these equations are provided in Section \ref{sec:solution}. The solutions are applied to three types of benchmark problems, and the results are shown in Section \ref{sec:results} with comparisons to DSMC solutions. The semi-linearization of the equations and the derivation of the boundary conditions are detailed in Section \ref{sec:model}. Finally, we give some concluding remarks in Section \ref{sec:conclusion}.

\section{R13 equations for flows between parallel plates} \label{sec:R13}
\begin{figure}[h]
\centering
\begin{tikzpicture}

\draw[-{Latex[scale=1.5]}] (2.5,0) -- (5,0);
\draw[-{Latex[scale=1.5]}] (2.5,0) -- (0,0);
\node[above] at (2.5,0) {$L$};

\draw [line width=1mm] (0,-3) -- (0,3);\draw [line width=1mm] (5,-3) -- (5,3);
\fill [pattern=north east lines] (0,-3) rectangle (-1,3);
\fill [pattern=north east lines] (5,-3) rectangle (6,3);
\node[below] at (-0.5,-3) {$\theta_W^l$};\node[below] at (5.5,-3) {$\theta_W^r$};
\node[above] at (-0.5,3) {$\Big\downarrow$};\node[above] at (5.5,3) {$\Big\uparrow$};
\node[above] at (-1,3) {$v_W^l$};\node[above] at (6,3) {$v_W^r$};

\node at (2.5,-3) {$\otimes$}; 
\draw[-{Latex[scale=1]}] (2.5,-3) -- (4.5,-3);
\draw[-{Latex[scale=1]}] (2.5,-3) -- (2.5,-1);
\node[below left] at (2.5,-3) {$x_3$};
\node[below right] at (4.5,-3) {$x_2$};
\node[above left] at (2.5,-1) {$x_1$};

\draw[-{Latex[scale=1]}] (2.5,1.5) -- (2.5,3);
\node[above] at (2.5,3) {$G_1$};

\end{tikzpicture}
\caption{In the Couette flow, the walls of the channel are moving. In the Fourier flow, two parallel walls have different temperatures. In the Poiseuille flow, there exists a body force $G_1$.}
  \label{fig:flow}
\end{figure}
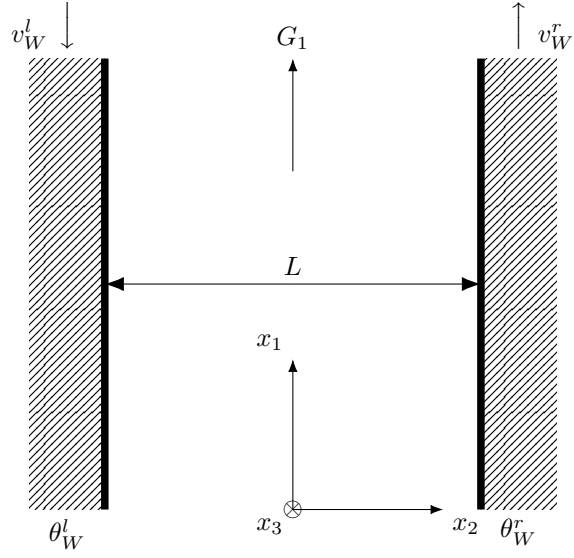

We are concerned about the single-species steady-state flows between two infinitely large parallel plates (see Figure \ref{fig:flow}). The distance of the two plates is $L$, and both plates are perpendicular to the $x_2$-axis. The temperatures of the left and right walls are assumed to be $\theta_W^l$ and $\theta_W^r$, respectively. Both walls can move inside their own plane, and we choose the reference frame and the coordinates such that both velocities are parallel to the $x_1$-axis. Under such settings, all the moments are functions of $x_2$ only, and the 13 moments can be reduced to eight variables including
\begin{itemize}
\item Equilibrium variables including density $\rho$, temperature $\theta$, and the velocity component parallel to the plates $v_1$;
\item Components of the stress tensor including the parallel stress $\sigma_{11}$, normal stress $\sigma_{22}$ and the shear stress $\sigma_{12}$;
\item Heat fluxes including the parallel heat flux $q_1$ and the normal heat flux $q_2$.
\end{itemize}

The governing equations of these quantities are derived based on the nonlinear R13 equations derived in \cite{Cai2020}. Here we focus mainly on the linear regime where the speed of flow is low. After nondimensionalization, dimension reduction and semi-linearization, eight equations are obtained. The first three of them come from the conservation laws of momentum and energy:
\begin{gather}
    \label{eq:v1} \frac{\dd \sigma_{12}}{\dd x_2} = G_1, \\%
    \label{eq:v2} \frac{\dd \theta}{\dd x_2} +  \frac{\dd \rho}{\dd x_2} +  \frac{\dd \sigma_{22}}{\dd x_2} = 0, \\%
    \label{eq:theta} \frac{\dd q_2}{\dd x_2} + \underline{  \sigma_{12} \frac{\dd v_1}{\dd x_2} }= 0.
\end{gather}
Here the underlined term highlights the nonlinearity we preserve during the semi-linearization of the equations, and such notation is adopted in all other equations in this section. As for the diagonal of the stress tensor $\sigma_{ij}$, we have $\sigma_{11} + \sigma_{22} + \sigma_{33} = 0$, and therefore only two of them need to be considered. Here we choose to consider the equations of $\sigma_{11}$ and $\sigma_{22}$. The equations of the stress tensor are related to the collision dynamics between gas molecules. Here we assume that the gas is monatomic, which is a simple case with no internal degrees of freedom, and the force between two molecules $F$ is determined by their distance $r$. More specifically, we consider only the inverse-power-law model, for which $F = \kappa r^{-\eta}$ with $\kappa$ and $\eta$ being  positive constants. After nondimensionalization, the constant $\kappa$ will be integrated into the Knudsen number $\Kn_0$. Then the equations for $\sigma_{11}$ and $\sigma_{22}$ can be written as
\begin{gather}
\label{eq:sigma11}
   \begin{split}
    \frac{ \alpha^{(\eta)}_{1,1}}{\Kn_0} \sigma_{11} +\alpha^{(\eta)}_{1,2} \frac{\dd q_2}{\dd x_2}  + \underline{  \alpha^{(\eta)}_{1,3}\sigma_{12} \frac{\dd v_1}{\dd x_2}   }+  \underline{  \alpha^{(\eta)}_{1,4} \Kn_0 \frac{\dd q_1}{\dd x_2} \frac{\dd v_1}{\dd x_2}   }+ \boxed{  \underline{  \alpha^{(\eta)}_{1,5}\Kn_0 \left(\frac{\dd v_1}{\dd x_2}\right)^2   }   }+ \boxed{   \alpha^{(\eta)}_{1,6} \Kn_0 \frac{\dd^2 \theta}{\dd x_2^2}   }& \\
    {}+   \boxed{   \alpha^{(\eta)}_{1,7}\Kn_0  \frac{\dd^2 \rho}{\dd x_2^2}   }   +  \underline{  \alpha^{(\eta)}_{1,8}\Kn_0 q_1 \frac{\dd^2 v_1}{\dd x_2^2}  } +  \alpha^{(\eta)}_{1,9}\Kn_0 \frac{\dd^2 \sigma_{11}}{\dd x_2^2} +  \alpha^{(\eta)}_{1,10}\Kn_0 \frac{\dd^2 \sigma_{22}}{\dd x_2^2} & = 0 ,
       \end{split} \\%
\label{eq:sigma22}
   \begin{split}
   \frac{ \alpha^{(\eta)}_{1,1} }{\Kn_0} \sigma_{22} +  \alpha^{(\eta)}_{3,1} \frac{\dd q_2}{\dd x_2}  +  \underline{    \alpha^{(\eta)}_{3,2} \sigma_{12} \frac{\dd v_1}{\dd x_2}   }+   \underline{   \alpha^{(\eta)}_{3,3} \Kn_0 \frac{\dd q_1}{\dd x_2} \frac{\dd v_1}{\dd x_2}   }+ \boxed{  \underline{   \alpha^{(\eta)}_{3,4}  \Kn_0 \left(\frac{\dd v_1}{\dd x_2}\right)^2  } }  - \boxed{  2  \alpha^{(\eta)}_{1,6} \Kn_0 \frac{\dd^2 \theta}{\dd x_2^2}  }& \\
   {}- \boxed{  2  \alpha^{(\eta)}_{1,7}\Kn_0  \frac{\dd^2 \rho}{\dd x_2^2}   }+  \underline{    \alpha^{(\eta)}_{3,5}\Kn_0 q_1 \frac{\dd^2 v_1}{\dd x_2^2}   } +  \alpha^{(\eta)}_{3,6}\Kn_0 \frac{\dd^2 \sigma_{22}}{\dd x_2^2}  &= 0.
  \end{split}
\end{gather}
Here all the coefficients $\alpha^{(\eta)}_{i,j}$ are constants dependent on the value of $\eta$, and their values for several collision models are given in Appendix \ref{sec:coefficients}. When $\eta = 5$, the specific inverse-power-law model is known as the model for Maxwell molecules, whose equations have been derived in \cite{Taheri2009}. Compared with the equations in \cite{Taheri2009}, the boxed terms in the above two equations are additional for general $\eta$. One can see from Table \ref{tab:alpha} that the coefficients in the boxes all equal zero when $\eta = 5$.

The equation of the only nonzero shear stress component $\sigma_{12}$ does not include nonlinear terms:
\begin{equation}
\label{eq:sigma12}
    \frac{ \alpha^{(\eta)}_{1,1} }{\Kn_0} \sigma_{12} +  \alpha^{(\eta)}_{2,1} \frac{\dd q_1}{\dd x_2} + \alpha^{(\eta)}_{2,2} \frac{\dd v_1}{\dd x_2} + \alpha^{(\eta)}_{2,3} \Kn_0 \frac{\dd^2 \sigma_{12}}{\dd x_2^2} = 0,
\end{equation}
and no boxes appear in the above equations, meaning that the general form of this equation is the same as Maxwell molecules. At last, the remaining two equations in the system are from the balance laws of the heat fluxes:
\begin{align}
\label{eq:q1}
    \frac{\alpha^{(\eta)}_{4,1}}{\Kn_0} q_1 +  \alpha^{(\eta)}_{4,2} \frac{\dd \sigma_{12}}{\dd x_2} +  \alpha^{(\eta)}_{4,3} \Kn_0  \frac{\dd^2 q_1}{\dd x_2^2} +  \boxed{   \alpha^{(\eta)}_{4,4} \Kn_0 \frac{\dd^2 v_1}{\dd x_2^2}  } &= 0 , \\%
\label{eq:q2}
   \begin{split}
   \frac{\alpha^{(\eta)}_{4,1}}{\Kn_0} q_2  + \alpha^{(\eta)}_{4,2} \frac{\dd \sigma_{22}}{\dd x_2}  +  \alpha^{(\eta)}_{5,1}\Kn_0 \frac{\dd^2 q_2}{\dd x_2^2} +   \underline{  \alpha^{(\eta)}_{5,2} \Kn_0 \sigma_{12} \frac{\dd^2 v_1}{\dd x_2^2}    } +  \alpha^{(\eta)}_{5,3} \frac{\dd \theta}{\dd x_2} \\
   {} + \underline{   \alpha^{(\eta)}_{5,4} q_1 \frac{\dd v_1}{ \dd x_2}    } + \underline{  \alpha^{(\eta)}_{5,5} \sigma_{12} \frac{\dd \sigma_{12}}{\dd x_2}  } + \underline{   \alpha^{(\eta)}_{5,6} \Kn_0 \frac{\dd v_1}{ \dd x_2} \frac{\dd \sigma_{12}}{\dd x_2}  } &= 0.
  \end{split}
\end{align}

The solution to the above equations can be determined only with boundary conditions. The boundary conditions are formulated based on the model proposed by Maxwell, where the velocity distribution of the reflected particles is the linear combination of specular reflection and diffusive reflection. Consider the left solid wall and assume that the proportion of diffusive reflection is $\chi^l \in [0,1]$. The following two boundary conditions are extensions of second-order slip boundary conditions for Navier-Stokes equations:
\begin{align}
  & q_2 = - \frac{\chi^l}{2 - \chi^l} \left(  \beta^{(\eta)}_{1,1} ( \theta-\theta_W^l )   +  \beta^{(\eta)}_{1,2} \sigma_{22} +  \beta^{(\eta)}_{1,3} \Kn_0 \frac{\dd q_2}{\dd x_2} +   \underline{  \beta^{(\eta)}_{1,4}(v_1 - v_W^l)^2   }+  \underline{ \beta^{(\eta)}_{1,5}\Kn_0 \sigma_{12} \frac{\dd v_1}{\dd x_2} }  \right), \label{eq:bc1} \\
 & \sigma_{12} = - \frac{\chi^l}{2 - \chi^l} \left( \beta^{(\eta)}_{4,1} q_1 +  \beta^{(\eta)}_{4,2} (v_1 - v_W^l) +  \beta^{(\eta)}_{4,3} \Kn_0 \frac{\dd \sigma_{12} }{\dd x_2}    \right).  \label{eq:bc4}
\end{align}
Additional boundary conditions come from higher-order moments of the distribution function:
\begin{align}
 &\underline{  \beta^{(\eta)}_{2,1}\Kn_0 q_1 \frac{\dd v_1}{\dd x_2}   } +  \beta^{(\eta)}_{2,2} \Kn_0\frac{\dd \sigma_{22}}{\dd x_2} \notag \\
& \hspace{40pt}  = - \frac{\chi^l}{2 - \chi^l} \left(  \beta^{(\eta)}_{2,3} (\theta - \theta_W^l ) + \beta^{(\eta)}_{2,4} \sigma_{22}+ \beta^{(\eta)}_{2,5} \Kn_0 \frac{\dd q_2}{\dd x_2} +  \underline{  \beta^{(\eta)}_{2,6} (v_1 - v_W^l)^2    }+  \underline{  \beta^{(\eta)}_{2,7}  \Kn_0 \sigma_{12}  \frac{\dd v_1}{\dd x_2}  }\right),
  \label{eq:bc2}  \\
 &  \boxed{ \beta^{(\eta)}_{3,1} \sigma_{12}  }+ \beta^{(\eta)}_{3,2}\Kn_0 \frac{\dd q_1}{\dd x_2} + \boxed{ \beta^{(\eta)}_{3,3} \Kn_0\frac{\dd v_1}{\dd x_2}  } = - \frac{\chi^l}{2-\chi^l} \left(  \beta^{(\eta)}_{3,4} q_1 +  \beta^{(\eta)}_{3,5} (v_1 - v_W^l) +  \beta^{(\eta)}_{3,6}\Kn_0 \frac{\dd \sigma_{12}}{\dd x_2} \right),  \label{eq:bc3}\\
& \underline{  \beta^{(\eta)}_{5,1} \Kn_0 q_1 \frac{\dd v_1}{\dd x_2} }+  \beta^{(\eta)}_{5,2} \Kn_0  \left( 2 \frac{\dd \sigma_{11}}{\dd x_2}  + \frac{\dd \sigma_{22}}{ \dd x_2}\right) = - \frac{\chi^l}{2-\chi^l} \left(  \underline{  \beta^{(\eta)}_{5,3} (v_1 - v_W^l)^2  }+  \beta^{(\eta)}_{5,4} (2\sigma_{11} + \sigma_{22}) + \underline{  \beta^{(\eta)}_{5,5} \Kn_0  \sigma_{12} \frac{\dd v_1}{\dd x_2} } \right).    \label{eq:bc5}
\end{align}
The coefficients $\beta_{i,j}^{(\eta)}$ for different $\eta$ are given in the Appendix \ref{sec:coefficients}. For boundary conditions on the right solid wall, we just need to make the following replacement:
\begin{displaymath}
q_2 \rightarrow -q_2, \quad \sigma_{12} \rightarrow -\sigma_{12}, \quad x_2 \rightarrow -x_2, \quad \chi^l \rightarrow \chi^r, \quad \theta_W^l \rightarrow \theta_W^r, \quad v_W^l \rightarrow v_W^r.
\end{displaymath}
Thus we have in total ten boundary conditions on both walls. To fully determine the solution, we need to specify the average mass. Here we set the range of $x_2$ to be $[-1/2, 1/2]$ (so that $L = 1$) and fix the total mass to be
\begin{equation} \label{eq:zero_mass}
\int_{-1/2}^{1/2} \rho(x_2) \,\mathrm{d}x_2 = 0.
\end{equation}
Details on the derivation of these equations and the boundary conditions will be discussed in Section \ref{sec:model}. Here we first consider the solutions to these equations.

\section{Analytical solutions to semi-linear R13 equations} \label{sec:solution}
The semi-linear R13 equations \eqref{eq:v1}--\eqref{eq:q2} should be solved following a proper ordering. Straightforwardly, one may obtain the shear stress $\sigma_{12}$ by solving the equation \eqref{eq:v1}:
\begin{equation}
\sigma_{12} = G_1 x_2 + \C_1,
\end{equation}
where $\C_1$ denotes the constant to be determined by the boundary conditions (similar notations for such constants are used later without further declaration). Now combining equations \eqref{eq:sigma12}, \eqref{eq:q1}, and plugging in the above expression for $\sigma_{12}$, we solve the heat flux $q_1$ as well as the velocity $v_1$ as
\begin{gather}
      q_1 = \gamma^{(\eta)}_{1,1} G_1 \Kn_0 +   A_1  ,  \\
   v_1 =    \gamma^{(\eta)}_{2,1}  \left(  \frac{G_1}{\Kn_0} x_2^2 + 2  \frac{\C_1}{\Kn_0} x_2 \right) - \gamma^{(\eta)}_{2,2} A_1 +\C_4,
\end{gather}
where
\begin{equation}
    A_1 =   \C_2 \ee^{\frac{\delta^{(\eta)}_{1}}{\Kn_0}\cdot x_2 } + \C_3 \ee^{-\frac{\delta^{(\eta)}_{1}}{\Kn_0}\cdot x_2 },
\end{equation}
and the corresponding values of coefficients $\gamma^{(\eta)}_{i,j}$ and $\delta^{(\eta)}_{k}$ for different $\eta$ are listed in Table \ref{tab:gamma}. At this point, we may insert the formulas for $\sigma_{12}$ and $v_1$ into equation \eqref{eq:theta} to get the normal heat flux \begin{equation}
     q_2 =    -\frac{2}{3}\gamma^{(\eta)}_{2,1} \left(   \frac{G_1^2}{\Kn_0} x_2^3 + \frac{3G_1\C_1}{\Kn_0} x_2^2 + \frac{3\C_1^2}{\Kn_0} x_2   \right) + \C_5
      + \gamma^{(\eta)}_{2,2} \left( \sigma_{12}A_1 -  \frac{ G_1 \Kn_0}{\delta^{(\eta)}_{1}} D \right),
\end{equation}
with
\begin{equation}
     D =  \C_2 \ee^{\frac{\delta^{(\eta)}_{1}}{\Kn_0}\cdot x_2 } - \C_3 \ee^{-\frac{\delta^{(\eta)}_{1}}{\Kn_0}\cdot x_2 }.
\end{equation}
The terms $A_1$ and $D$ contribute to the Knudsen layers in the velocity and the normal heat flux. The constant $\delta_1^{(\eta)}$ gives the thickness of the Knudsen layer, which depends on the gas model. Note that for Maxwell molecules, the thickness of the Knudsen layer is solely determined by the equation \eqref{eq:q1}, while for other gases, the boxed term in \eqref{eq:q1} coming from collision of molecules couples the parallel heat flux with the parallel velocity, so that the balance law for the shear stress also plays a role in determining the thickness of the boundary layer.

Next, we solve the normal stress $\sigma_{22}$, temperature $\theta$, and mass density $\rho$ together using equations \eqref{eq:v2}, \eqref{eq:sigma22}, and \eqref{eq:q2}. In specific, we first write $\rho$ in terms of $\theta$ and $\sigma_{22}$ according to the equation \eqref{eq:v2}:
\begin{equation}
     \rho = \C_9  - \theta - \sigma_{22}
\end{equation}
and then combine and solve the two equations \eqref{eq:sigma22} and \eqref{eq:q2} based on the above formula together with the expressions for $v_1$, $q_1$, $q_2$, and $\sigma_{12}$ obtained by previous calculation to arrive at
\begin{gather}
    \sigma_{22} = \gamma^{(\eta)}_{3,1}G_1^2 \Kn_0^2 + \gamma^{(\eta)}_{3,2} \sigma_{12}^2  +   \gamma^{(\eta)}_{3,3} \C_2 \C_3     + \gamma^{(\eta)}_{3,4}G_1 \Kn_0 A_1 + \gamma^{(\eta)}_{3,5} \sigma_{12} D +\gamma^{(\eta)}_{3,6} B + A_2,\\
    \begin{split}
&\theta =    \gamma^{(\eta)}_{4,1}\frac{G_1}{\Kn_0^2}(G_1 x_2^4 + 4\C_1 x_2^3) + \left( \gamma^{(\eta)}_{4,2} G_1^2 + \gamma^{(\eta)}_{4,3} \frac{\C_1^2}{\Kn_0^2} \right) x_2^2 + \left( 2\gamma^{(\eta)}_{4,2} G_1 \C_1 +  \gamma^{(\eta)}_{4,4} \frac{\C_5}{\Kn_0}\right) x_2 + \C_8 + \\
& \hspace{220pt} + \gamma^{(\eta)}_{4,5} G_1 \Kn_0  A_1 + \gamma^{(\eta)}_{4,6} \sigma_{12} D + \gamma^{(\eta)}_{4,7} B + \gamma^{(\eta)}_{4,8} A_2,
     \end{split}
\end{gather}
where
\begin{equation}
     A_2 =  \C_6 \ee^{\frac{\delta^{(\eta)}_{2}}{\Kn_0}\cdot x_2 } + \C_7 \ee^{-\frac{\delta^{(\eta)}_{2}}{\Kn_0}\cdot x_2 }, \quad
     B = \C_2^2 \ee^{\frac{2\delta^{(\eta)}_{1}}{\Kn_0}\cdot x_2 } + \C_3^2 \ee^{-\frac{2\delta^{(\eta)}_{1}}{\Kn_0}\cdot x_2 }.
\end{equation}
The mass density then can be solved following the condition \eqref{eq:zero_mass}. Regardless of boundary terms, the temperature $\theta$ is a quartic function of $x_2$. Such a structure agrees with the analysis for the BGK model studied in \cite{Tij1994}, where $G_1$ is considered as a small parameter. Our result corresponds to the solution in \cite{Tij1994} expanded up to $O(G_1^2)$, and some boundary effect can also be captured by our solution.

Finally, we solve the parallel stress $\sigma_{11}$ using the equation \eqref{eq:sigma11} upon knowing all other quantities:
\begin{equation}
    \sigma_{11} = \gamma^{(\eta)}_{5,1} G_1^2 \Kn_0^2 +\gamma^{(\eta)}_{5,2} \sigma_{12}^2 +   \gamma^{(\eta)}_{5,3} \C_2 \C_3   +  \gamma^{(\eta)}_{5,4}G_1 \Kn_0 A_1 + \gamma^{(\eta)}_{5,5} \sigma_{12} D +\gamma^{(\eta)}_{5,6}  B+\gamma^{(\eta)}_{5,7} A_2 + A_3,
\end{equation}
where
\begin{equation}
     A_3 =  \C_{10} \ee^{\frac{\delta^{(\eta)}_{3}}{\Kn_0}\cdot x_2 } + \C_{11} \ee^{-\frac{\delta^{(\eta)}_{3}}{\Kn_0}\cdot x_2 }.
\end{equation}
The solution shows that when $G = 0$, besides boundary layers, $\sigma_{11}$ is a constant in the bulk; when $G$ is a constant, $\sigma_{11}$ is quadratic in the bulk. Note that the bulk solutions of $\sigma_{11}$ and $\sigma_{22}$ cannot be predicted by Navier-Stokes-Fourier equations.

% \begin{center}
% \begin{varwidth}{\textwidth}
% \begin{enumerate}[label=(\Roman*)]
%  \item Solve Eq \eqref{eq:v1} to get $\sigma_{12}$
%   \item Solve Eq \eqref{eq:sigma12}\eqref{eq:q1} to get $q_1$, $v_1$
%   \item Solve Eq \eqref{eq:theta} to get $q_2$
%   \item Solve Eq \eqref{eq:v2}\eqref{eq:sigma22}\eqref{eq:q2} to get $\sigma_{22}$, $\theta$, $\rho$
%   \item Solve Eq \eqref{eq:sigma11} to get $\sigma_{11}$
% \end{enumerate}
% \end{varwidth}
% \end{center}

% Some relationships between coefficients $\alpha^{(\eta)}_{i,j}$ and $\gamma^{(\eta)}_{i,j}$ are given by
% \begin{gather}
%      \gamma^{(\eta)}_{1,1} = \frac{\alpha^{(\eta)}_{1,1}\alpha^{(\eta)}_{4,4}-\alpha^{(\eta)}_{2,2}\alpha^{(\eta)}_{4,2}}{\alpha^{(\eta)}_{2,2}\alpha^{(\eta)}_{4,1}}, \quad   \gamma^{(\eta)}_{1,2} = \left( \frac{\alpha^{(\eta)}_{2,2}\alpha^{(\eta)}_{4,1}}{\alpha^{(\eta)}_{2,1}\alpha^{(\eta)}_{4,4} - \alpha^{(\eta)}_{2,2}\alpha^{(\eta)}_{4,3}}\right)^{\frac{1}{2}};\\
%      \gamma^{(\eta)}_{2,1} = -\frac{\alpha^{(\eta)}_{1,1}}{2\alpha^{(\eta)}_{2,2}}, \quad \gamma^{(\eta)}_{2,2} = \frac{\alpha^{(\eta)}_{2,1}}{\alpha^{(\eta)}_{2,2}}.
% \end{gather}

The above exact solutions will be considered under three different settings, which are
\begin{itemize}
\item Couette flow, which represents the case with moving plates;
\item Fourier flow, which represents the case with a temperature difference of both walls;
\item Force-drive Poiseuille flow, which represents the case with a body force.
\end{itemize}
The settings of these three cases will be detailed in the following subsections.

\subsection{Couette flow}
The Couette flow refers to the case where $G_1=0$ and $\theta_W^l = \theta_W^r$, and we choose the frame of reference such that $v_W^l = - v_W^r$. Due to such symmetry, the stationary Couette flow satisfies the properties $v_1(x_2)= -v_1(-x_2)$ and $q_1(x_2) = -q_1(-x_2)$, resulting in $\C_4 = 0$ and $\C_2 +\C_3=0$. Therefore, the general solutions for the velocity $v_1$, shear stress $\sigma_{12}$, and heat flux $q_1$ reduce to
\begin{equation}\label{eq:couette-sol-vq}
   v_1 = 2 \gamma^{(\eta)}_{2,1}  \frac{\C_1}{\Kn_0} x_2 - 2 \gamma^{(\eta)}_{2,2} \C_2 \sinh\left(\frac{\delta^{(\eta)}_1}{\Kn_0} x_2\right) , \quad  \sigma_{12} = \C_1, \quad q_1 = 2\C_2 \sinh\left(\frac{\delta^{(\eta)}_1}{\Kn_0} x_2\right).
\end{equation}

As for the temperature, the Couette flow has the symmetry $\theta(x_2) = \theta(-x_2)$, which gives $\C_5 = 0$ and $\C_6 = \C_7$. Consequently, we have
\begin{equation}
\begin{split}
    & \theta =     \gamma^{(\eta)}_{4,3} \frac{\C_1^2}{\Kn_0^2}  x_2^2  + \C_8  + 2\gamma^{(\eta)}_{4,6} \C_1   \C_2 \cosh\left(\frac{\delta^{(\eta)}_1}{\Kn_0} x_2\right) + 2 \gamma^{(\eta)}_{4,7} \C_2^2  \cosh\left(2\frac{\delta^{(\eta)}_1}{\Kn_0} x_2\right) + 2 \gamma^{(\eta)}_{4,8} \C_6 \cosh\left(\frac{\delta^{(\eta)}_2}{\Kn_0} x_2\right),  \\
    &  \sigma_{22} = \gamma^{(\eta)}_{3,2} \C_1^2  -   \gamma^{(\eta)}_{3,3} \C_2^2      + 2\gamma^{(\eta)}_{3,5} \C_1   \C_2 \cosh\left(\frac{\delta^{(\eta)}_1}{\Kn_0} x_2\right) +2 \gamma^{(\eta)}_{3,6} \C_2^2  \cosh\left(2\frac{\delta^{(\eta)}_1}{\Kn_0} x_2\right) + 2  \C_6 \cosh\left(\frac{\delta^{(\eta)}_2}{\Kn_0} x_2\right),\\
    & q_2 =   -2\gamma^{(\eta)}_{2,1}  \frac{\C_1^2}{\Kn_0} x_2
      + 2\gamma^{(\eta)}_{2,2} \C_1 \C_2 \sinh\left(\frac{\delta^{(\eta)}_1}{\Kn_0} x_2\right).
\end{split}
\end{equation}

\subsection{Fourier flow}
For the Fourier flow, we again have $G_1 = 0$. Meanwhile we assume that the plates are stationary so that $v_W^l = v_W^r = 0$. The flow structure is generated by the temperature difference between walls. Here we assume that $\theta_W^l < \theta_W^r$. In this scenario, the parallel velocity vanishes: $v_1 = 0$, which gives us $\C_1 = \C_2 = \C_3 = \C_4 = 0$. As a result,
\begin{equation}
     v_1 = 0, \quad \sigma_{12} = 0, \quad q_1 = 0.
\end{equation}
Meanwhile, the normal stress has the symmetry $\sigma_{22}(x_2) = -\sigma_{22}(-x_2)$, yielding that $\C_6 + \C_7 =0$. Thus we have
\begin{equation}
 \begin{split}
&\theta =    \gamma^{(\eta)}_{4,4} \frac{\C_5}{\Kn_0}  x_2 + \C_8  +  2 \gamma^{(\eta)}_{4,8} \C_6 \sinh\left(\frac{\delta^{(\eta)}_2}{\Kn_0} x_2\right), \\
&  \sigma_{22} =  2 \C_6\sinh\left(\frac{\delta^{(\eta)}_2}{\Kn_0} x_2\right) , \quad q_2 =   \C_5.
 \end{split}
\end{equation}

\subsection{Force-driven Poiseuille flow}
This is the case where $\theta_W^l = \theta_W^r$ and $v_l^W = v_l^W = 0$, while the body force $G_1$ is nonzero. For the Poiseuille flow, both the temperature and velocity are symmetric, i.e, $\theta(x_2)= \theta(-x_2)$ and $v_1(x_2) = v_1(-x_2)$. Therefore, we have $\C_2 = \C_3$, $\C_6 = \C_7$, and $\C_1 = \C_5 = 0$, which eventually lead to the solution
\begin{equation} \label{eq:Poiseuille}
\begin{split}
&  v_1 =    \gamma^{(\eta)}_{2,1}   \frac{G_1}{\Kn_0} x_2^2  - 2\gamma^{(\eta)}_{2,2} \C_2 \cosh\left(\frac{\delta^{(\eta)}_1}{\Kn_0} x_2\right)  +\C_4, \quad \sigma_{12} = G_1 x_2, \quad   q_1 = \gamma^{(\eta)}_{1,1} G_1 \Kn_0 +   2\C_2  \cosh\left(\frac{\delta^{(\eta)}_1}{\Kn_0} x_2\right) ,  \\ &\theta =   \gamma^{(\eta)}_{4,1}\frac{G_1^2}{\Kn_0^2} x_2^4  + \gamma^{(\eta)}_{4,2} G_1^2  x_2^2 + \C_8 + 2 \gamma^{(\eta)}_{4,5} \C_2 G_1 \Kn_0 \cosh\left(\frac{\delta^{(\eta)}_1}{\Kn_0} x_2\right) +\\
&\hspace{70pt}+2\gamma^{(\eta)}_{4,6} \C_2 G_1 x_2 \sinh\left(\frac{\delta^{(\eta)}_1}{\Kn_0} x_2\right)  + 2\gamma^{(\eta)}_{4,7} \C_2^2 \cosh\left(2\frac{\delta^{(\eta)}_1}{\Kn_0} x_2\right) + 2\gamma^{(\eta)}_{4,8} \C_6 \cosh\left(\frac{\delta^{(\eta)}_2}{\Kn_0} x_2\right), \\
&   \sigma_{22} = \gamma^{(\eta)}_{3,1}G_1^2 \Kn_0^2 + \gamma^{(\eta)}_{3,2} G_1^2 x_2^2  +   \gamma^{(\eta)}_{3,3} \C_2^2      + 2\gamma^{(\eta)}_{3,4}\C_2 G_1 \Kn_0 \cosh\left(\frac{\delta^{(\eta)}_1}{\Kn_0} x_2\right) +\\
&\hspace{70pt}+2\gamma^{(\eta)}_{3,5} \C_2 G_1  x_2  \sinh\left(\frac{\delta^{(\eta)}_1}{\Kn_0} x_2\right) +2\gamma^{(\eta)}_{3,6} \C_2^2 \cosh\left(2\frac{\delta^{(\eta)}_1}{\Kn_0} x_2\right) + 2\C_6 \cosh\left(\frac{\delta^{(\eta)}_2}{\Kn_0} x_2\right),\\
&   q_2 =    -\frac{2}{3}\gamma^{(\eta)}_{2,1}  \frac{G_1^2}{\Kn_0} x_2^3   + 2\gamma^{(\eta)}_{2,2} \C_2\left[ G_1 x_2 \cosh\left(\frac{\delta^{(\eta)}_1}{\Kn_0} x_2\right) -  \frac{ G_1 \Kn_0}{\delta^{(\eta)}_{1}} \sinh\left(\frac{\delta^{(\eta)}_1}{\Kn_0} x_2\right) \right].
\end{split}
\end{equation}

\section{Results} \label{sec:results}
The results of the above analytical solutions are illustrated and compared to DSMC simulations in this section. For Couette and Fourier flows, the DSMC solutions are obtained with Bird's code \cite{Bird1994}, where the molecular mass and the average density of the gas are set to $\mathfrak{m}=\SI{6.63e-26}{\kg}$ and $\rho_0=\SI{9.282e-6}{\kg\per\meter\cubed}$, respectively. At the reference temperature $T_0=\SI{273.15}{K}$, the mean free path can be fixed as $\lambda_0=\SI{9.2456e-3}{\meter}$. Thus the distance between two plates is given by $L = \lambda_0 / \Kn$, where $\Kn$ is the Kundsen number. Given the viscosity index $\omega$, the viscosity coefficient $\mu_0$ can be then obtained from \eqref{eq:mfp}. To match our solutions with the DSMC results, the same $\mu_0$ is employed in the semi-linear R13 equations instead of \eqref{eq:mu_omega}. In more detail, the parameter $\Kn_0$ in previous sections is related to the Knudsen number $\Kn$ and the viscosity index $\omega$ by $\Kn_0 = \sqrt{\dfrac{\pi}{2}} \dfrac{15 \Kn}{(5-2\omega)(7-2\omega)}$. As for force-driven Poiseuille flow, the dimensionless DSMC results are obtained for a hard sphere gas from \cite{Garcia}. In addition, both plates are assumed to be completely diffusive, that is, the accommodation coefficients are $\chi^l=\chi^r=1,$ if not specified in the simulations.

\subsection{Couette flow}
For the Couette flow, we consider the case that the plates move with $v_W^r=-v_W^l=\SI{47.6998}{\meter\per\second}$ (the corresponding dimensionless wall velocities are $\pm 0.2$) at the reference temperature $T_0$.

\begin{figure}[!htb]
% \colorbox{black}
  \centering
  {\includegraphics[width=0.32\textwidth,clip]{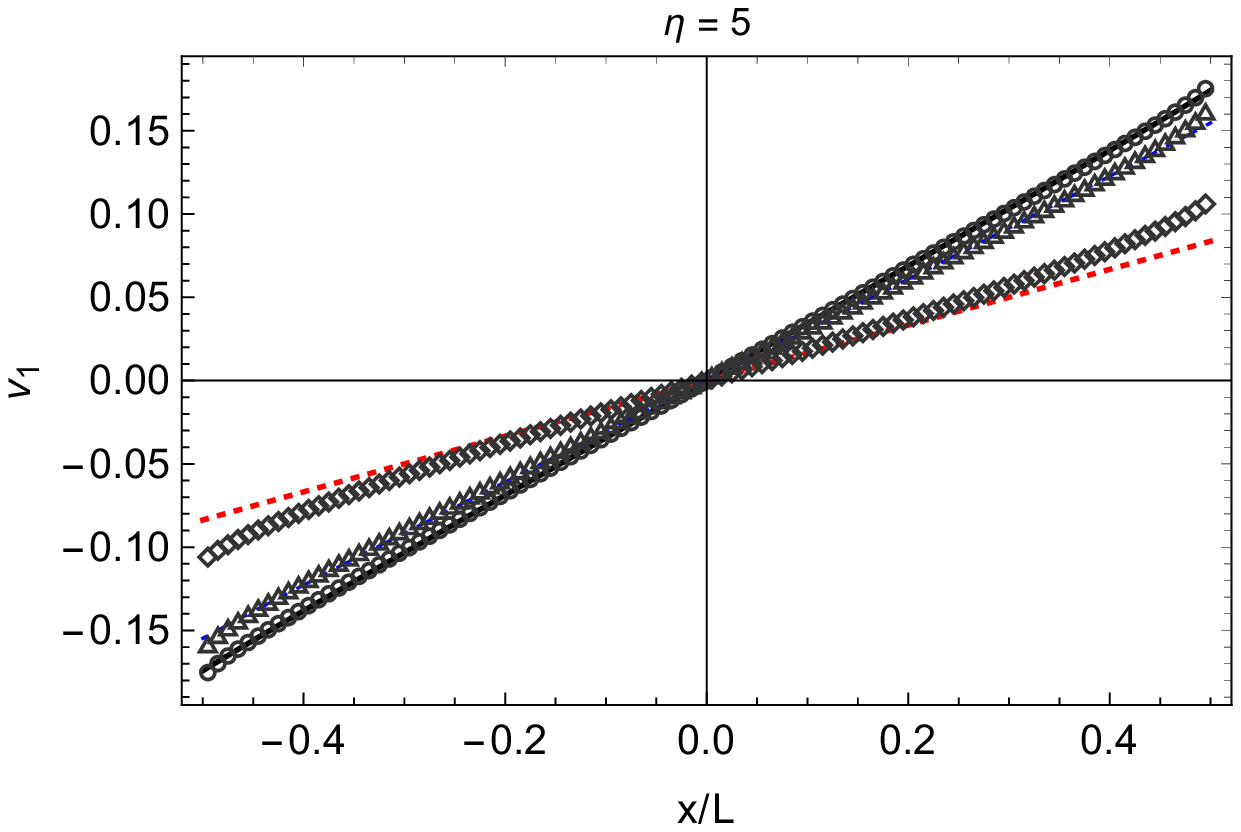}}
  {\includegraphics[width=0.32\textwidth,clip]{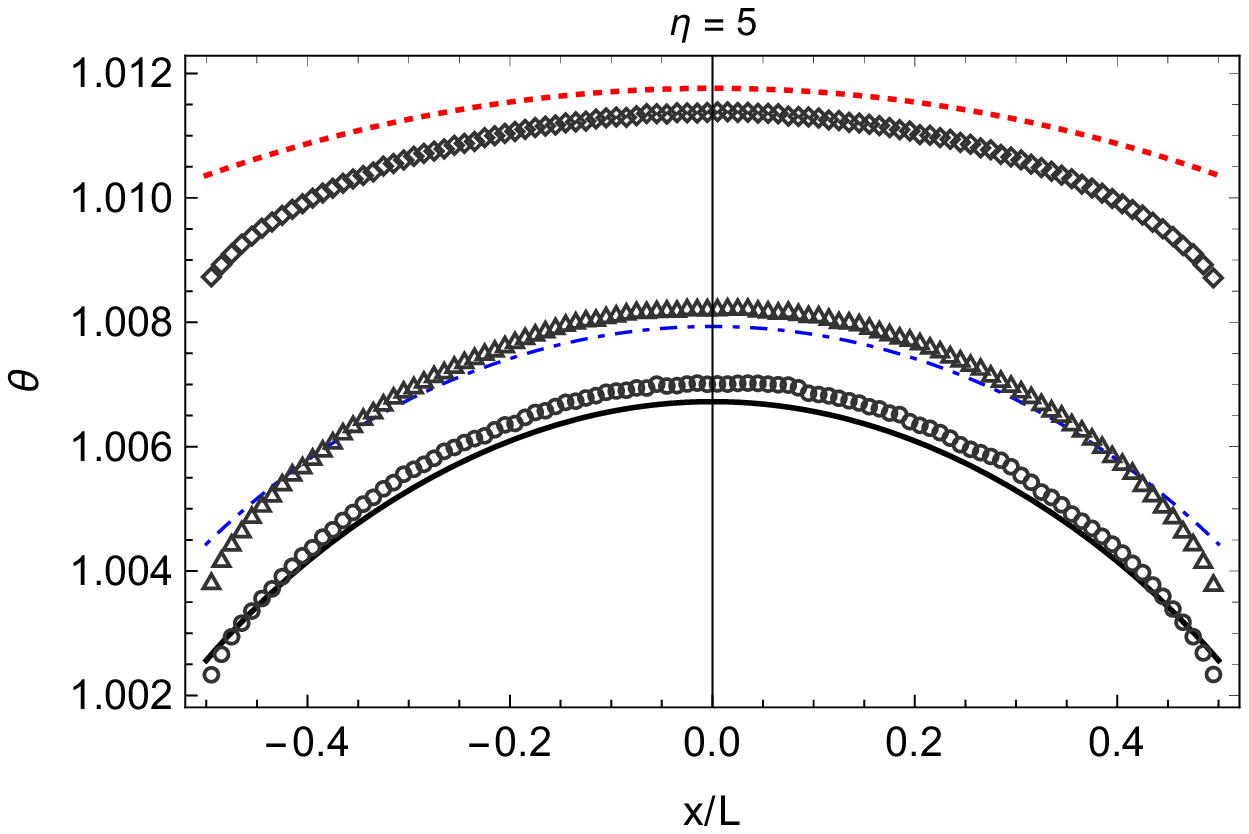}}
  {\includegraphics[width=0.32\textwidth,clip]{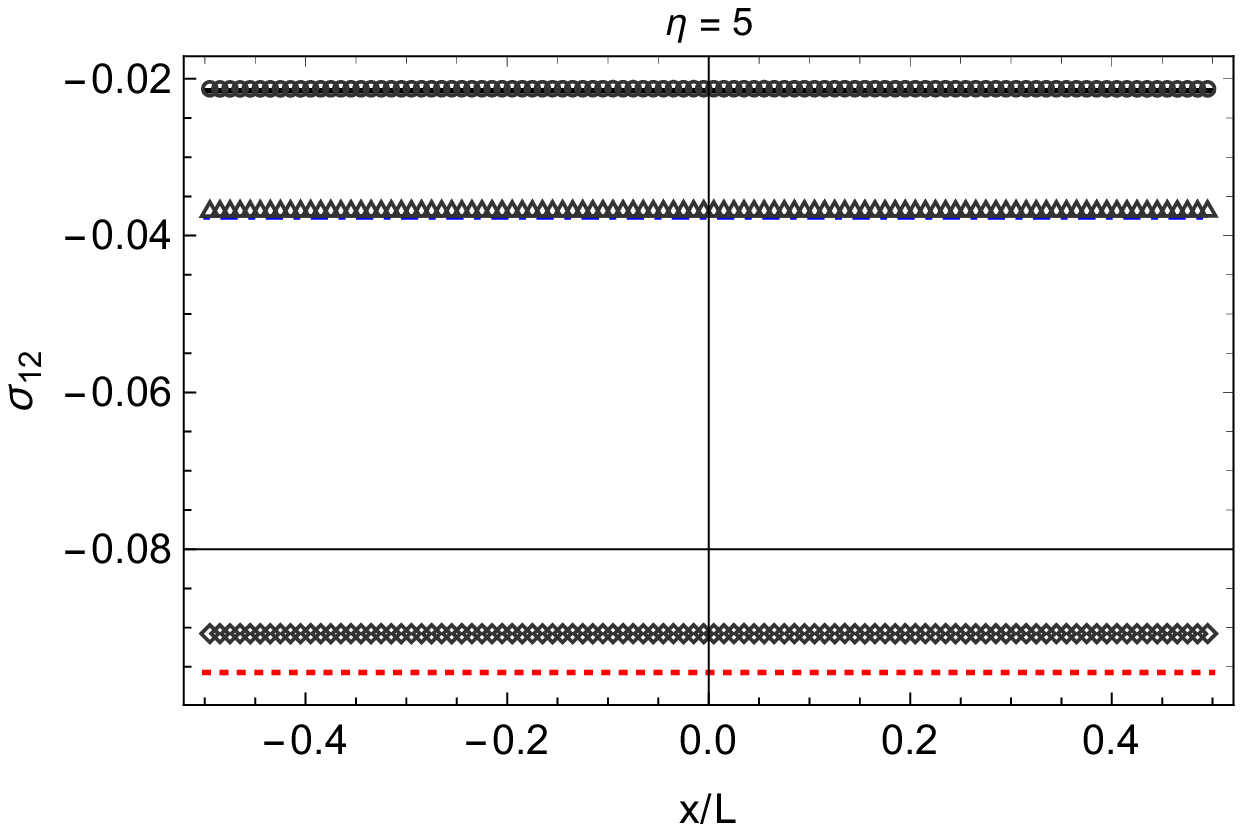}}\\
  {\includegraphics[width=0.32\textwidth,clip]{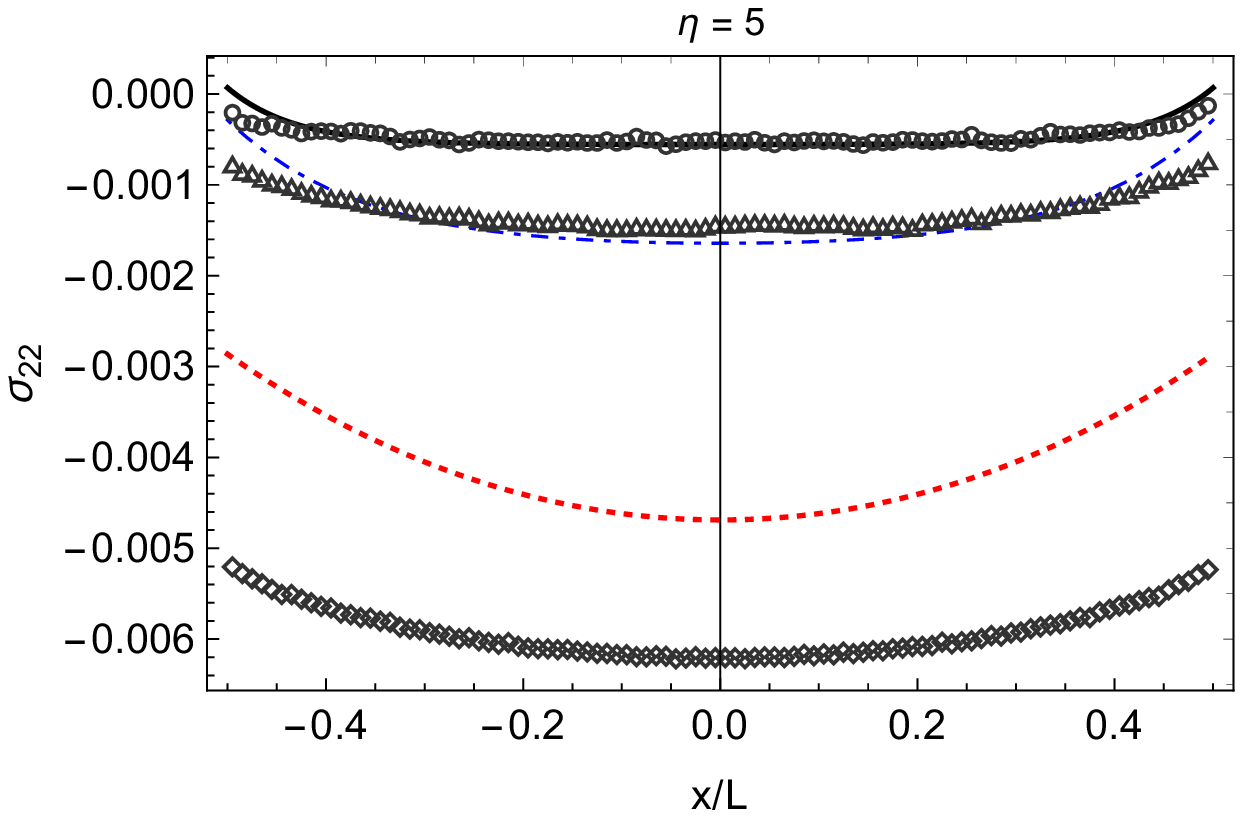}}
  {\includegraphics[width=0.32\textwidth,clip]{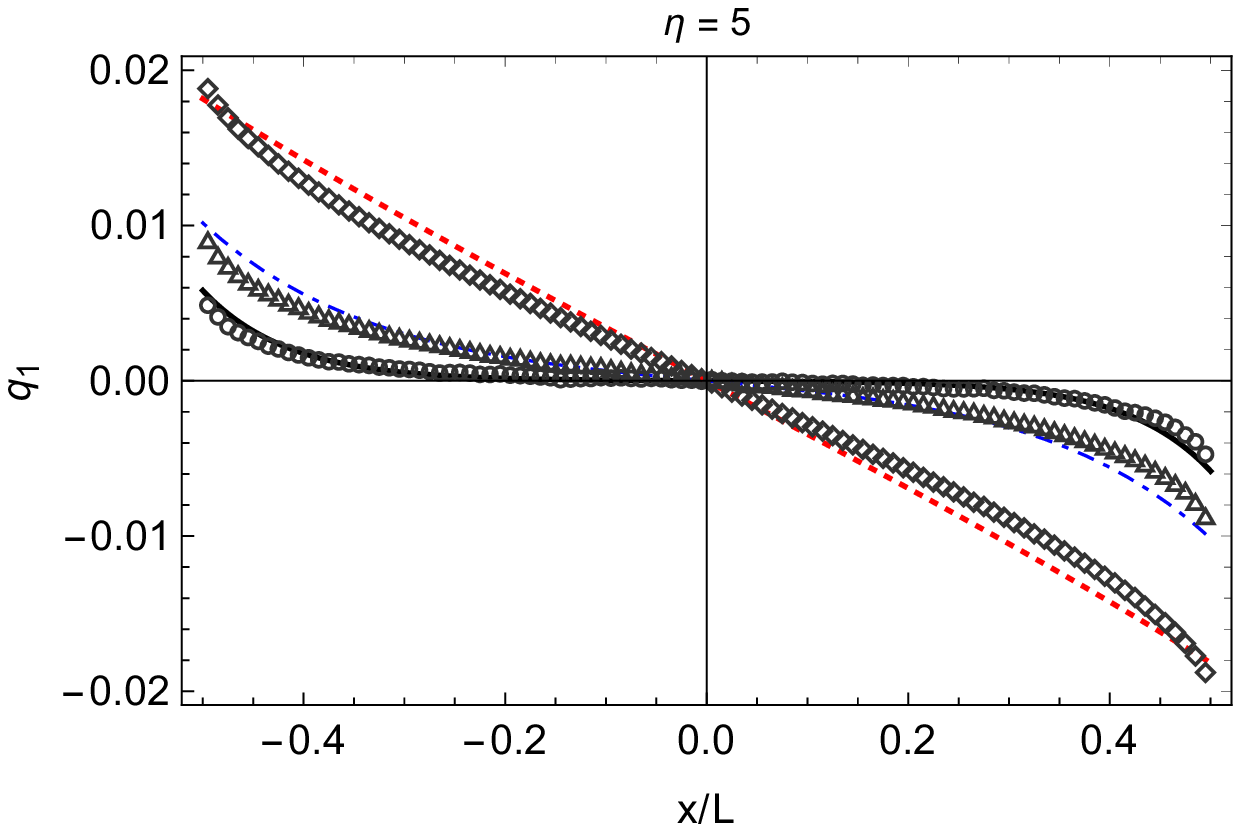}}
  {\includegraphics[width=0.32\textwidth,clip]{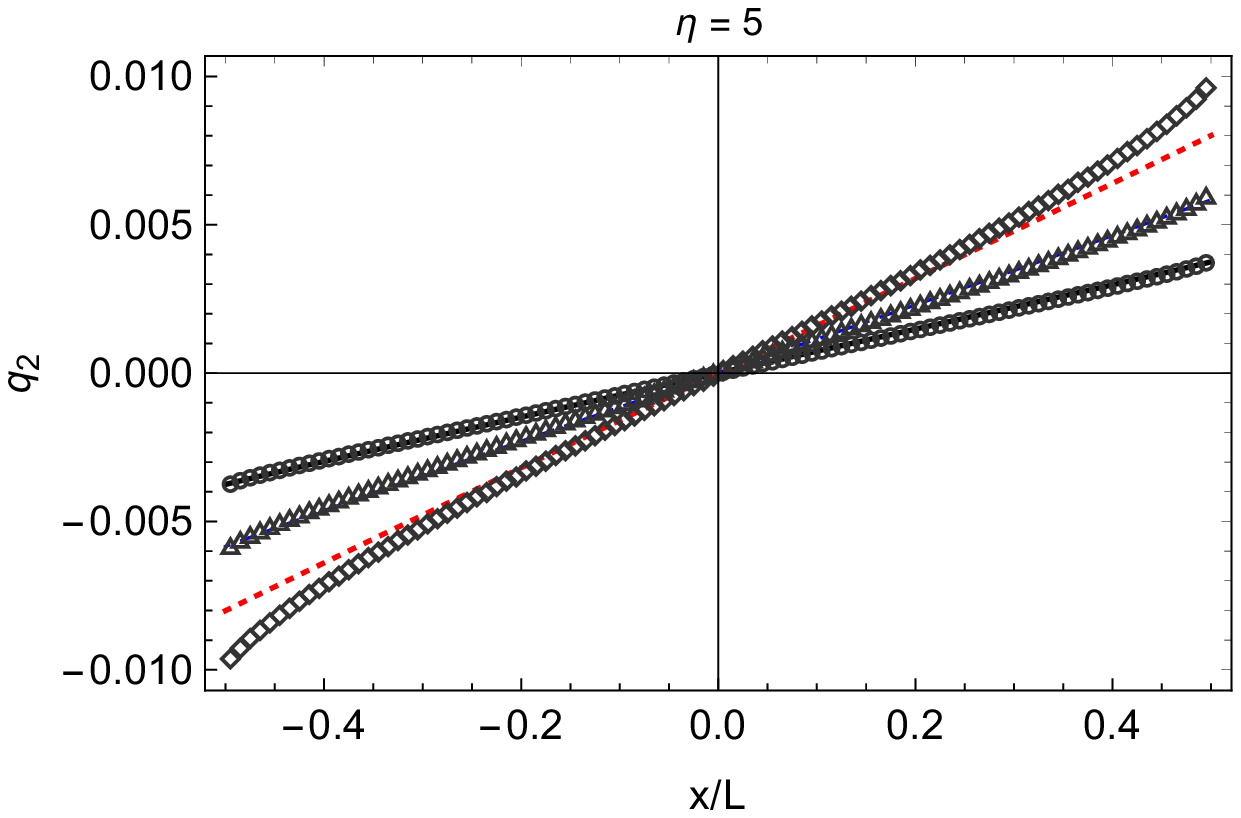}}
  \caption{Couette flow of the semi-linear R13 equations for $\eta=5$ (Maxwell molecules). The dimensionless semi-linear R13 solutions for $\Kn=0.05$, $0.1$, and $0.5$ are plotted in black solid line, blue dashed-dotted line, and red dotted line, respectively. The corresponding DSMC solutions are marked by circle, triangle, and diamond.}
  \label{fig:couette-eta5}
\end{figure}

\begin{figure}[!htb]
% \colorbox{black}
  \centering
  {\includegraphics[width=0.32\textwidth,clip]{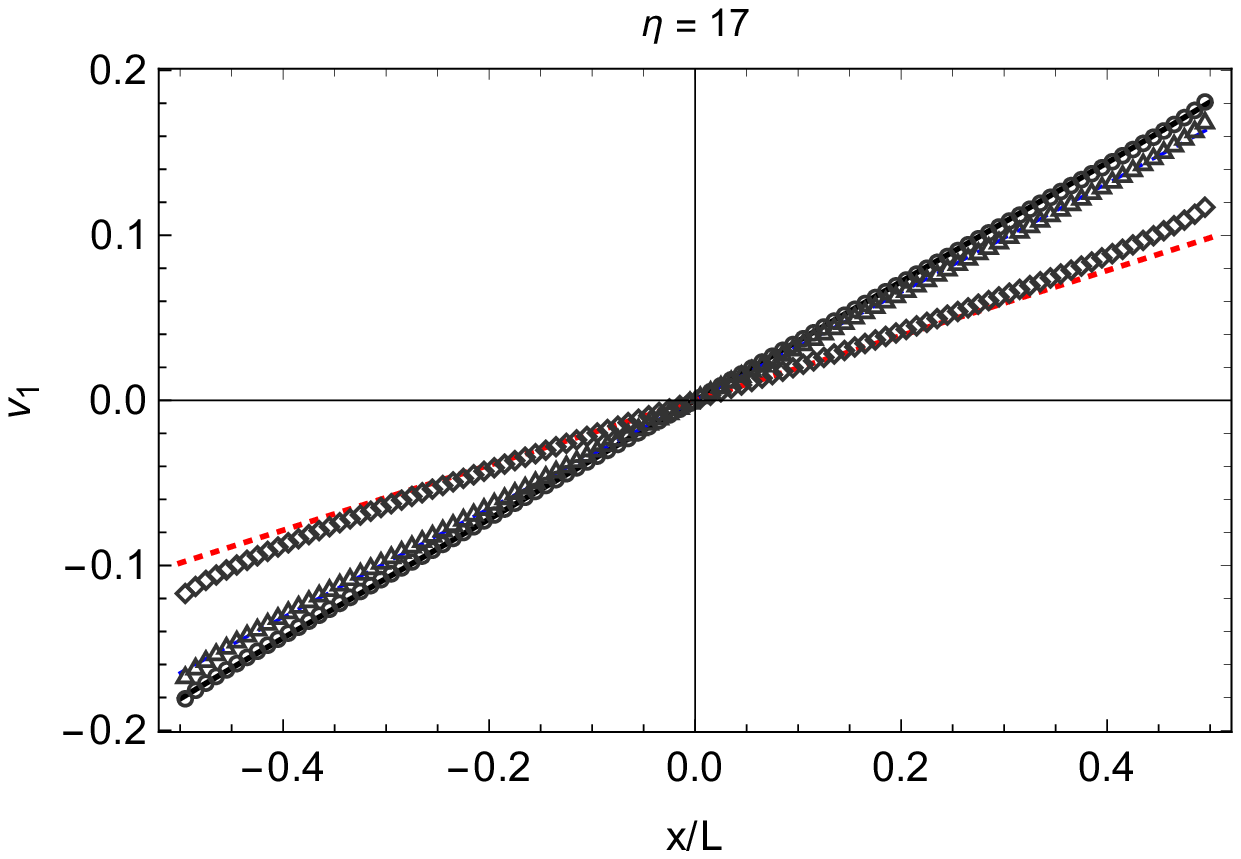}}
  {\includegraphics[width=0.32\textwidth,clip]{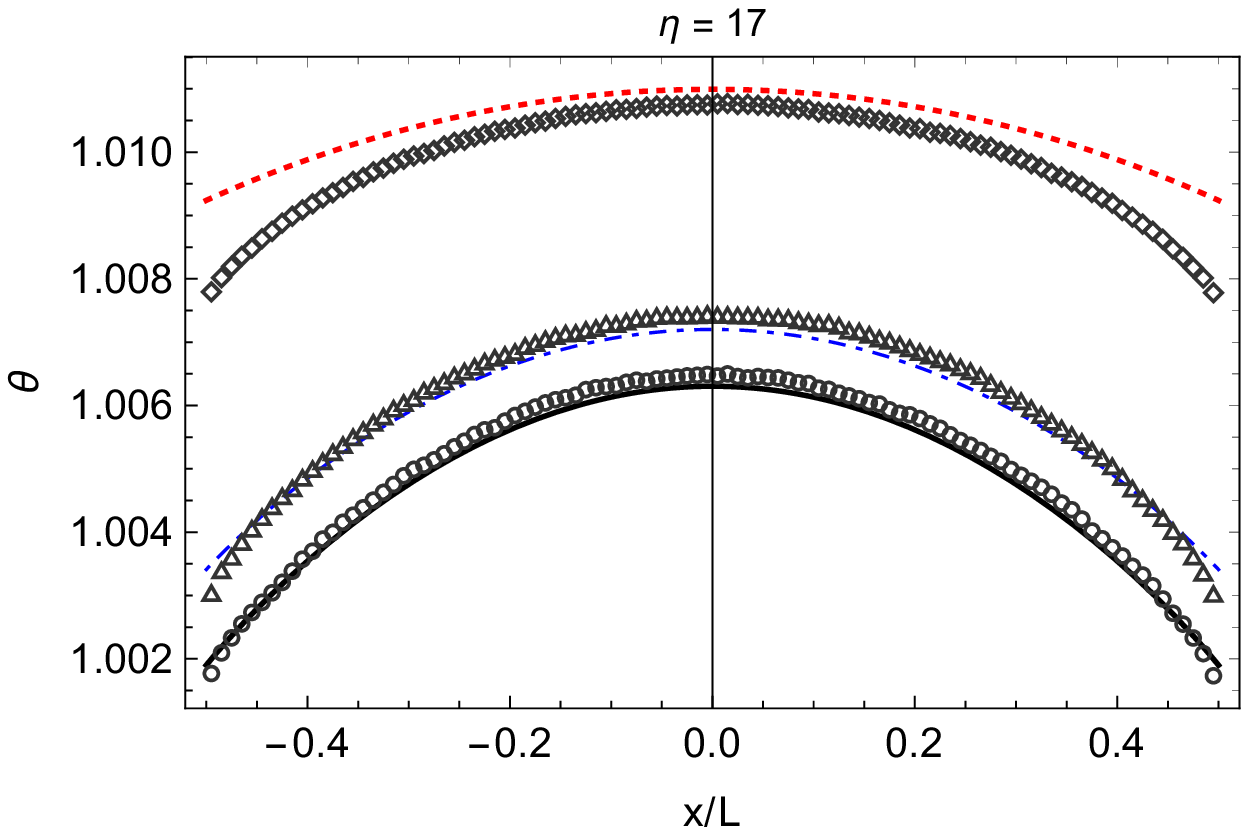}}
  {\includegraphics[width=0.32\textwidth,clip]{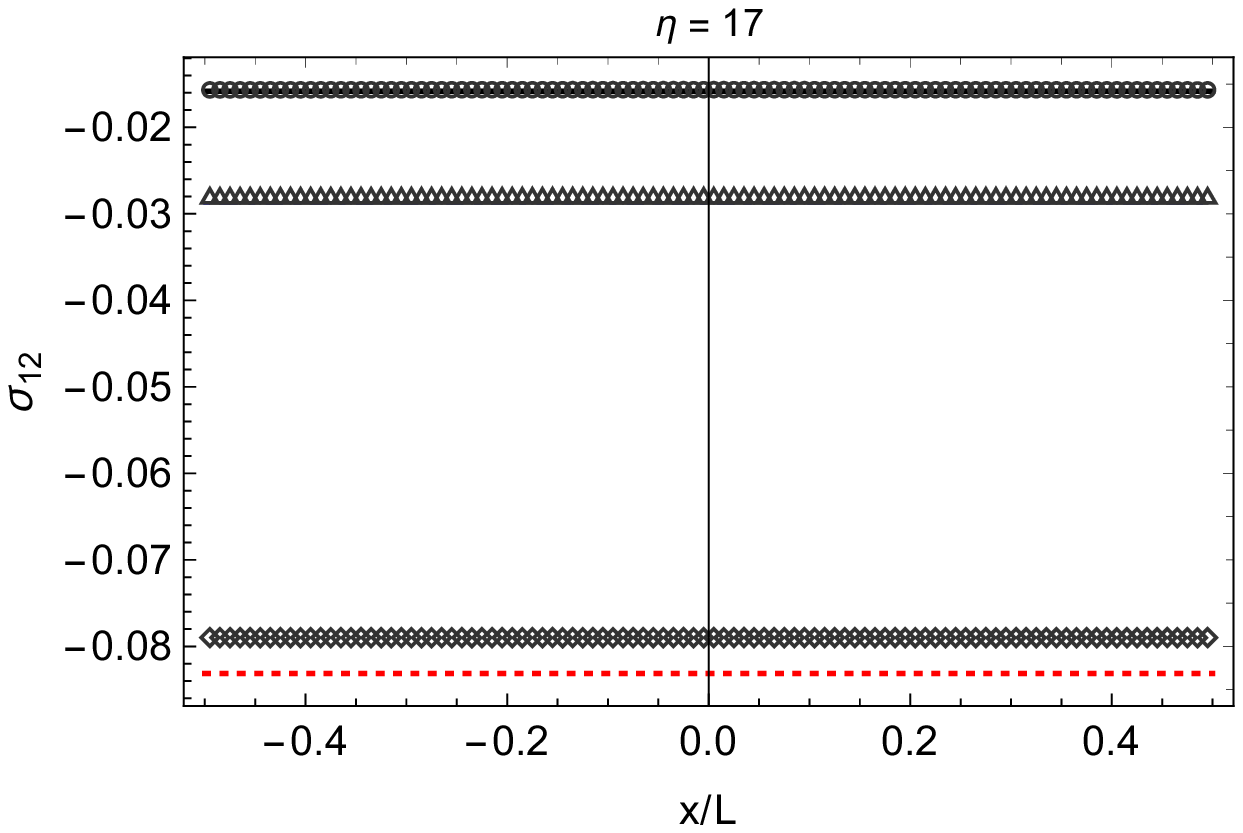}}\\
  {\includegraphics[width=0.32\textwidth,clip]{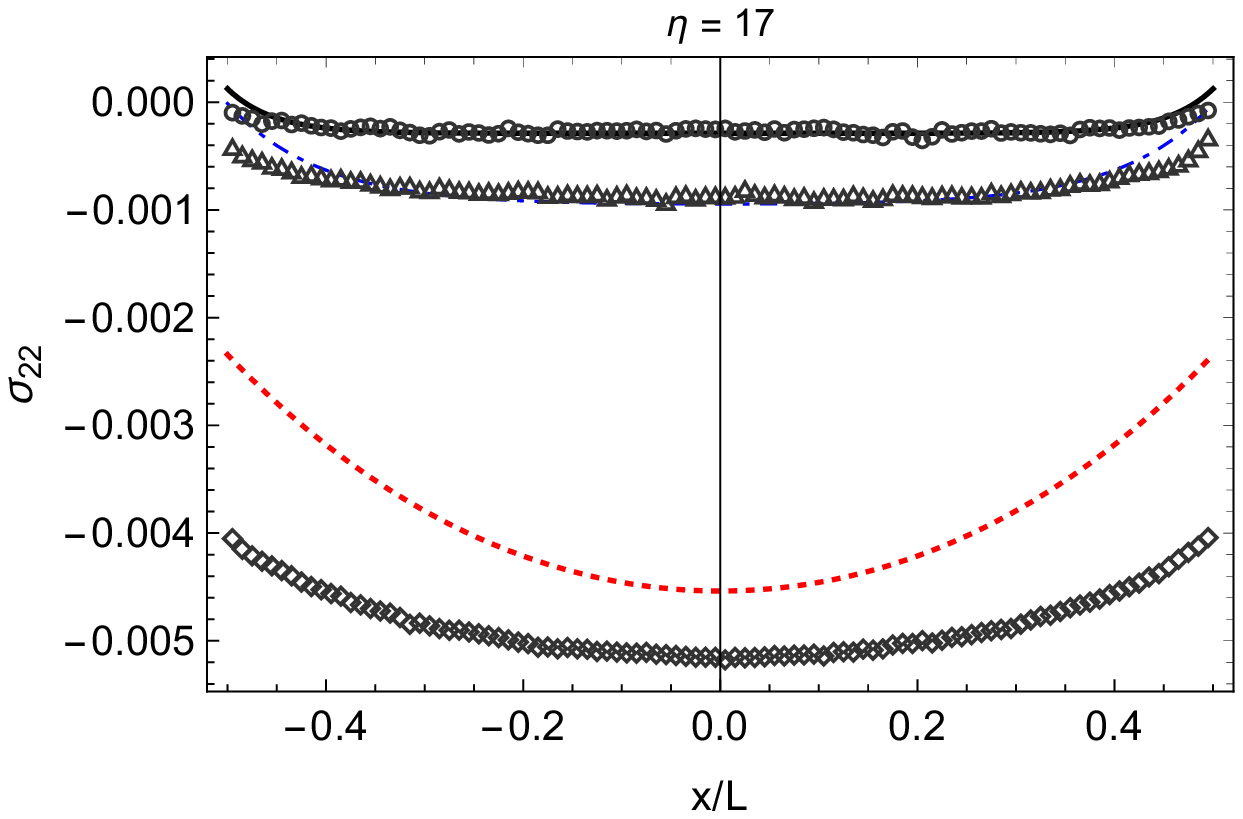}}
  {\includegraphics[width=0.32\textwidth,clip]{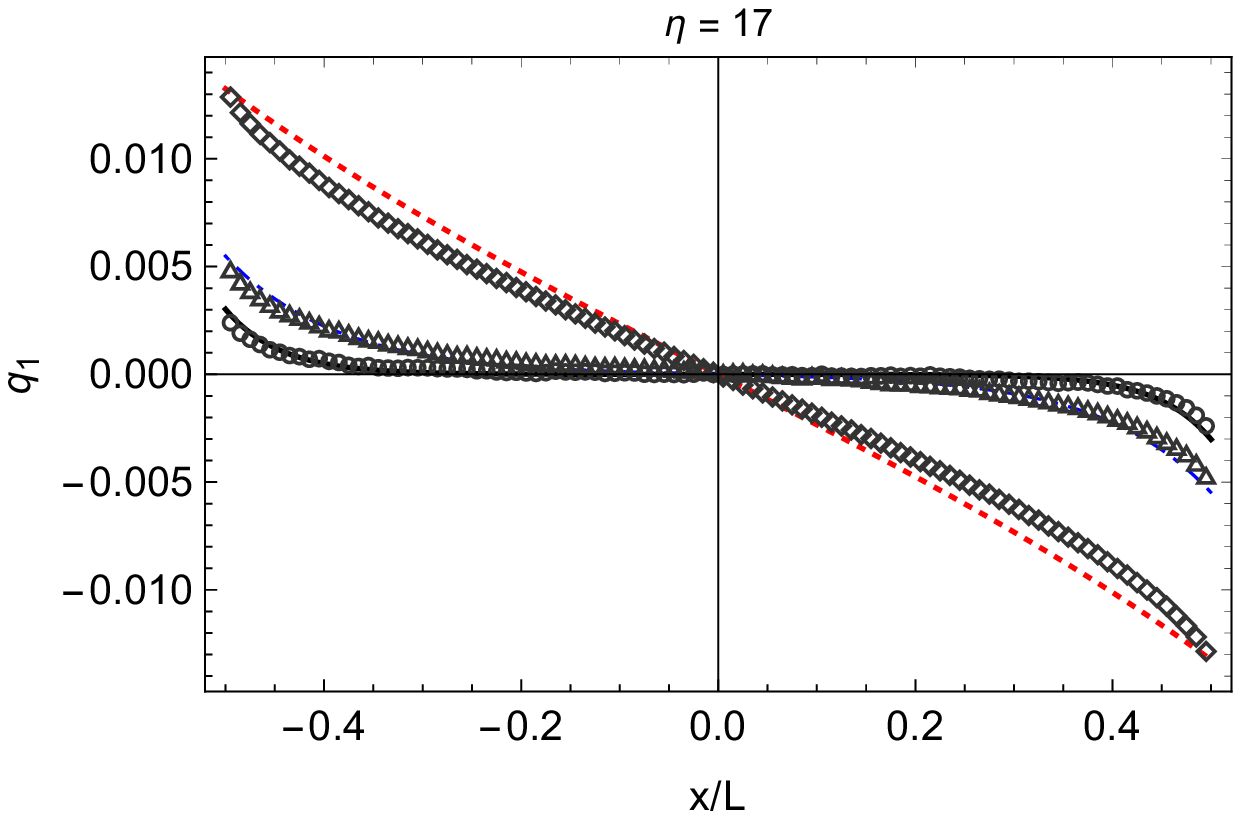}}
  {\includegraphics[width=0.32\textwidth,clip]{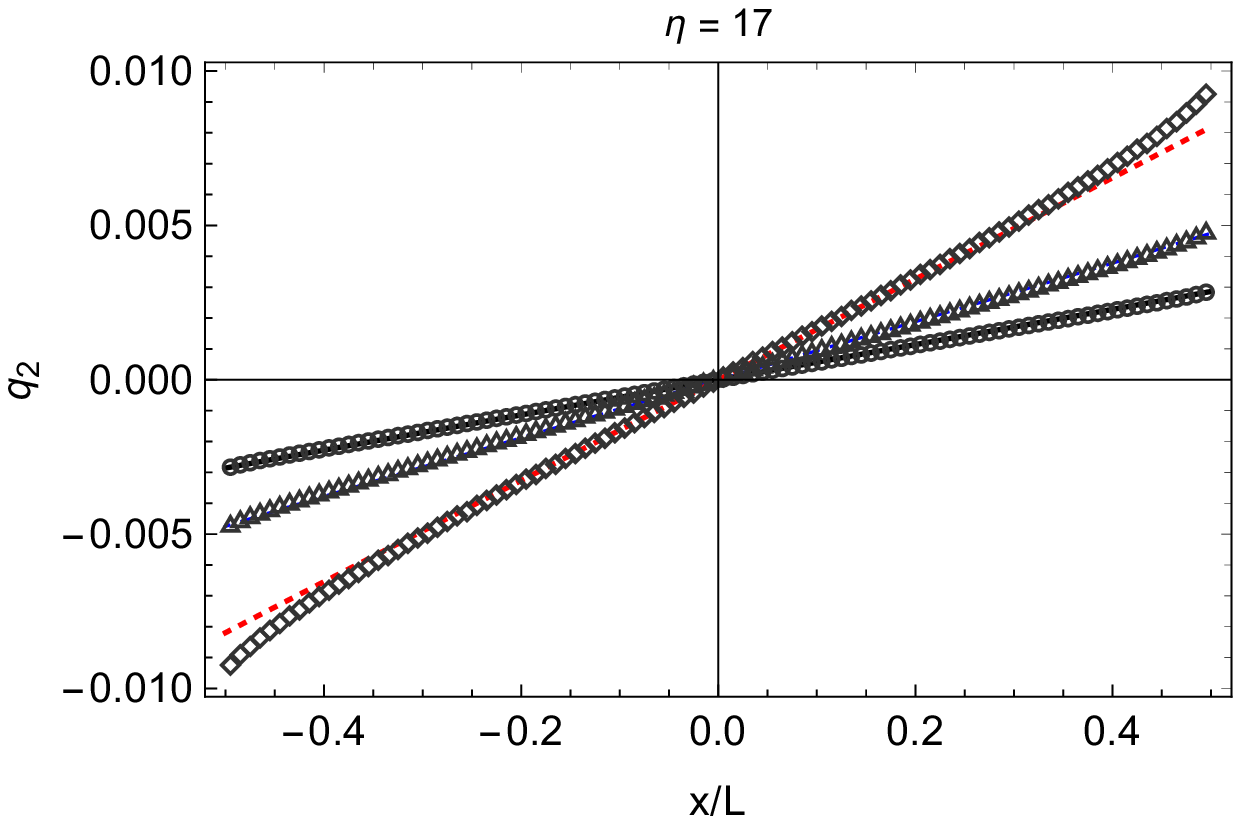}}
  \caption{Couette flow of the semi-linear R13 equations for $\eta=17$. The dimensionless semi-linear R13 solutions for $\Kn=0.05$, $0.1$, and $0.5$ are plotted in black solid line, blue dashed-dotted line, and red dotted line, respectively. The corresponding DSMC solutions are marked by circle, triangle, and diamond.}
  \label{fig:couette-eta17}
\end{figure}

We first investigate the solutions of the semi-linear R13 equations for various $\eta$ from Maxwell molecules ($\eta=5$) to hard sphere molecules ($\eta=\infty$) at $\Kn=0.05$, $0.1$, and $0.5$. The dimensionless results for $\eta=5$, $7$, $10$, $17$, and $\infty$ are
%% DELETE
% respectively shown in \figurename~\ref{fig:couette-eta5}--\ref{fig:couette-etaHS}, where each solution is
compared to the DSMC result with the corresponding viscosity index $\omega$, determine by $\eta$ via \eqref{eq:mu_omega}, that is, $\omega=1.0$, $0.83$, $0.72$, $0.625$, and $0.5$ for $\eta=5$, $7$, $10$, $17$, and $\infty$, respectively. Since the results behave very similarly, only the solutions with $\eta=5$ and $17$ are shown in \figurename~\ref{fig:couette-eta5} and \ref{fig:couette-eta17}, respectively, to avoid redundancy. As can be seen, all profiles are in good agreement with the DSMC results for small Knudsen numbers. For larger Knudsen number such as $\Kn=0.5$, some deviations away from the DSMC results are observed. The main reason might be that the rarefaction effects in such a case are too strong to be sufficiently described by the semi-linear R13 equations. Consequently, the higher-order moments should be taken into account in such a situation. For further comparison, the mean relative deviations of the semi-linear R13 solution from the DSMC solution for various $\eta$ and $\Kn$ are presented in \tablename~\ref{tab:meanRE}. It can be seen that even for $\Kn=0.5$, the mean relative deviations of velocity $v_1$, temperature $\theta$, shear stress $\sigma_{12}$, and heat flux $q_1$ and $q_2$ are quite small. For the normal stress $\sigma_{22}$, the deviations are much larger, especially when $\Kn=0.05$, in which case the maximum of the absolute value of $\sigma_{22}$ are too small. Actually, the deviations of $\sigma_{22}$ are mainly from the region near the boundaries, where the semi-linear R13 equations may be insufficient due to the strong nonequilibrium near the boundary, as can be observed in \figurename~\ref{fig:couette-eta5} and \ref{fig:couette-eta17}. For $\Kn \leqslant 0.1$, the quantitative result in the center of the domain is still reliable.

\begin{table}[!htb]
  \centering
\begin{tabular}{ll@{\hspace{10pt}}c@{\hspace{10pt}}c@{\hspace{10pt}}c@{\hspace{10pt}}c@{\hspace{10pt}}c@{\hspace{10pt}}c}
\hline
\hline
$\Kn$ & $\eta$ & $v_1$ & $\theta$ & $\sigma_{22}$ & $\sigma_{12}$ & $q_1$ & $q_2$ \\
\hline
% Kn=0.05
% 0.504314,0.0248885,36.148,1.24408,4.38944,1.05243
% 0.477283,0.0172404,50.3771,1.18266,4.18146,1.01452
% 0.474931,0.0139842,95.5228,1.18753,4.17887,1.02058
% 0.42257,0.0154783,39.9675,1.08356,4.85913,0.892859
% 0.410876,0.0116896,203.078,1.12309,3.6683,0.954018
% Kn=0.1
% 0.611883,0.0220108,18.1215,2.33579,5.23698,1.32005
% 0.584592,0.0243622,19.2104,2.17789,4.36241,1.32524
% 0.580271,0.0211703,19.4069,2.03625,4.13173,1.29113
% 0.569731,0.0169785,22.1239,1.99617,3.70424,1.29128
% 0.545932,0.0191938,21.5139,1.91065,4.18662,1.27181
% Kn=0.5
% 6.54581,0.0670141,35.1178,5.46227,5.34948,4.43476
% 5.47165,0.0603237,29.0008,5.38051,6.60053,3.32292
% 4.89548,0.0539227,26.0099,5.3085,6.68775,2.77192
% 4.46119,0.0511077,24.8355,5.25203,6.32153,2.31181
% 3.94784,0.044887,24.215,5.14289,5.18853,1.82963
0.05 & 5        & 0.504\% & 0.025\% & 36.15\% & 1.244\% & 4.389\% & 1.052\% \\
     & 7        & 0.477\% & 0.017\% & 50.38\% & 1.183\% & 4.181\% & 1.015\% \\
     & 10       & 0.475\% & 0.014\% & 95.52\% & 1.188\% & 4.179\% & 1.021\% \\
     & 17       & 0.423\% & 0.015\% & 39.97\% & 1.084\% & 4.859\% & 0.893\% \\
     & $\infty$ & 0.411\% & 0.012\% & 203.1\% & 1.123\% & 3.668\% & 0.954\% \\
\hline
0.1  & 5        & 0.612\% & 0.022\% & 18.12\% & 2.336\% & 5.237\% & 1.320\% \\
     & 7        & 0.585\% & 0.024\% & 19.21\% & 2.178\% & 4.362\% & 1.325\% \\
     & 10       & 0.580\% & 0.021\% & 19.41\% & 2.036\% & 4.132\% & 1.291\% \\
     & 17       & 0.570\% & 0.017\% & 22.12\% & 1.996\% & 3.704\% & 1.291\% \\
     & $\infty$ & 0.546\% & 0.019\% & 21.51\% & 1.911\% & 4.187\% & 1.272\% \\
\hline
0.5  & 5        & 6.546\% & 0.067\% & 35.12\% & 5.462\% & 5.349\% & 4.435\% \\
     & 7        & 5.472\% & 0.060\% & 29.00\% & 5.381\% & 6.601\% & 3.323\% \\
     & 10       & 4.895\% & 0.054\% & 26.01\% & 5.309\% & 6.688\% & 2.772\% \\
     & 17       & 4.461\% & 0.051\% & 24.84\% & 5.252\% & 6.322\% & 2.312\% \\
     & $\infty$ & 3.948\% & 0.045\% & 24.22\% & 5.143\% & 5.189\% & 1.830\% \\
\hline
\hline
\end{tabular}
\caption{Mean relative deviations of the semi-linear R13 solution from the DSMC solution for Couette flow with various $\Kn$ and $\eta$.}
 \label{tab:meanRE}
\end{table}

In \cite{Taheri2009}, the semi-linear R13 equations for Maxwell molecules are used to simulate the gas with viscosity index $\omega \neq 1.0$. To this end, the coefficients of the equations including $\alpha_{i,j}^{(\eta)}$ and $\beta_{i,j}^{(\eta)}$ remain the same as those for $\eta=5$, while the viscosity index $\omega$, which determines the value of $\Kn_0$, is adjusted according to the simulated gas. In other words, $\eta$ and $\omega$ are viewed as two independent parameters, i.e., $\omega \neq \frac{1}{2}+\frac{2}{\eta-1}$, in the simulation. We would now like to study the difference between such a strategy and our solution of the ``genuine'' R13 equations for the general inverse power law models. In our test, we fix $\omega$ to be $0.5$, and consider the solutions of R13 equations for $\eta = 5$, $7$, $10$, $17$, and $\infty$. The solutions are compared with the DSMC solutions with viscosity index $\omega = 0.5$.
%With the same strategy, we investigate the variation of the solutions for the semi-linear R13 equations in terms of $\eta$ when the viscosity index is fixed to $\omega=0.5$, to show that if the semi-linear R13 equations with an altered viscosity index could always give satisfactory results. %\zc{It seems that we have not clearly described the models used to solve this lines. Maybe we should provide some formulas here? Add a few words on the purpose of such a comparison? I think it is because the R13 model for $\eta=5$ is used in the previous literature with altered viscosity index to simulate other gas molecules.}

\begin{figure}[!htb]
% \colorbox{black}
  \centering
  {\includegraphics[width=0.32\textwidth,clip]{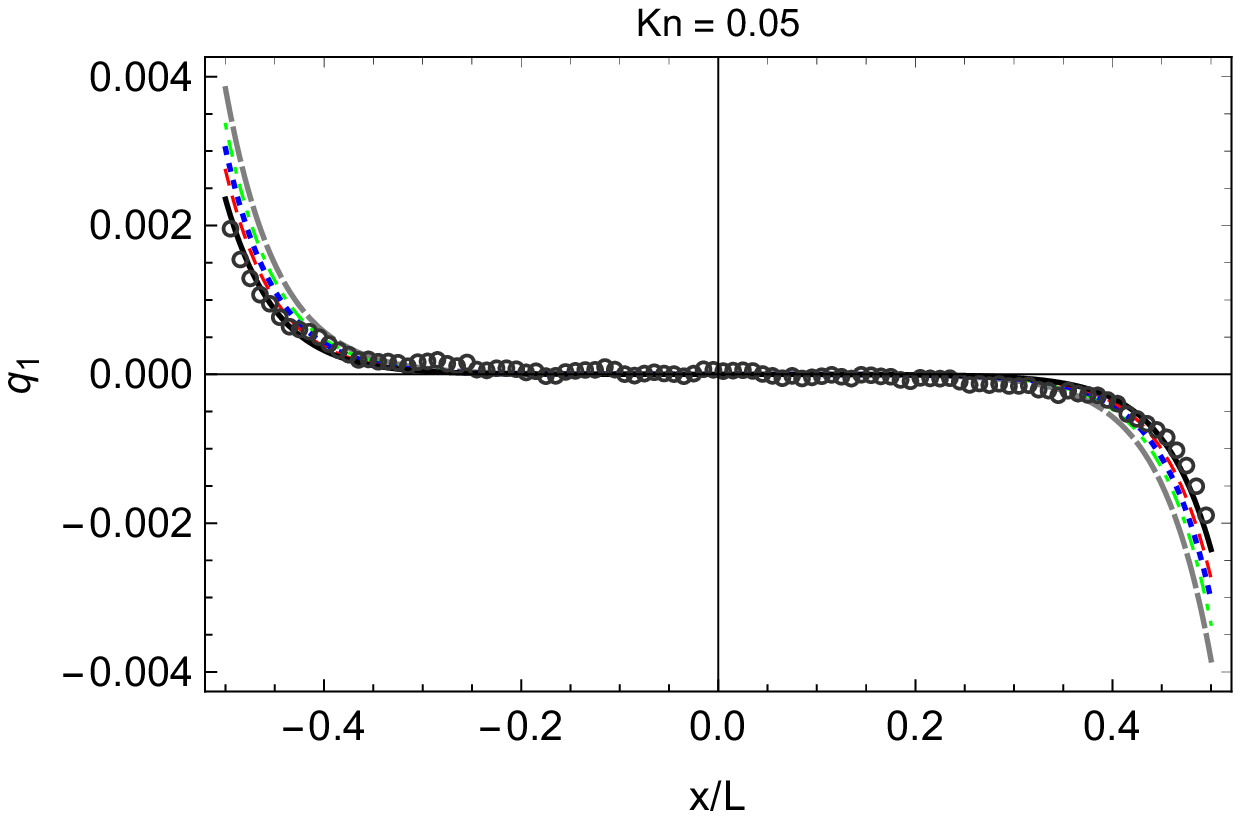}}
  {\includegraphics[width=0.32\textwidth,clip]{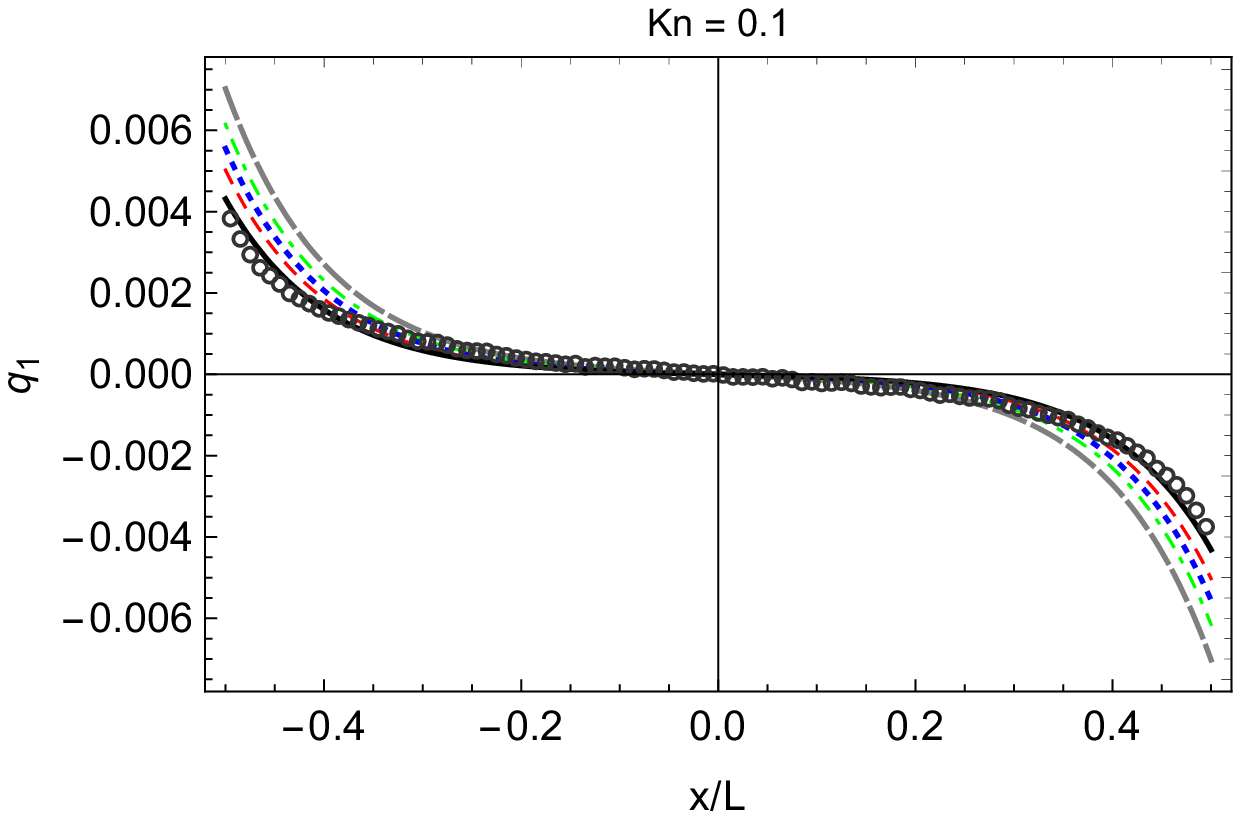}}
  {\includegraphics[width=0.32\textwidth,clip]{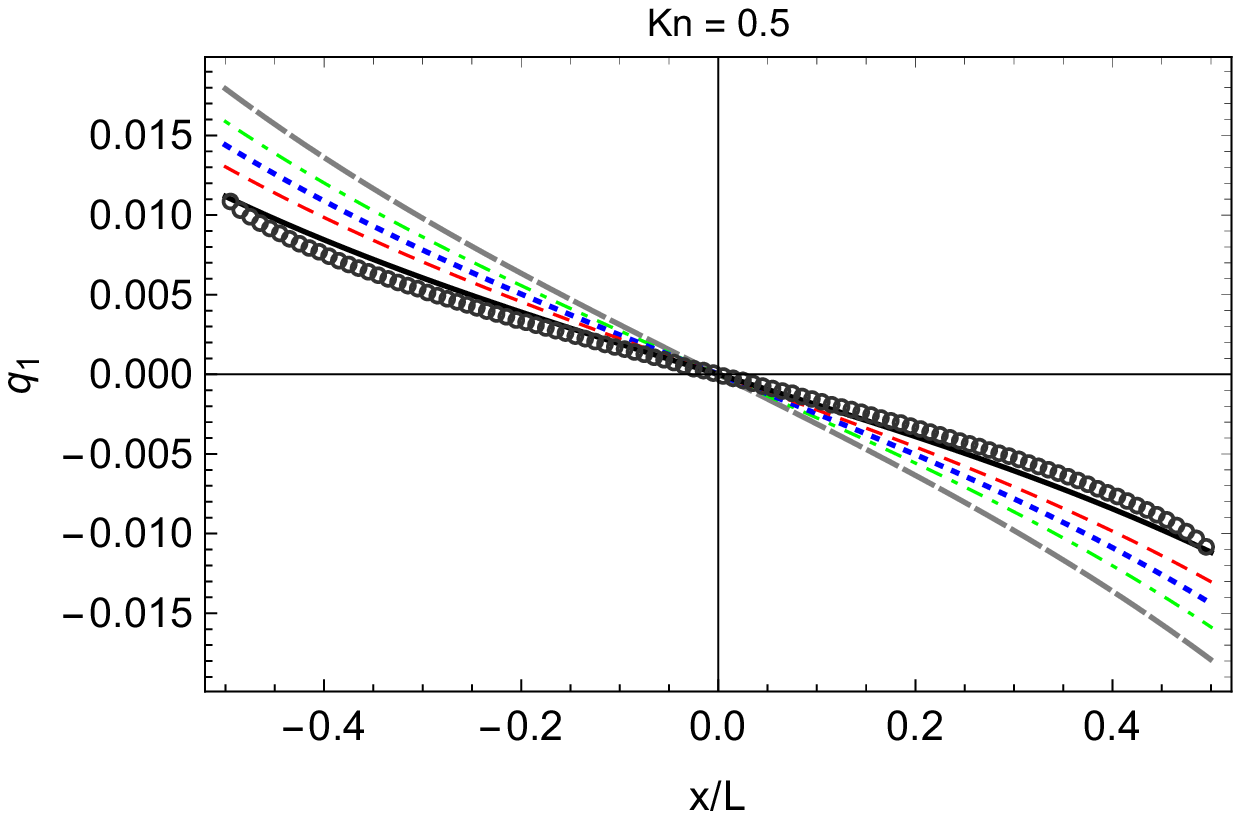}}
  \caption{Dimensionless heat flux $q_1$ for Couette flow with $\omega=0.5$. The semi-linear R13 solutions are plotted in gray dashed line, green dashed-dotted line, blue dotted line, red dashed line, and black solid line for $\eta=5, 7, 10, 17$, and $\infty$, respectively. The DSMC solutions are marked by circle.}
  \label{fig:couette-veta}
\end{figure}

For a number of quantities, the results for different $\eta$ are very close to each other. To avoid massive plots, we only provide profiles for the heat flux $q_1$, which is the moment that shows the most significant difference. The dimensionless plots for $\Kn=0.05$, $0.1$, and $0.5$ are presented in \figurename~\ref{fig:couette-veta}, where the DSMC results with $\omega=0.5$ are provided for comparison. The disagreement between different models is especially obvious near the boundaries, where the R13 equations for Maxwell molecules predict heat flux about 1.8 times as large as the DSMC result, while the R13 equations for hard-sphere molecules can provide satisfactory accuracy. This phenomenon is mainly due to the variation of the coefficient $\C_2$ in \eqref{eq:couette-sol-vq}, which can be determined from the linear system of $\C_1$ and $\C_2$ by inserting \eqref{eq:couette-sol-vq} into \eqref{eq:bc4} and \eqref{eq:bc3}. It is found that with the fixed viscosity index $\C_2$ decreases as $\eta$ increases. In particular, $\C_2$ for $\eta=\infty$ is reduced by more than a half compared with the value of $\C_2$ for $\eta=5$. Noting that the DSMC code is implemented for variable-hard-sphere gas and $\omega=0.5$ corresponds to the hard sphere molecules ($\eta=\infty$) exactly from \eqref{eq:mu_omega}, we expect that the semi-linear R13 solutions with $\eta=\infty$ gives the best results in comparison to the DSMC results. Indeed, it is observed from \figurename~\ref{fig:couette-veta} that the profiles of $q_1$ become closer to the DSMC solutions as $\eta$ increases from $5$ to $\infty$ for the given three Knudsen numbers. This experiment confirms the advantage of using R13 equations derived from the Boltzmann equation with original collision terms.

\begin{figure*}[!htb]
% \colorbox{black}
  \centering
  {\includegraphics[width=0.3\textwidth,clip]{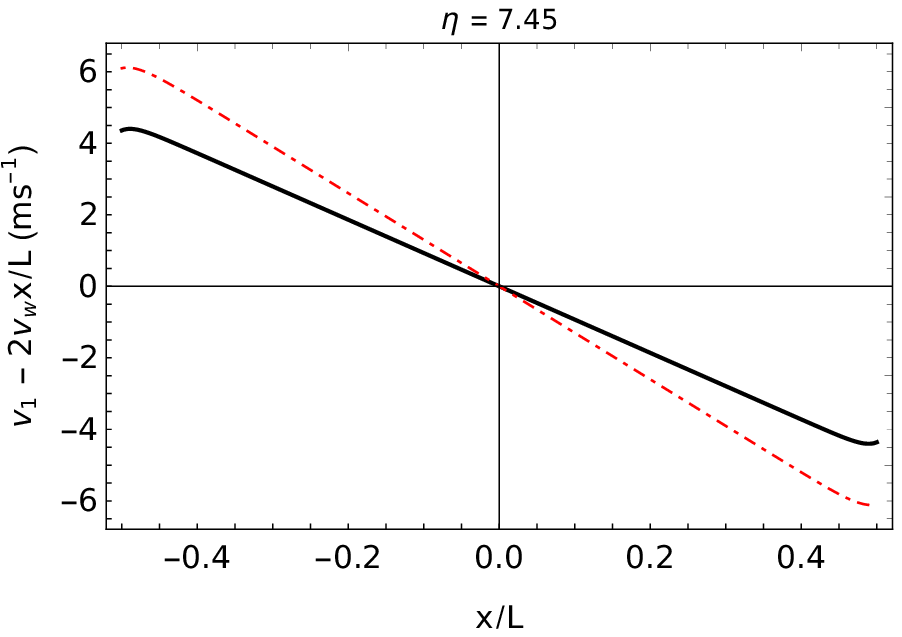}}
  {\includegraphics[width=0.305\textwidth,clip]{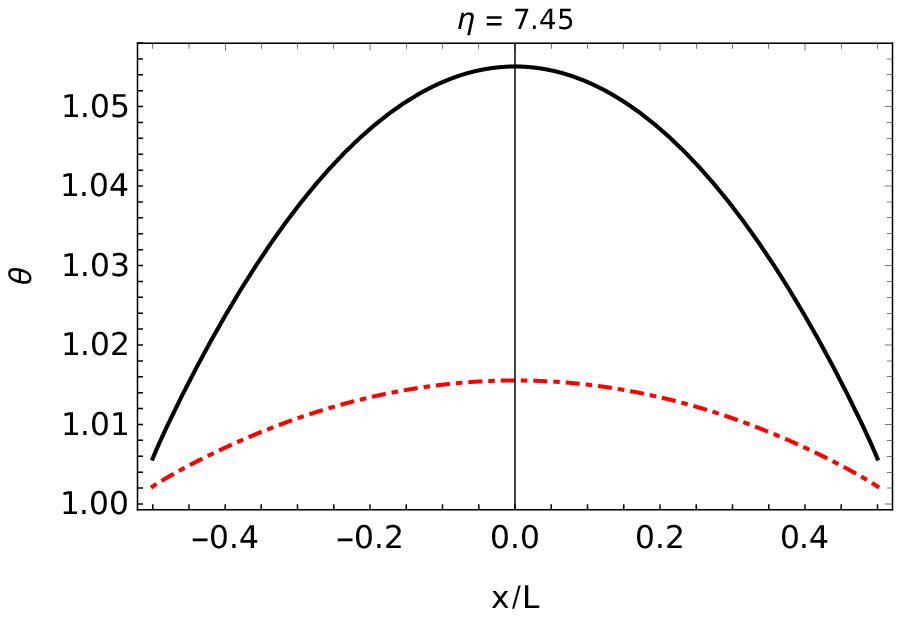}}
  {\includegraphics[width=0.32\textwidth,clip]{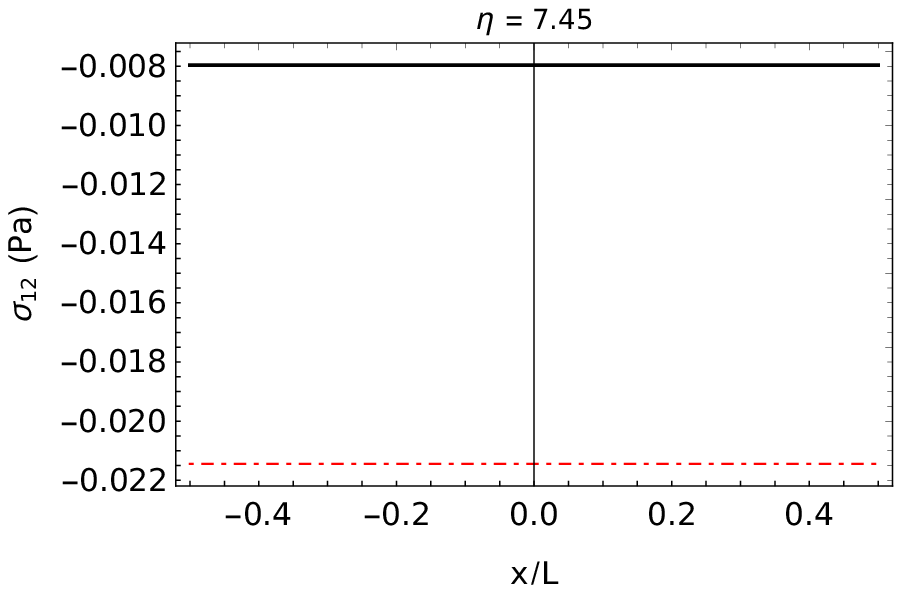}}
  \caption{The results of the semi-linear R13 equations for the Couette flow Case 1 (black solid line) and Case 3 (red dashed-dotted line).}
  \label{fig:VHS}
\end{figure*}

To study the behavior of the semi-linear R13 equations with larger wall speeds, we consider two cases in \cite[Section IV]{Weaver2014}, where both the variable hard sphere (VHS) model and the Lennard-Jones polynomial approximation (LJPA) model are considered. We set our parameters to be the same as Case 1 and Case 3 in \cite{Weaver2014}, which are detailed in Table \ref{tab:VHS}. In Figure. \ref{fig:VHS}, we plot the results in the same dimensions as in Ref. \cite{Weaver2014} so that the readers can make comparisons conveniently. In general, our results are close the VHS results computed in \cite{Weaver2014}, while the difference from the LJPA model is significant especially for the high-temperature case. Such deviation is expected to be from the lack of attractive potential in the inverse power law model. Fixing this requires further development of R13 equations for Lennard-Jones models.

  \begin{table}[!htb]
  \centering
\begin{tabular}{l@{\hspace{10pt}} c@{\hspace{10pt}} c}
\hline
\hline
    Quantities & Case 1 & Case 3  \\
\hline
     Wall temperature, $T_w$ (K) & 273    & 1000 \\
     Initial number density, $n$ ($\text{m}^{-3}$) & $1.4 \times 10^{20}$ & $1.4 \times 10^{20}$\\
     Left wall velocity, $v_W^l$ ($\text{ms}^{-1}$) & 150 & 150\\
     Right wall velocity, $v_W^r$ ($\text{ms}^{-1}$) & $-150$ & $-150$\\
     Mach number, $M$ & 0.97 & 0.51 \\
     Knudsen number, $\Kn$ & 0.012 & 0.017\\
     Wall separation, $L$ (m) & 1 & 1\\
     $\omega$ & 0.81 & 0.81\\
     $\eta$ & 7.45 & 7.45\\
\hline
\hline
\end{tabular}
\caption{Parameter settings for Couette flows.}
 \label{tab:VHS}
\end{table}

% \begin{figure}[!htb]
% % \colorbox{black}
%   \centering
%   {\includegraphics[width=0.32\textwidth,clip]{couette_etaHS_uW0p2_v1.eps}}
%   {\includegraphics[width=0.32\textwidth,clip]{couette_etaHS_uW0p2_theta.eps}}
%   {\includegraphics[width=0.32\textwidth,clip]{couette_etaHS_uW0p2_sigma12.eps}}\\
%   {\includegraphics[width=0.32\textwidth,clip]{couette_etaHS_uW0p2_sigma22.eps}}
%   {\includegraphics[width=0.32\textwidth,clip]{couette_etaHS_uW0p2_q1.eps}}
%   {\includegraphics[width=0.32\textwidth,clip]{couette_etaHS_uW0p2_q2.eps}}
%   \caption{Couette flow of the semi-linear R13 equations for $\eta=\infty$ (hard sphere molecules). The dimensionless semi-linear R13 solutions for $\Kn=0.05$, $0.1$, and $0.5$ are plotted in black solid line, blue dashed-dotted line, and red dotted line, respectively. The corresponding DSMC solutions are marked by circle, triangle, and diamond.}
%   \label{fig:couette-etaHS}
% \end{figure}

\subsection{Fourier flow}
For the Fourier flow, the gas is driven solely by the difference of wall temperatures. In our simulations, the wall temperatures are set as $T_W^l=T_0$ and $T_W^r=1.2 T_0$, i.e., the corresponding dimensionless wall temperatures are $\theta_W^l=1$ and $\theta_W^r=1.2$ respectively.

\begin{figure}[!htb]
% \colorbox{black}
  \centering
  {\includegraphics[width=0.32\textwidth,clip]{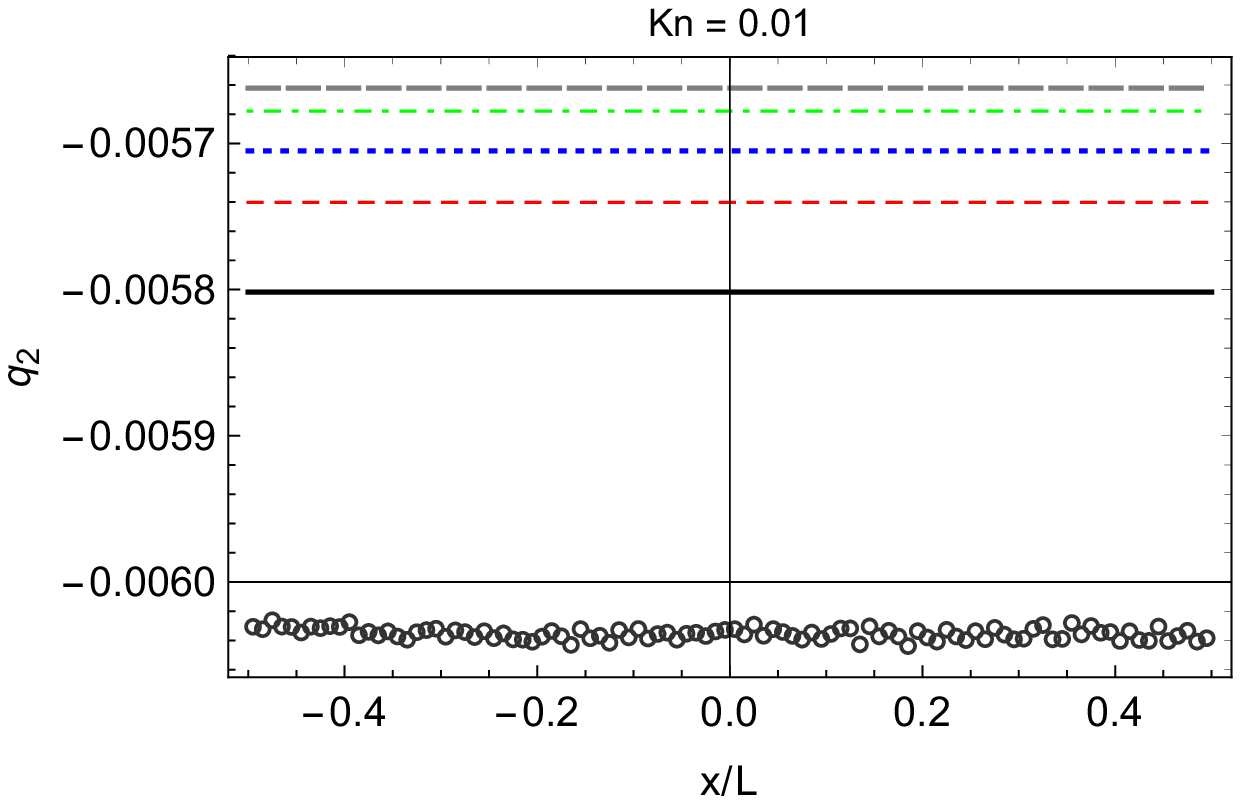}}
  {\includegraphics[width=0.32\textwidth,clip]{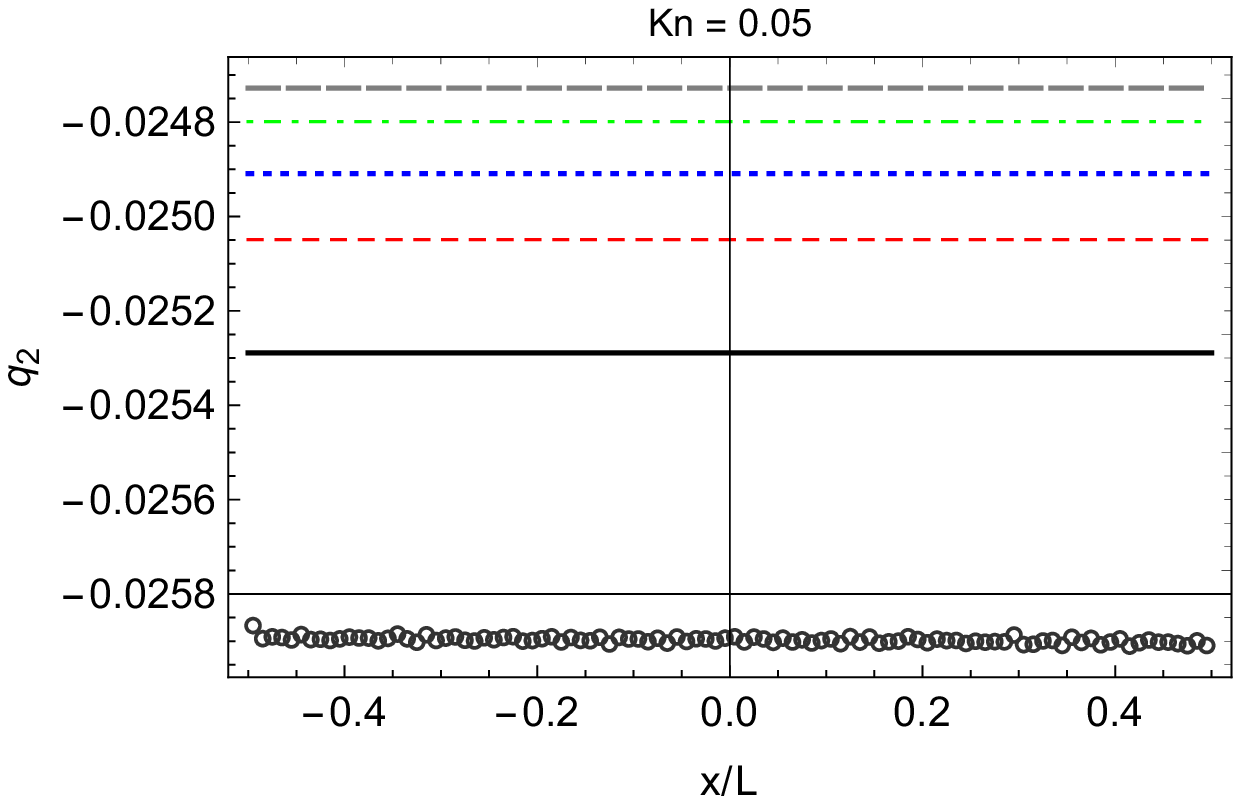}}
  {\includegraphics[width=0.32\textwidth,clip]{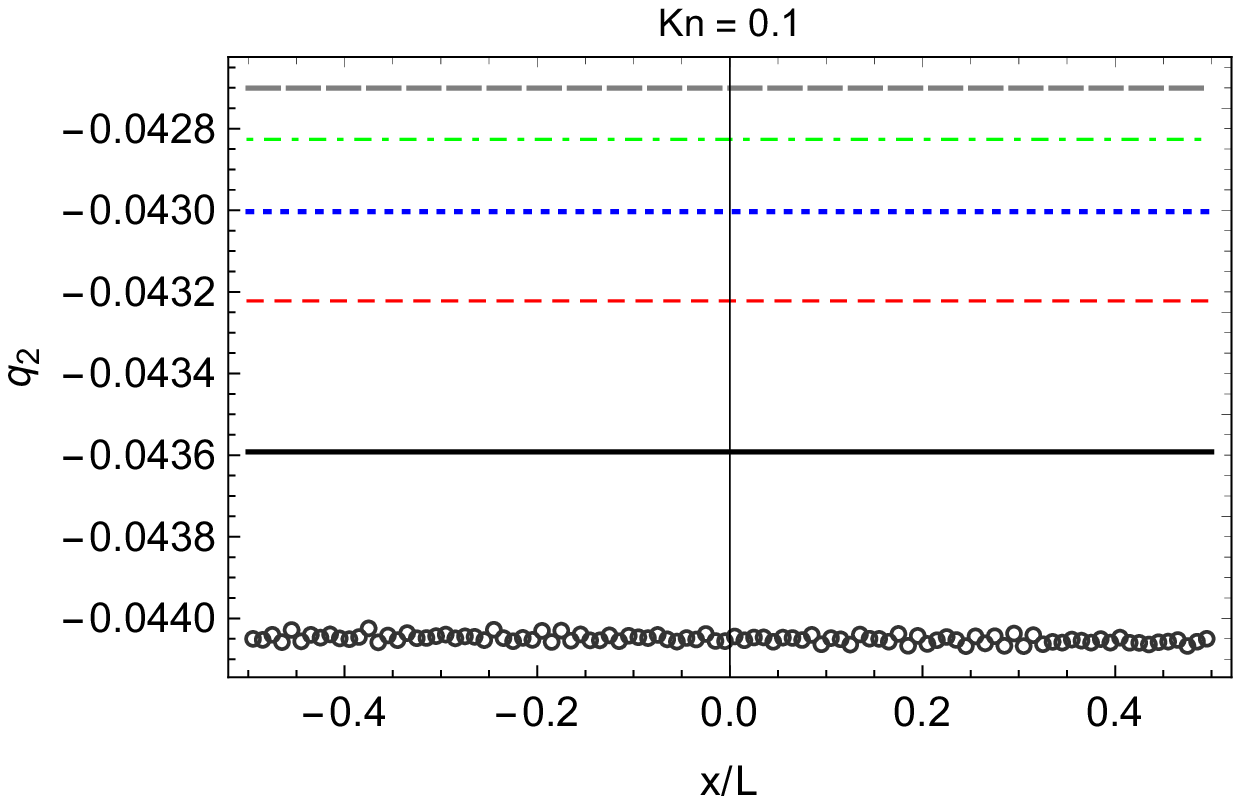}}
  \caption{Dimensionless heat flux $q_2$ for Fourier flow with $\omega=0.5$. The semi-linear R13 solutions are plotted in gray dashed line, green dashed-dotted line, blue dotted line, red dashed line, and black solid line for $\eta=5, 7, 10, 17$, and $\infty$, respectively. The DSMC solutions are marked by circle.}
  \label{fig:fourier-veta}
\end{figure}

For this example, we first repeat the experiment that fixes the viscosity index $\omega$ and compares the solutions of R13 equation for different $\eta$. Again we choose $\omega = 0.5$ and $\eta=5$, $7$, $10$, $17$, and $\infty$, but the Knudsen numbers are selected as $\Kn=0.01$, $0.05$, and $0.1$. Due to the simplicity of the flow structure, the solutions for different models are quite close to each other. Here in \figurename~\ref{fig:fourier-veta}, we plot the dimensionless profiles for $q_2$ as well as the DSMC results for the hard-sphere model. It is observed that compared with the model for Maxwell molecules, the hard-sphere R13 equations reduce the error of $q_2$ by a half for $\Kn = 0.05$ or two thirds for $\Kn = 0.1$. This again implies that for $\eta$ close to $5$, we can use the R13 equations for Maxwell gases, which have a much simpler form, to approximate the dynamics for the inverse power law gases, while when $\eta$ is large, such approximation may lead to significant modelling errors.

\begin{figure}[!htb]
% \colorbox{black}
  \centering
  {\includegraphics[width=0.32\textwidth,clip]{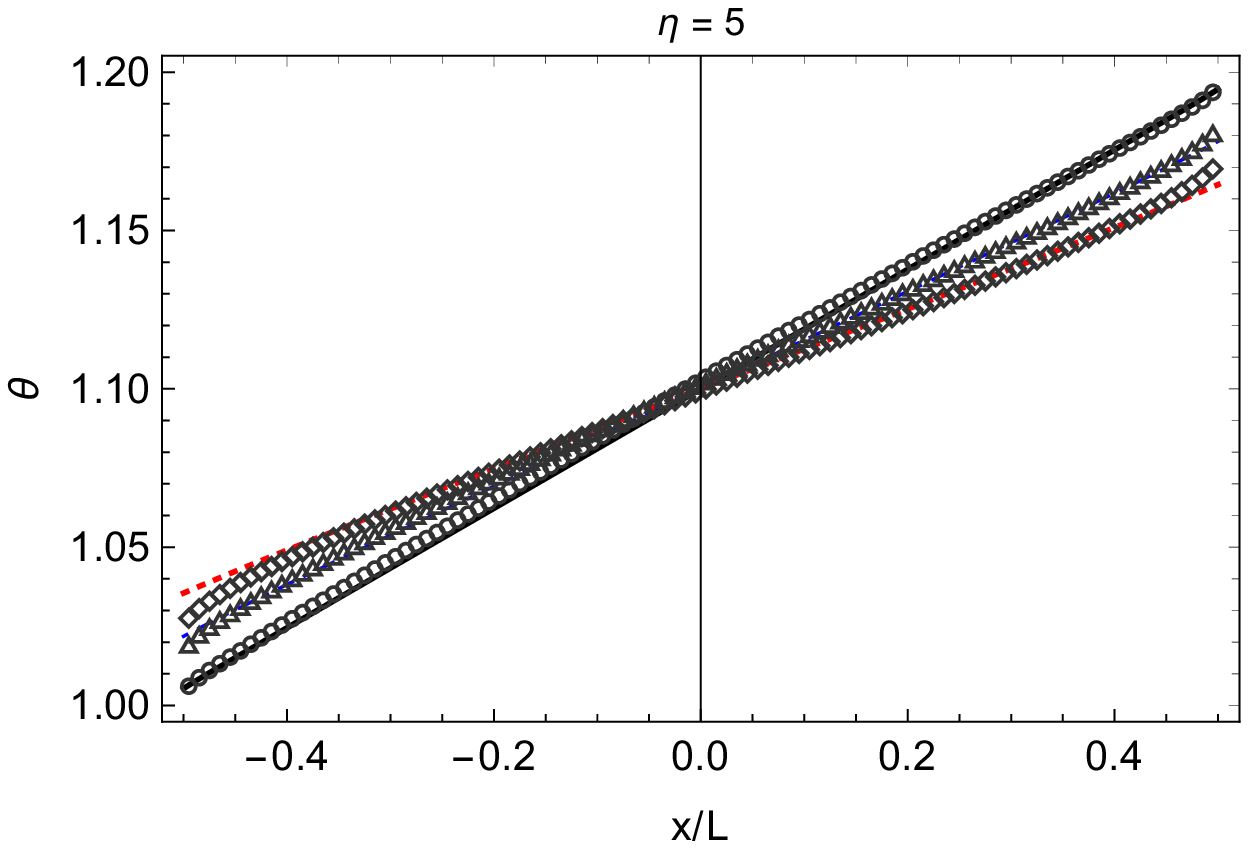}}
  {\includegraphics[width=0.32\textwidth,clip]{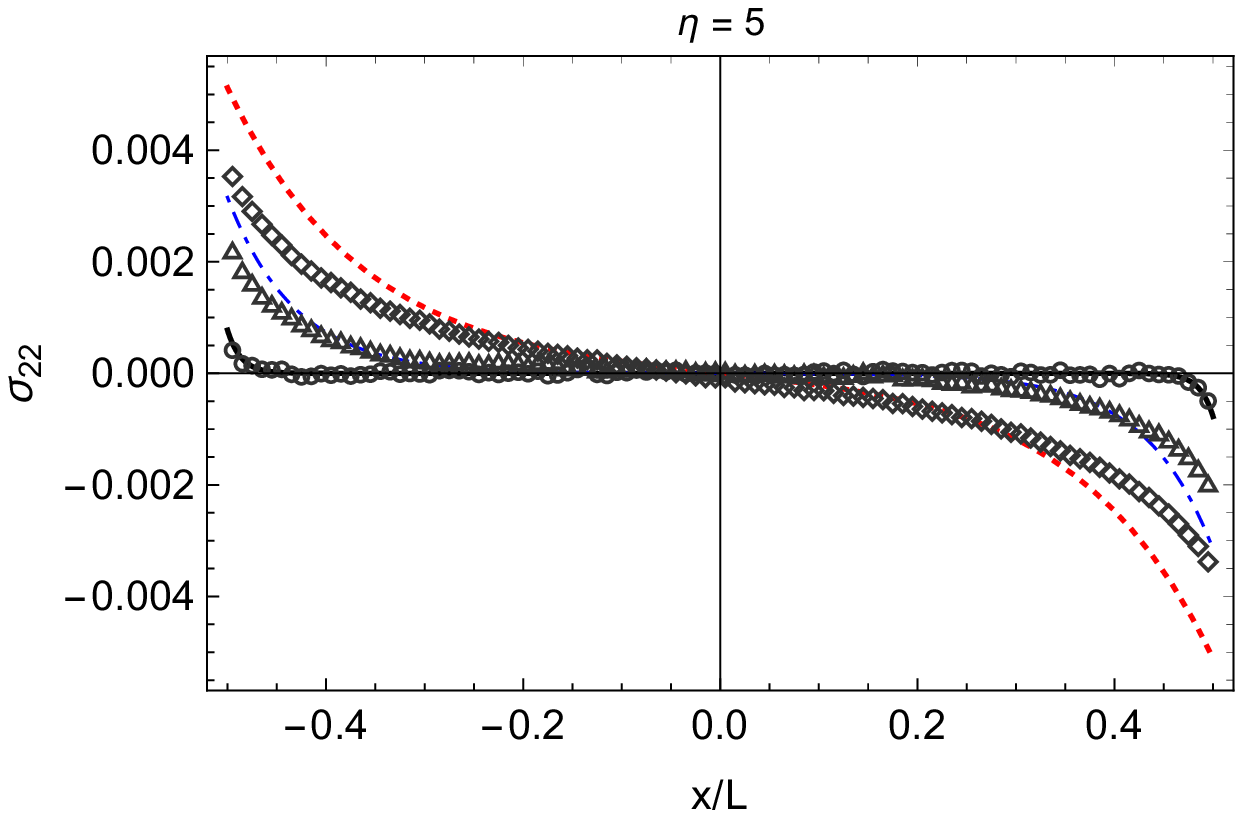}}
  {\includegraphics[width=0.32\textwidth,clip]{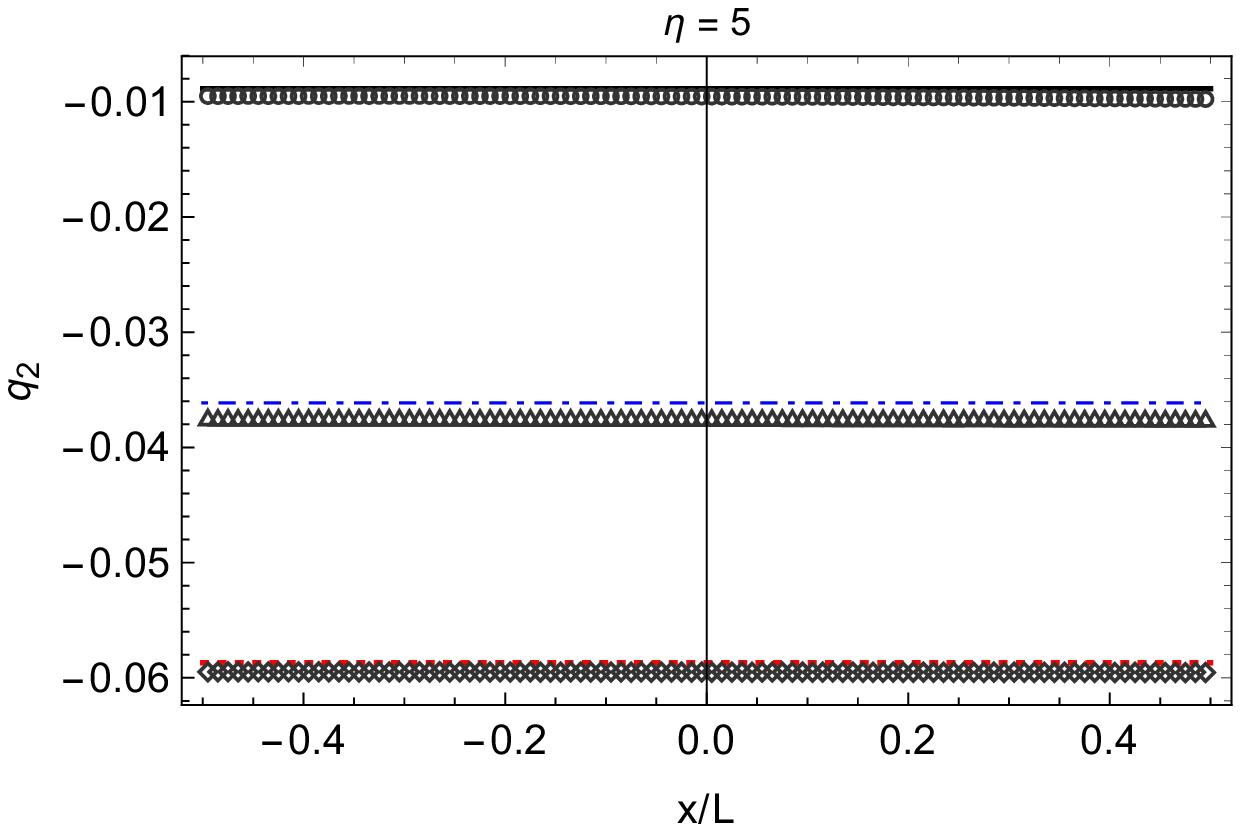}}
  \caption{Fourier flow of the semi-linear R13 equations with $\eta=5$. The dimensionless semi-linear R13 solutions for $\Kn=0.01$, $0.05$, and $0.1$ are plotted in black solid line, blue dashed-dotted line, and red dotted line, respectively. The corresponding DSMC solutions are marked by circle, triangle, and diamond.}
  \label{fig:fourier-eta5}
\end{figure}

\begin{figure}[!htb]
% \colorbox{black}
  \centering
  {\includegraphics[width=0.32\textwidth,clip]{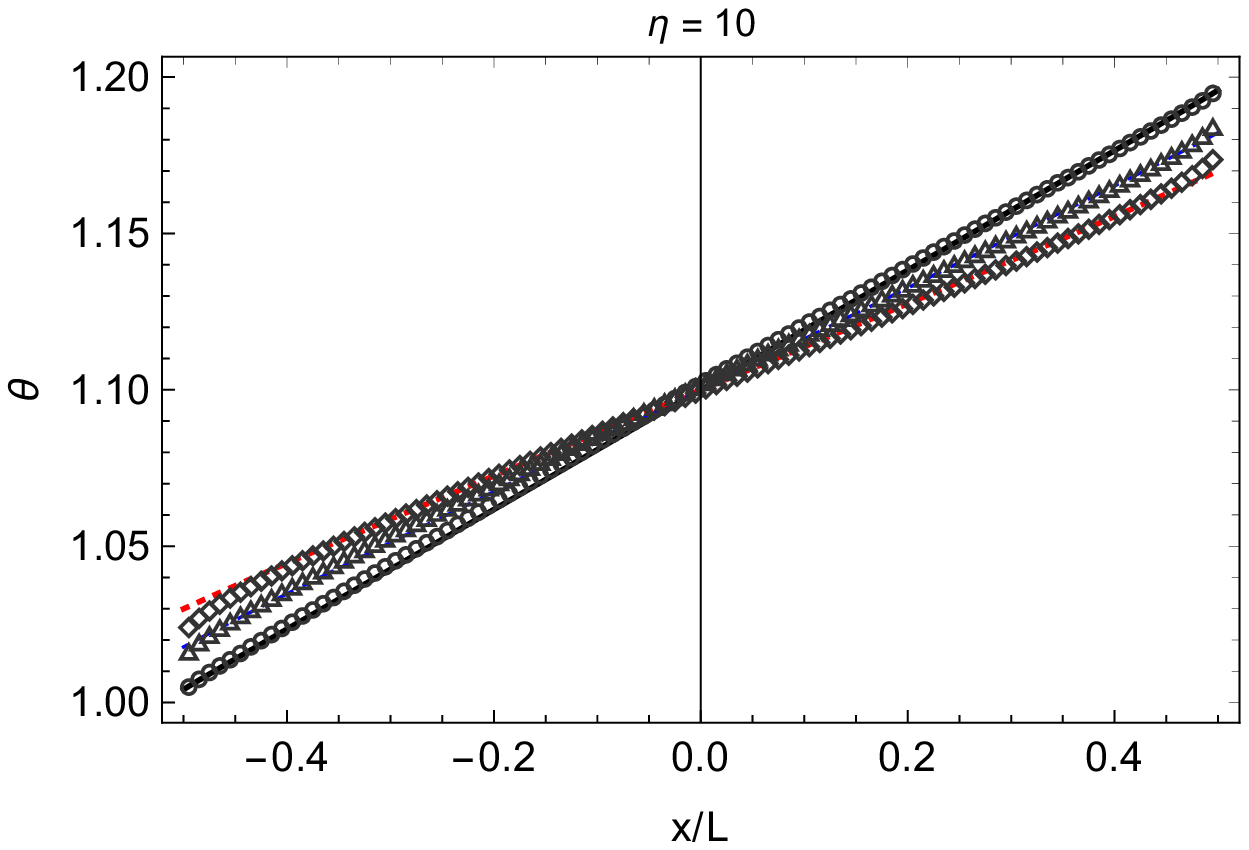}}
  {\includegraphics[width=0.32\textwidth,clip]{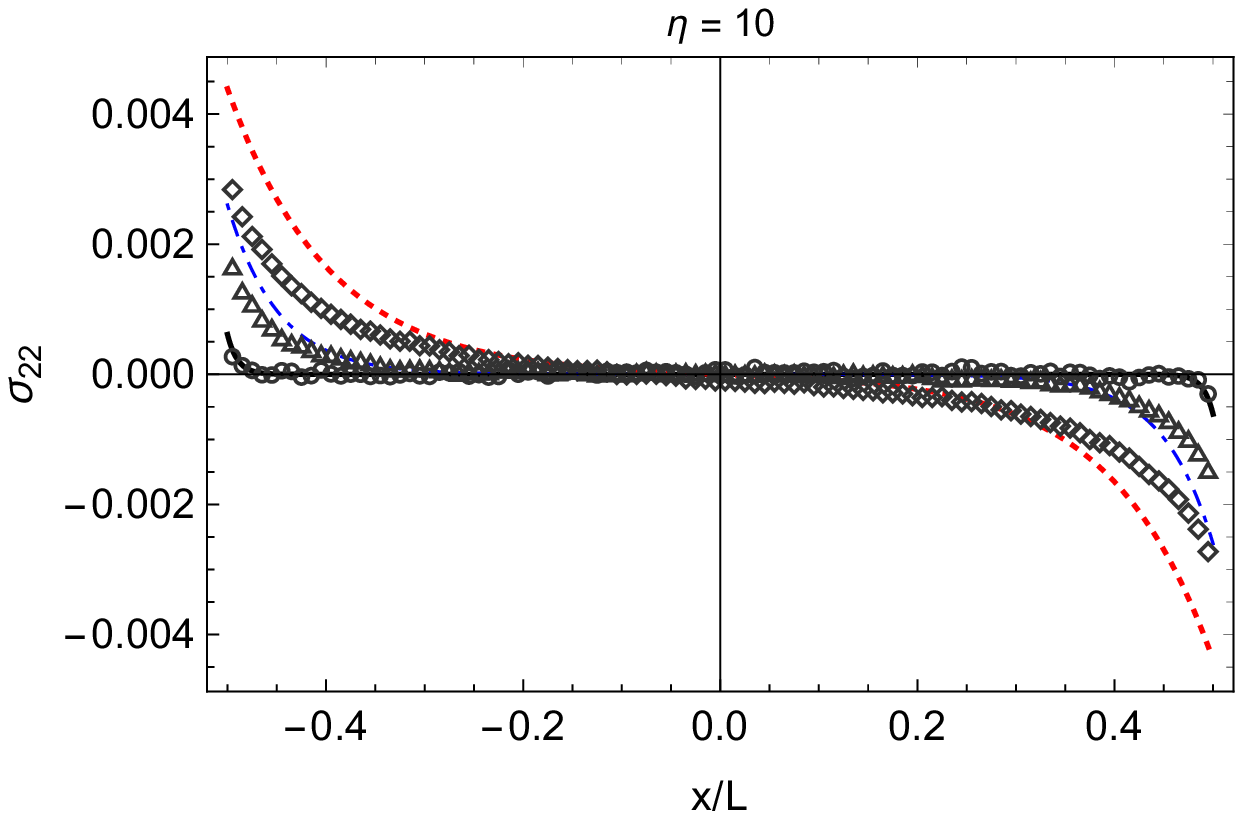}}
  {\includegraphics[width=0.32\textwidth,clip]{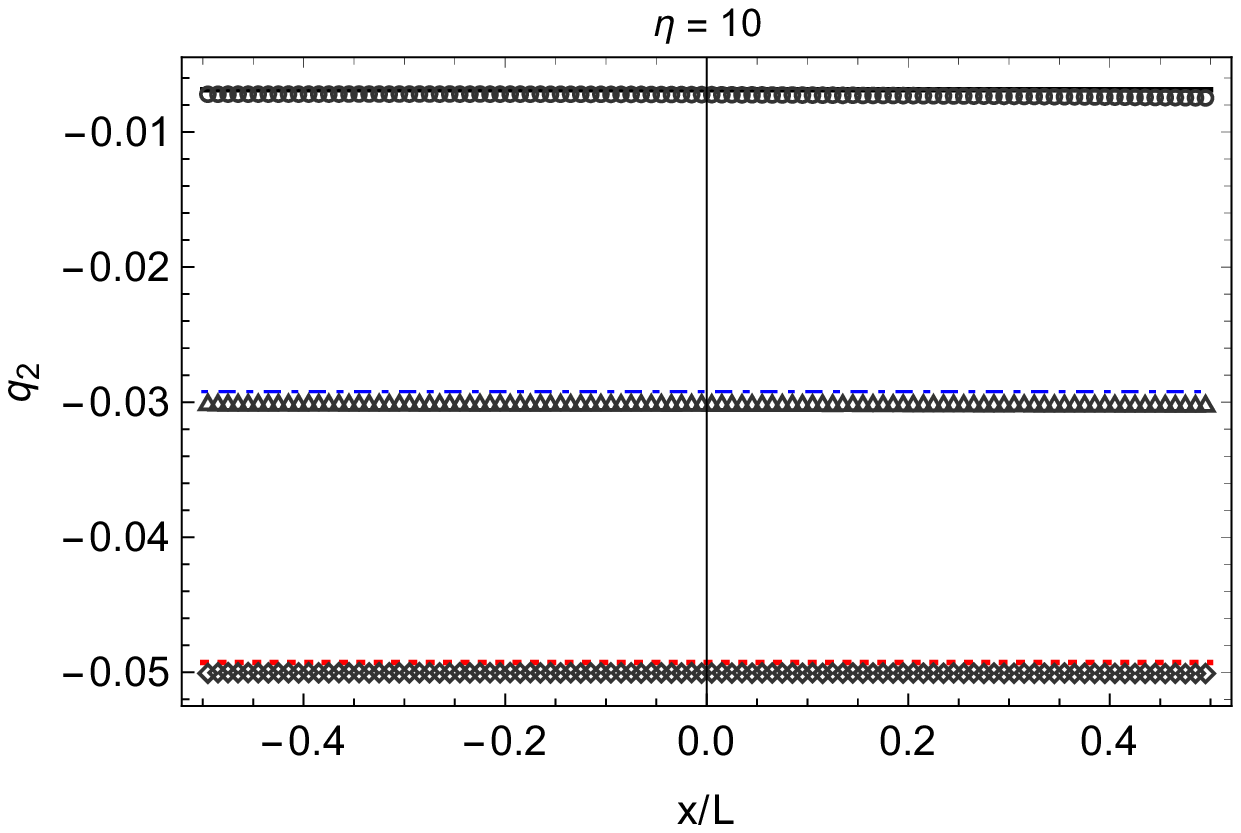}}
  \caption{Fourier flow of the semi-linear R13 equations with $\eta=10$. The dimensionless semi-linear R13 solutions for $\Kn=0.01$, $0.05$, and $0.1$ are plotted in black solid line, blue dashed-dotted line, and red dotted line, respectively. The corresponding DSMC solutions are marked by circle, triangle, and diamond.}
  \label{fig:fourier-eta10}
\end{figure}

\begin{figure}[!htb]
% \colorbox{black}
  \centering
  {\includegraphics[width=0.32\textwidth,clip]{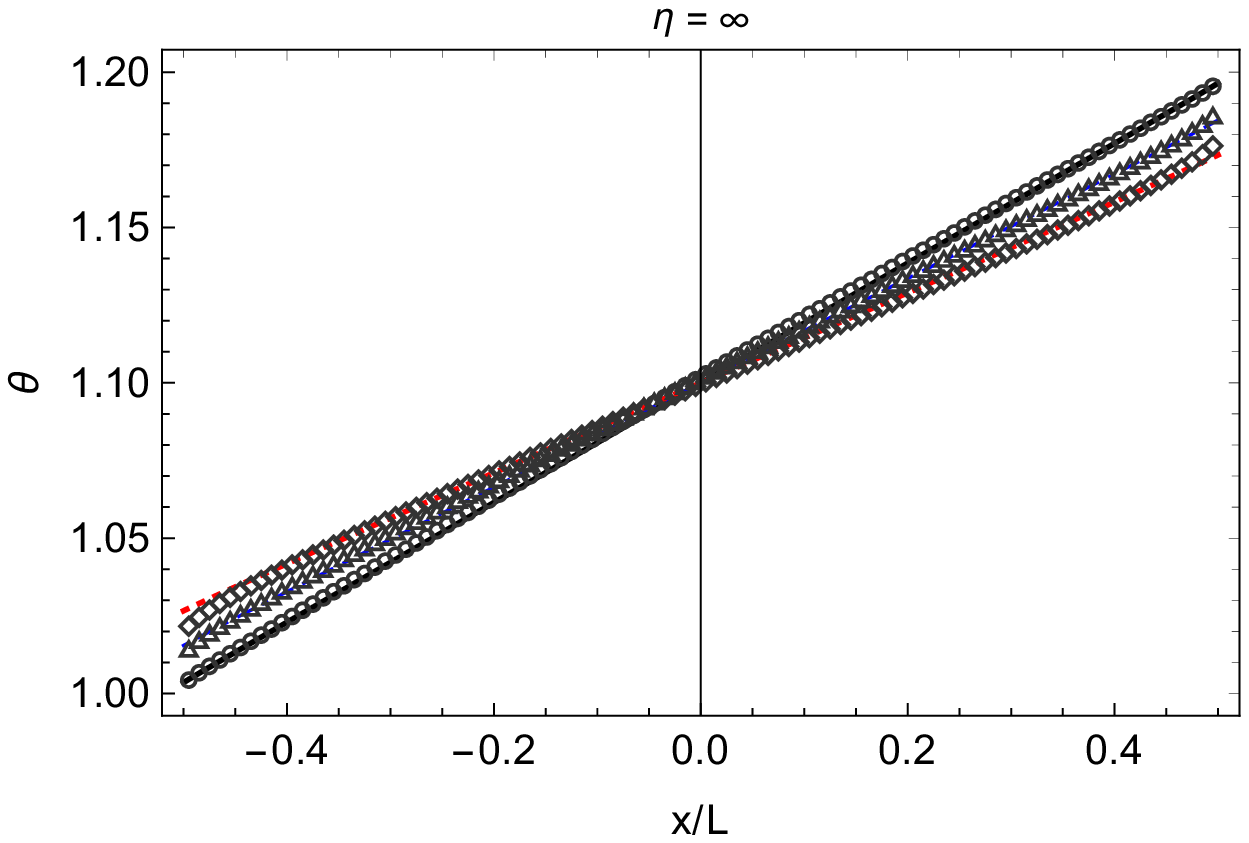}}
  {\includegraphics[width=0.32\textwidth,clip]{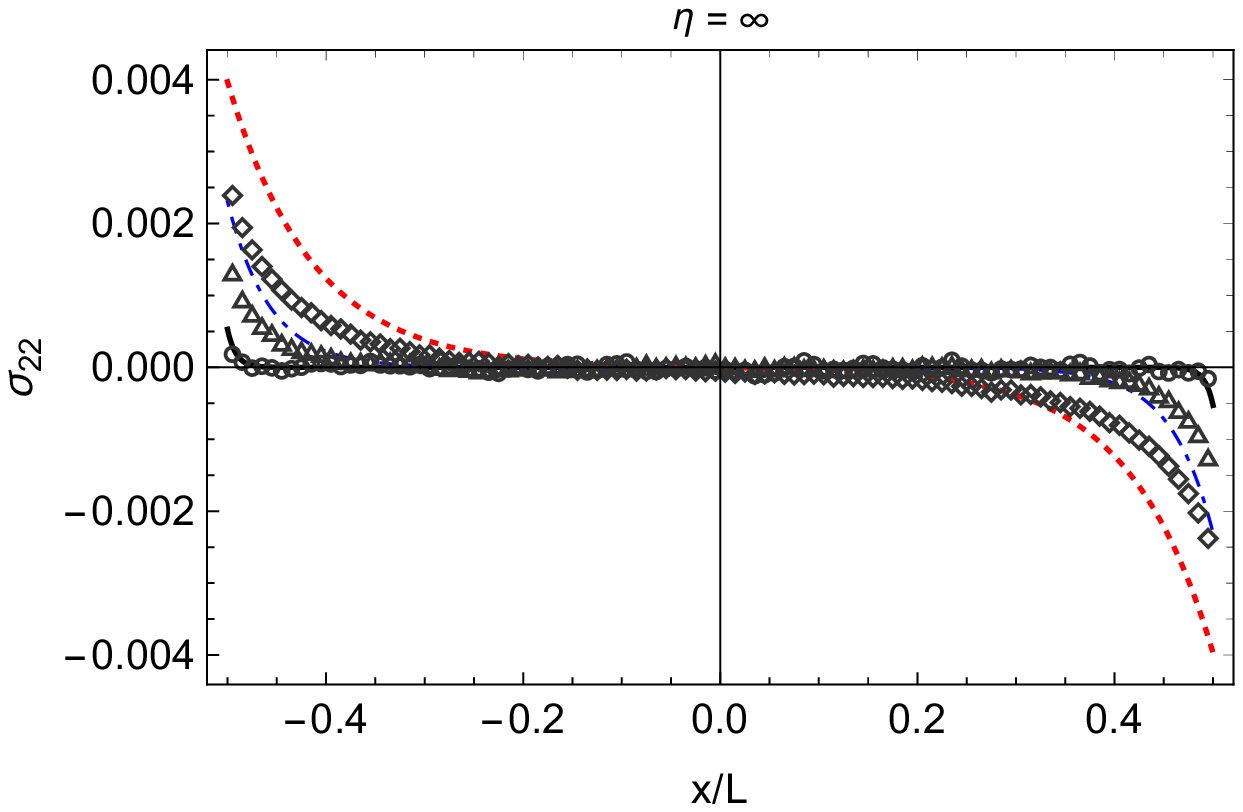}}
  {\includegraphics[width=0.32\textwidth,clip]{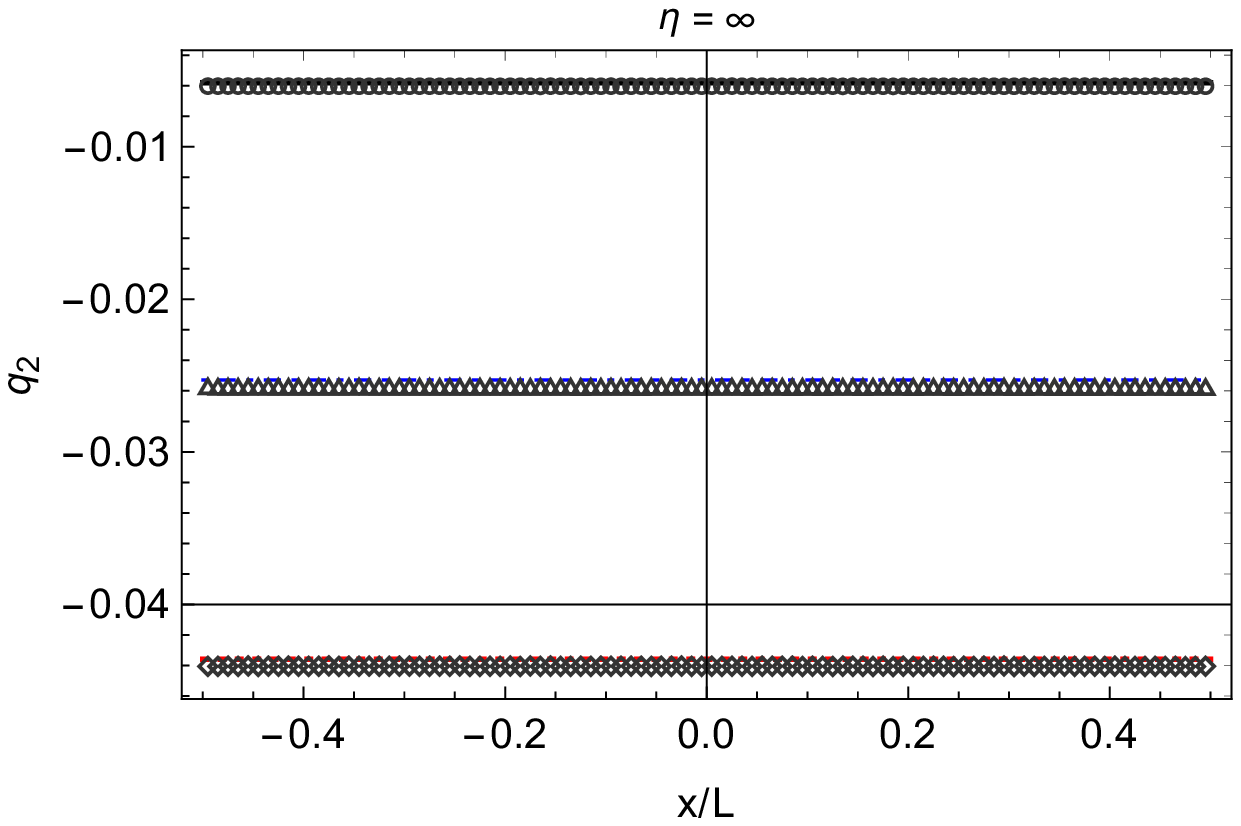}}
  \caption{Fourier flow of the semi-linear R13 equations with $\eta=\infty$. The dimensionless semi-linear R13 solutions for $\Kn=0.01$, $0.05$, and $0.1$ are plotted in black solid line, blue dashed-dotted line, and red dotted line, respectively. The corresponding DSMC solutions are marked by circle, triangle, and diamond.}
  \label{fig:fourier-hs}
\end{figure}

Then, with the viscosity index determined by $\eta$ via \eqref{eq:mu_omega}, i.e., $\omega=\frac{1}{2}+\frac{2}{\eta-1}$, we compare the semi-linear R13 solutions for $\eta=5$, $10$, and $\infty$ with the DSMC solutions obtained with the corresponding $\omega$ in \figurename~\ref{fig:fourier-eta5}-\ref{fig:fourier-hs}, respectively. It is observed that the profiles are in good agreement with the DSMC solutions, although some deviations can be found for the normal stress $\sigma_{22}$ near the boundaries. In the boundary layer, the nonlinear effects are stronger, so that the semi-linear R13 equations may fail to predict the solutions accurately.

\begin{figure*}[!htb]
% \colorbox{black}
  \centering
  {\includegraphics[width=0.32\textwidth,clip]{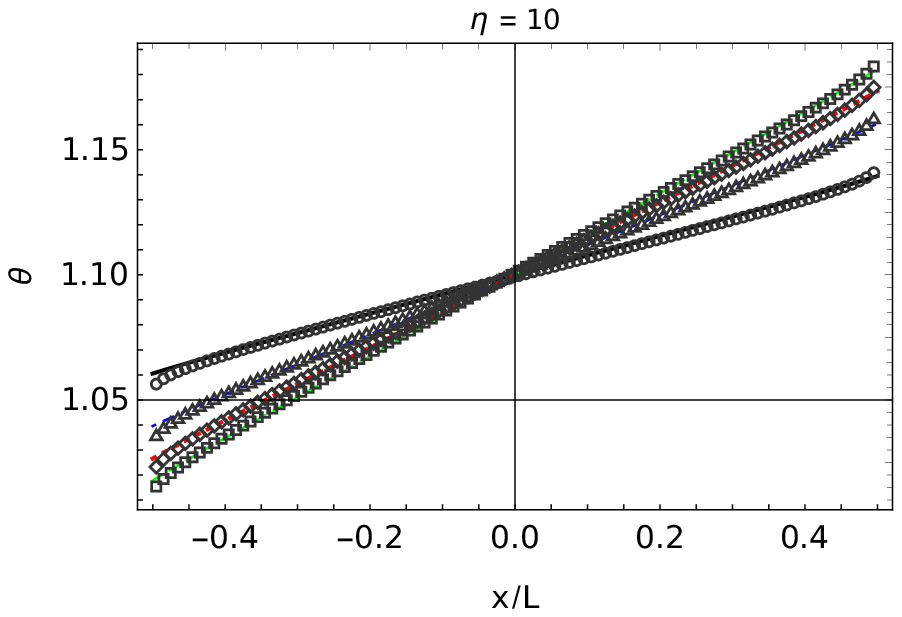}}
  {\includegraphics[width=0.32\textwidth,clip]{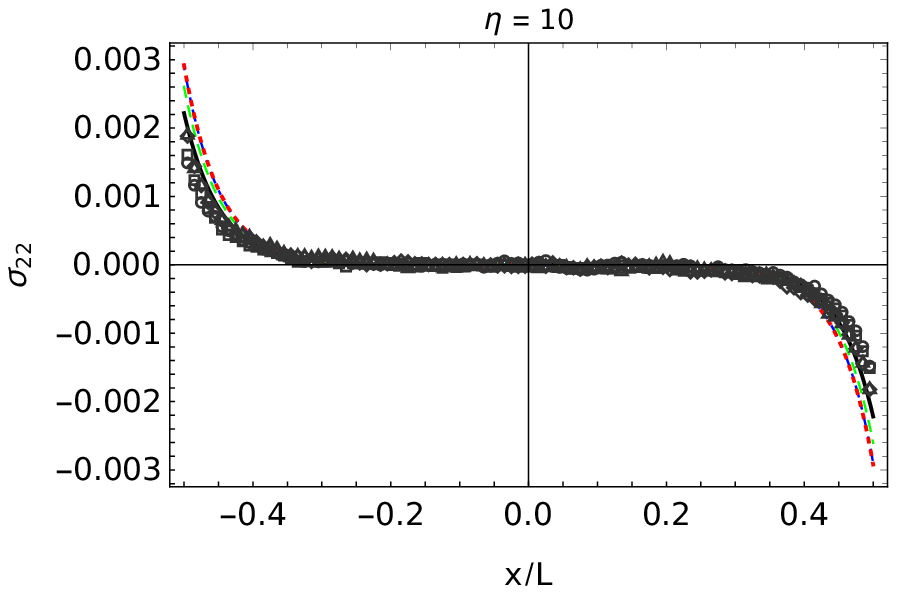}}
  {\includegraphics[width=0.32\textwidth,clip]{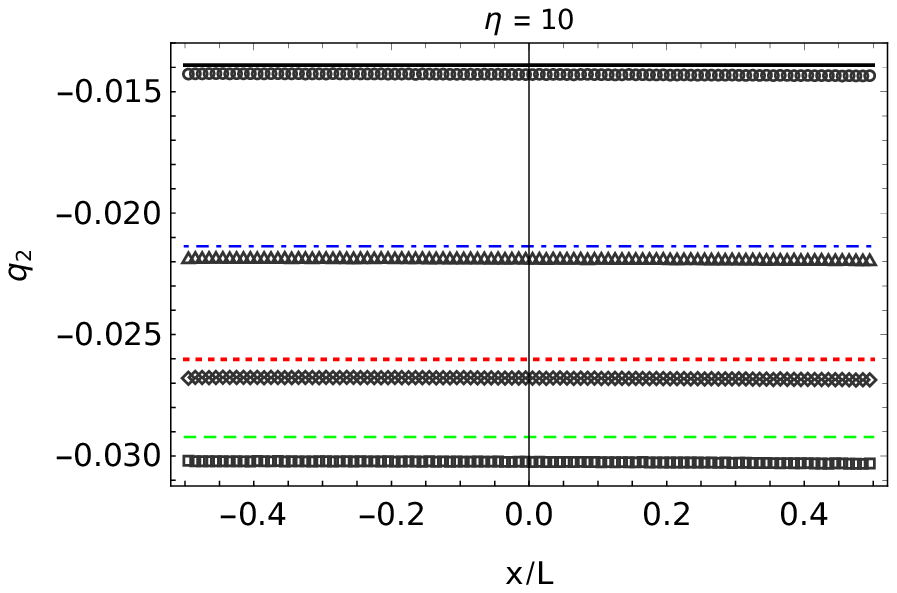}}
  \caption{Fourier flow of the semi-linear R13 equations with $\eta=\infty$ and $\Kn=0.05$ for various accommodation coefficients. The dimensionless semi-linear R13 solutions for $\chi=0.25$, $0.5$, $0.75$ and $1$ are plotted in black solid line, blue dashed-dotted line, red dotted line, and green dashed line, respectively. The corresponding DSMC solutions are marked by circles, triangles, diamonds, and squares.}
  \label{fig:fourier-chi}
\end{figure*}

To end this section, we consider the effect of the accommodation coefficient $\chi$ on the solutions from semi-linear R13 equations for the Fourier flow. We fix $\eta=10$ and $\Kn=0.05$, and Figure. \ref{fig:fourier-chi} plots the results of the semi-linear R13 solutions for $\chi=0.25$, $0.5$, $0.75$ and $1$ as well as the corresponding DSMC solutions. Our results are almost identical to the DSMC solutions under various $\chi$ for the temperature $\theta$. As for the other two quantities as $\sigma_{22}$ and $q_2$, the semi-linear R13 solutions are considerably closer to the DSMC solutions for smaller accommodation coefficients. This is possibly due to the fact that when $\chi$ is closer to zero, the distribution functions on the boundaries have smaller discontinuity, and therefore can be better approximated by the first few moments.

%\figurename~\ref{fig:fourier-veta-kn0p05} shows the variation of the semi-linear R13 solutions in terms of $\eta$ at $\Kn=0.05$. The DSMC solutions with $\omega=0.5$ are again presented as the reference. Slight differences of the heat flux $q_2$ for different $\eta$ are observed, while the profiles of temperature $\theta$ and normal stress $\sigma_{22}$ for different $\eta$ are very close to each other. It can also be seen that the semi-linear R13 solutions with $\eta=\infty$ match the DSMC solutions in an acceptable accuracy.

%\begin{figure}[!htb]
% %\colorbox{black}
%  \centering
%  {\includegraphics[width=0.32\textwidth,clip]{fourier_kn0p05_twr1p2_var_theta.eps}}
%  {\includegraphics[width=0.32\textwidth,clip]{fourier_kn0p05_twr1p2_var_sigma22.eps}}
%  {\includegraphics[width=0.32\textwidth,clip]{fourier_kn0p05_twr1p2_var_q2.eps}}
%  \caption{Fourier flow of the semi-linear R13 equations at $\Kn=0.05$. The semi-linear R13 solutions are plotted in black solid line, blue dashed-dotted line, red dotted line, green dashed line, and gray solid line for $\eta=5, 7, 10, 17$, and $\infty$, respectively. The reference DSMC solutions (circles) are obtained with $\omega=0.5$. }
%  \label{fig:fourier-veta-kn0p05}
%\end{figure}

\subsection{Force-driven Poiseuille flow}
This example studies the effect of the body force. Both plates in the force-driven Poiseuille flow are stationary, i.e., $v_W^l=v_W^r=0$, and we consider the case where the dimensionless wall temperatures and body force are set to be $\theta_W^l=\theta_W^r=1$ and $G_1=0.2555$ respectively.

\begin{figure}[!htb]
% \colorbox{black}
  \centering
  {\includegraphics[width=0.49\textwidth,clip]{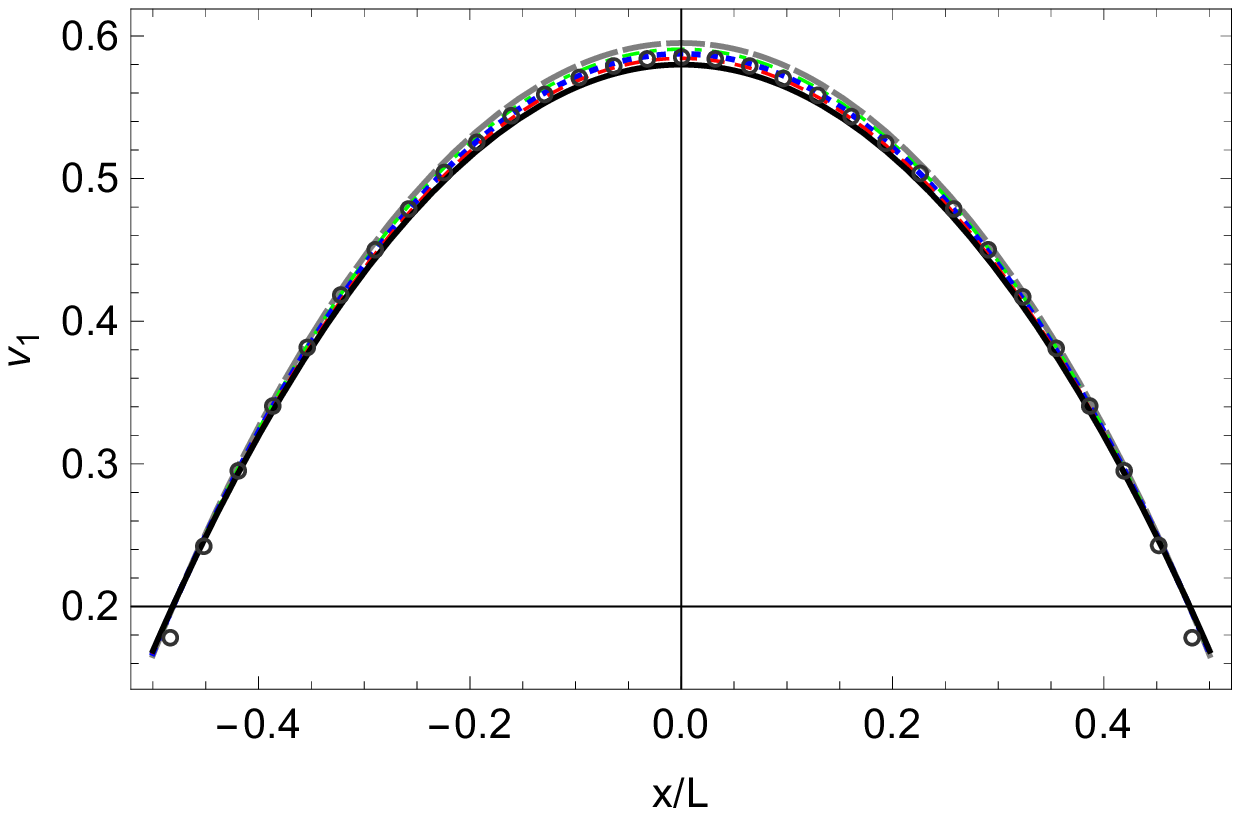}}\hfill
  {\includegraphics[width=0.49\textwidth,clip]{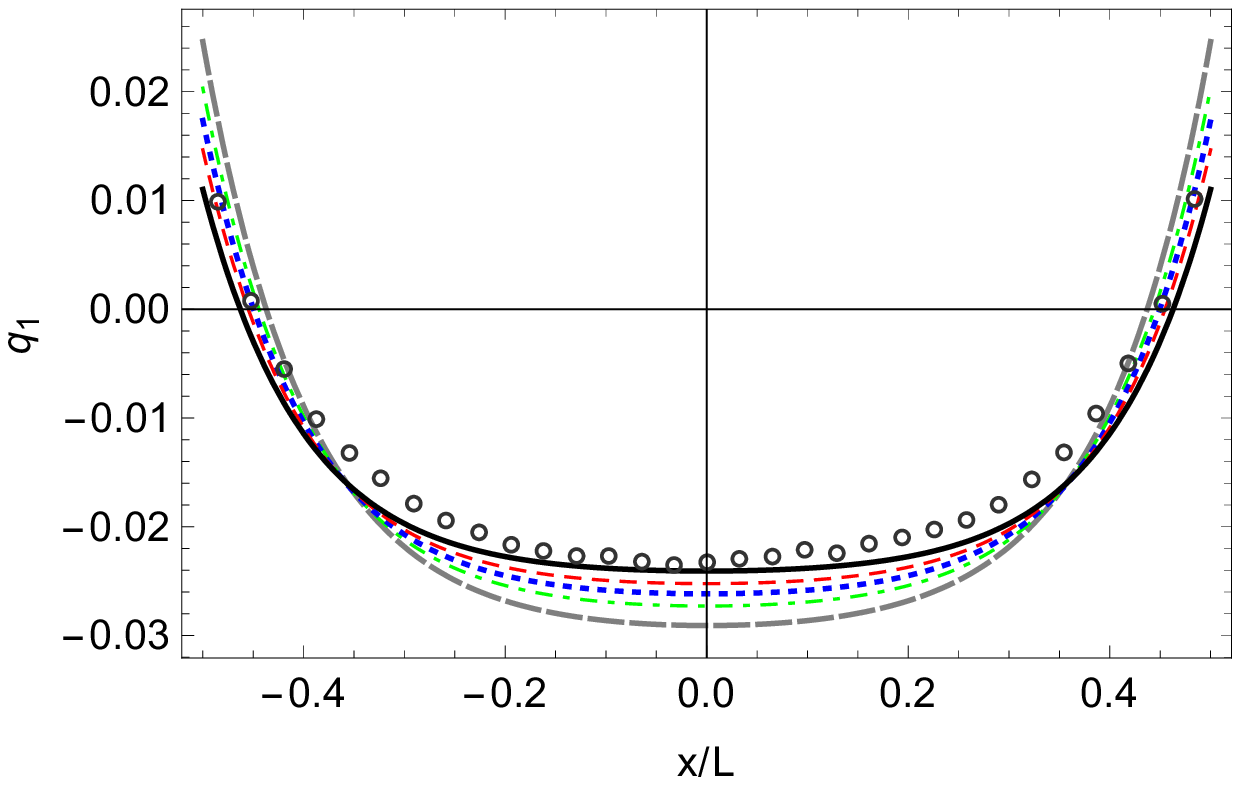}}
  \caption{Dimensionless profiles of velocity $v_1$ and heat flux $q_1$ for force-driven Poiseuille flow with $G_1=0.2555$ at $\Kn=0.1$. Comparison between the semi-linear R13 equations for $\eta=5$ (gray dashed line), $7$ (green dashed-dotted line), $10$ (blue dotted line), $17$ (red dashed line), and $\infty$ (black solid line) with the same $\omega=0.5$ and the DSMC results (circles) are presented.}
  \label{fig:poiseuille-veta}
\end{figure}

First, the results of the semi-linear R13 equations for $\eta=5,7,10,17$, and $\infty$ with the same viscosity index $\omega=0.5$ are compared to the DSMC results at $\Kn=0.1$. The results for the velocity $v_1$ and the heat flux $q_1$ are shown in \figurename~\ref{fig:poiseuille-veta} for a careful comparison. It can be observed that the deviations away from the DSMC results are reduced remarkably as $\eta$ increases especially for heat flux $q_1$. Again, the semi-linear R13 equations of hard sphere gas ($\eta=\infty$) give the solutions in best agreement with the DSMC results as expected. Furthermore, the behavior of $q_1$ in terms of $\eta$ in the bulk is mainly affected by the coefficient $\gamma_{1,1}^{(\eta)}$, which increases as $\eta$ increases. While near the boundaries, the different solutions of $q_1$ for different $\eta$ are mainly due to the large variation of the coefficient $\C_2$, which can be solved from the linear system of $\C_2$ and $\C_4$ by inserting the analytical expression of $v_1$, $\sigma_{12}$, and $q_1$ into \eqref{eq:bc4} and \eqref{eq:bc3}. Again, it is found that with the fixed viscosity index, $\C_2$ decreases as $\eta$ increases, and the value of $\C_2$ for $\eta=\infty$ is almost a half of its value for $\eta=5$.

\begin{figure}[!htb]
% \colorbox{black}
  \centering
  {\includegraphics[width=0.32\textwidth,clip]{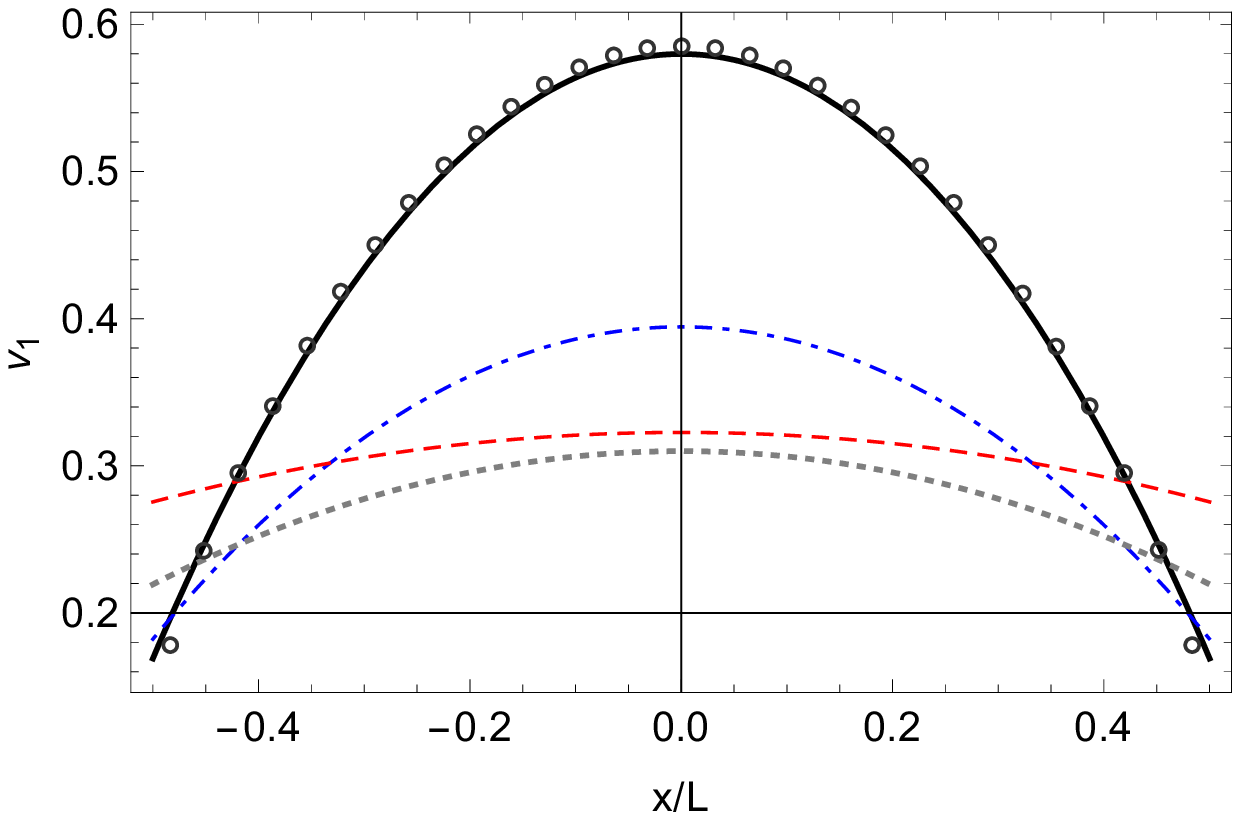}}
  {\includegraphics[width=0.32\textwidth,clip]{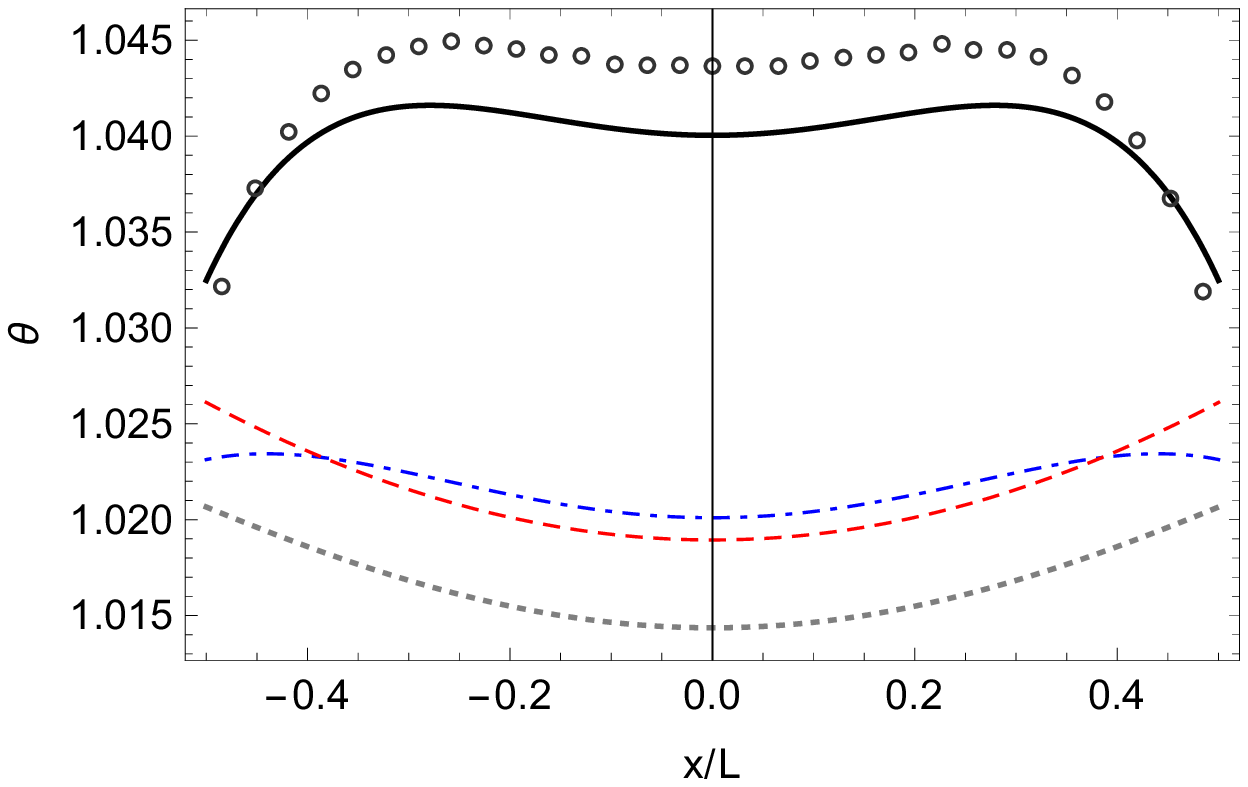}}
  {\includegraphics[width=0.32\textwidth,clip]{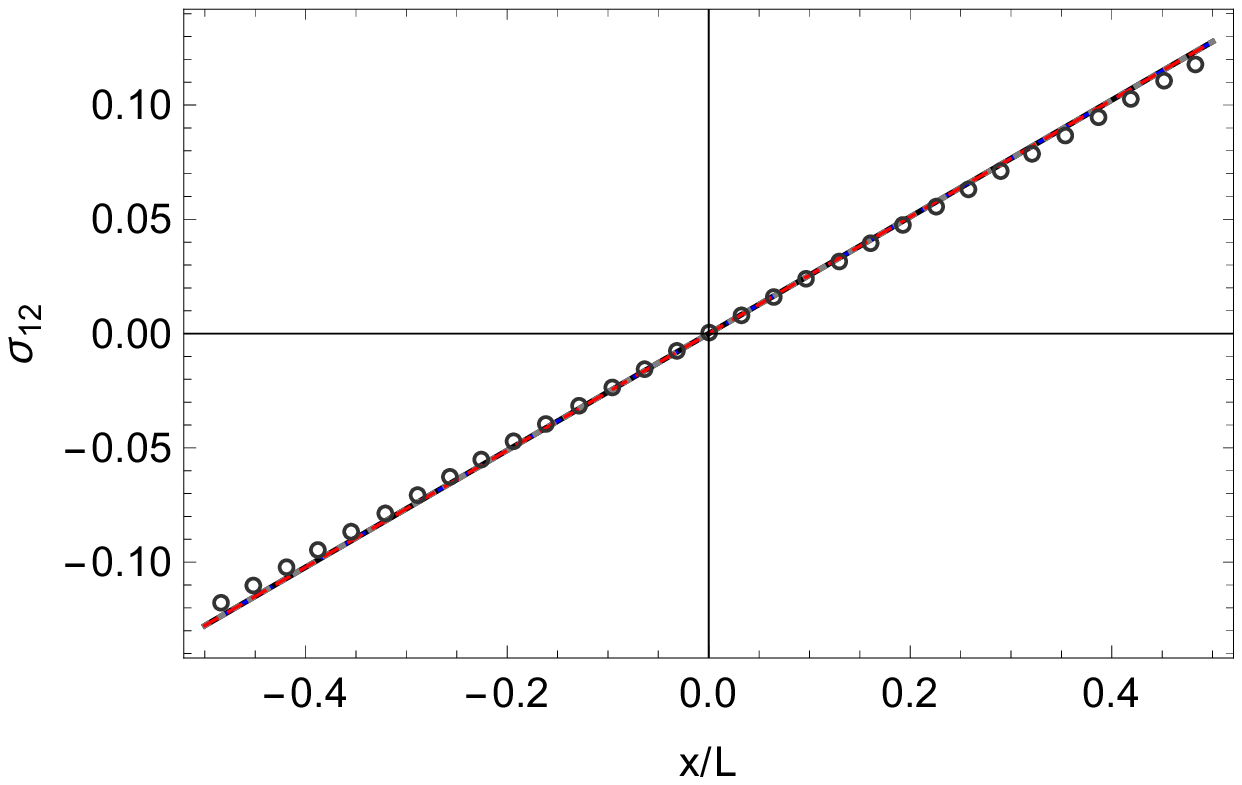}}\\
  {\includegraphics[width=0.32\textwidth,clip]{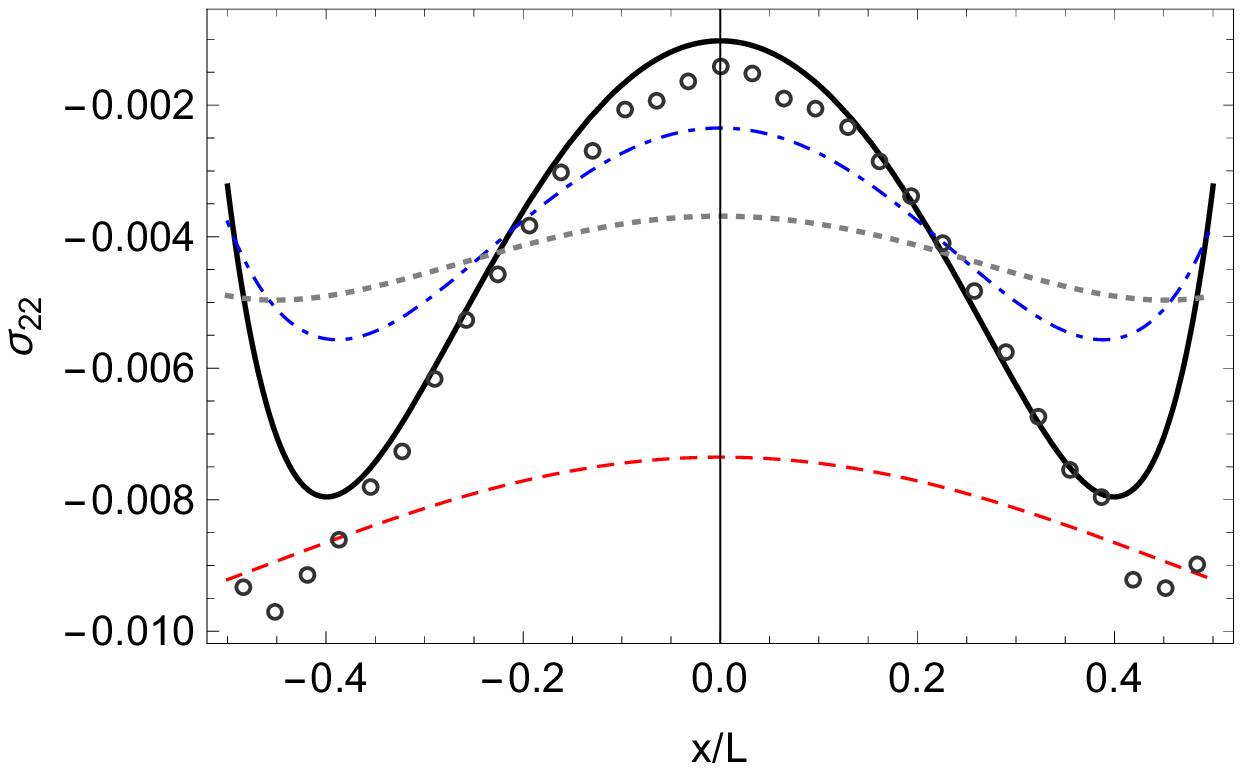}}
  {\includegraphics[width=0.32\textwidth,clip]{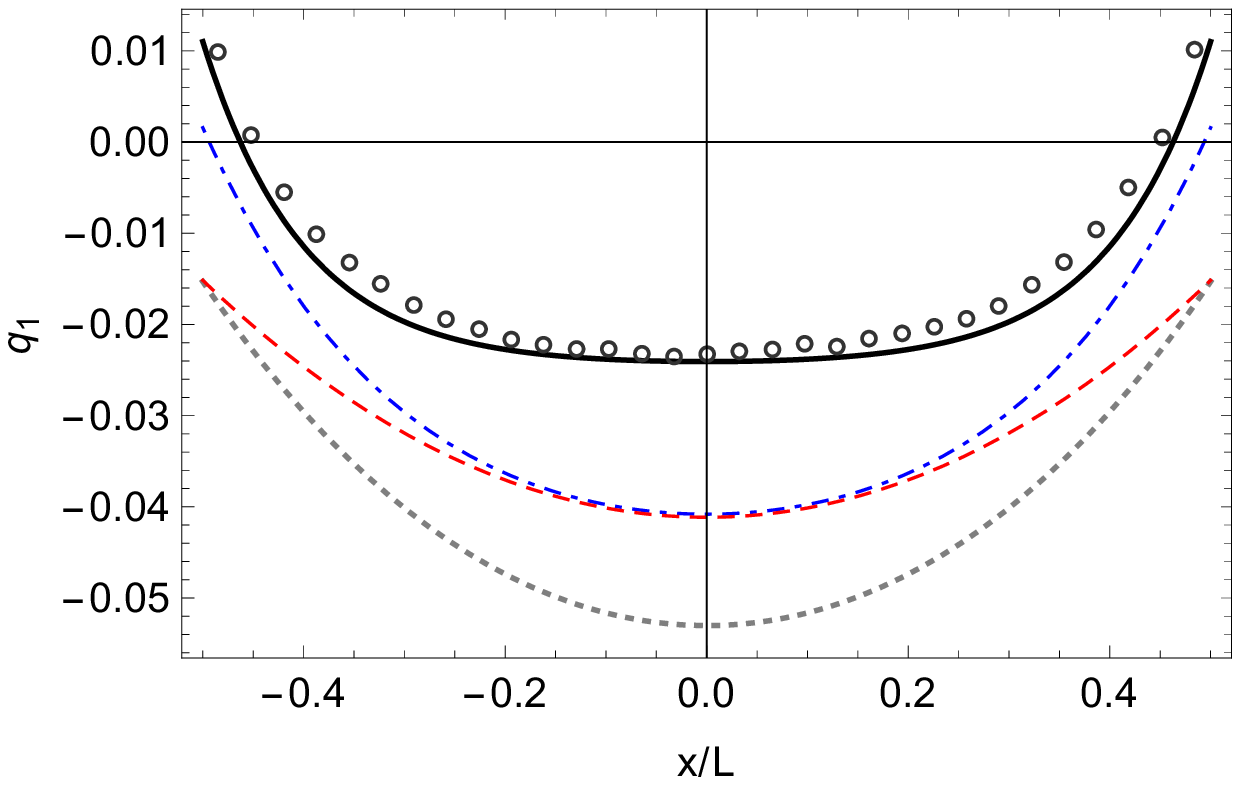}}
  {\includegraphics[width=0.32\textwidth,clip]{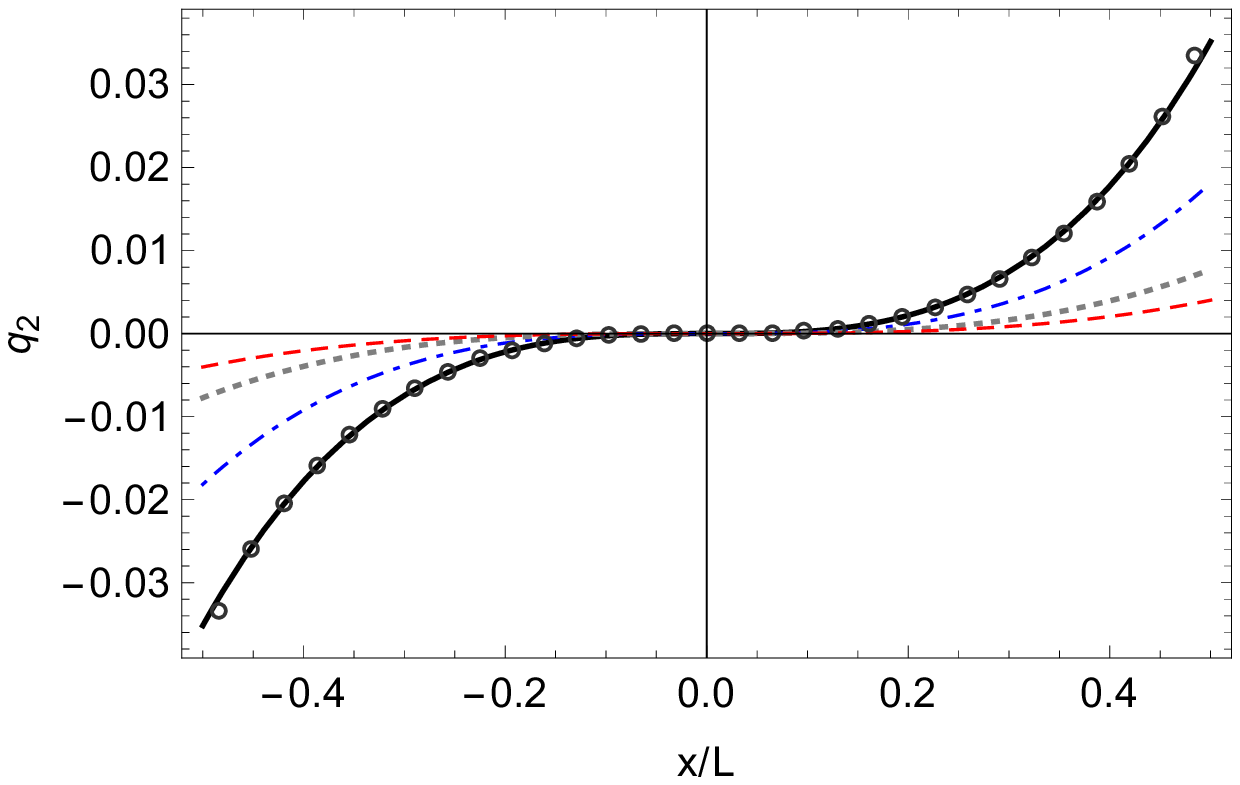}}
  \caption{Dimensionless profiles of force-driven Poiseuille flow with $G_1=0.2555$. The results of the semi-linear R13 equations with $\eta=\infty$ for $\Kn=0.1$ (black solid line), $0.2$ (blue dashed-dotted line), $0.5$ (gray dotted line), and $1.0$ (red dashed line), as well as the DSMC results (circles) for $\Kn=0.1$ are presented.}
  \label{fig:poiseuille-hs}
\end{figure}

\figurename~\ref{fig:poiseuille-hs} shows more details of the solutions obtained by the semi-linear R13 equations with $\eta=\infty$ at $\Kn=0.1, 0.2, 0.5$, and $1.0$. The reference DSMC solutions at $\Kn=0.1$ are also presented for comparison. As can be seen, the profiles of our solutions are generally consistent with the DSMC results. The R13 equations predict lower temperature in the bulk, but the relative error is only within $0.5\%$. The most significant error appears in the normal stress $\sigma_{22}$, especially for the flow near the boundary. The reason is likely to be the lack of higher-order moments in the system. As the Knudsen number increases, the general trends of the solutions agree with the results in \cite{Taheri2009}. In particular, for the temperature, the curve no longer bends down near the boundary for high Knudsen number. This can also be observed from the expression of the temperature \eqref{eq:Poiseuille}. The quartic coefficient $\gamma_{4,1}^{(\eta)} G_1^2 / \Kn_0^2$ is negative, which contributes to the concaveness of the profile. When $\Kn_0$ gets larger, the contribution of this term becomes smaller. When $\Kn_0$ is sufficiently large, the concaveness will be transcended by the convex quadratic term $\gamma_{4,2}^{(\eta)} G_1^2 x_2^2$, resulting in the reversion near the boundary.

\section{Semi-linearization of R13 equatoins and derivation of boundary conditions} \label{sec:model}
We now return to the derivation of the semi-linear R13 equations. In general, our semi-linearization agrees with the semi-linearization of the R13 equations for Maxwell molecules introduced in \cite{Taheri2009}, and the boundary conditions are derived based on Maxwell's boundary condition for the Boltzmann equation. To begin with, we first review the general idea of the derivation in \cite{Cai2020}, whose intermediate steps will be useful for the formulation of boundary conditions.

\subsection{Review of the derivation of R13 equations}
For general inverse-power-law models, the R13 equations are derived in \cite{Cai2020}. For the convenience of our further discussion, we would first like to review the general idea of the derivation of the R13 equations. The derivation is based on the Boltzmann equation with linearized collision operator:
\begin{displaymath}
\frac{\partial f}{\partial t} + \nabla_{\bx} \cdot (\bxi f) = \mathcal{L}[f],
\end{displaymath}
where $f(\bx,\bxi,t)$ denotes the distribution function of the gas molecules at time $t$, position $\bx \in \mathbb{R}^3$, and the variable $\bxi \in \mathbb{R}^3$ stands for the velocity of the gas molecules. The right-hand side $\mathcal{L}[f]$ is the Boltzmann collision operator linearized about the local Maxwellian $\mathcal{M}(\bx, \bxi, t)$, which is defined by
\begin{displaymath}
\mathcal{M}(\bx, \bxi, t) = \frac{\rho(\bx,t)}{\mathfrak{m}[2\pi\theta(\bx,t)]^{3/2}}
  \exp \left(-\frac{|\bxi - \bv(\bx,t)|^2}{2\pi \theta(x,t)} \right),
\end{displaymath}
where $\mathfrak{m}$ denotes the mass of a single gas molecule. The derivation of the R13 equations is based on the method of ``order of magnitude'' developed in \cite{Struchtrup2005, Struchtrup}. In \cite{Cai2020}, such a method is applied to the following series expansion of $f(\bx,\bxi,t)$:
\begin{equation} \label{eq:expansion}
f(\bx,\bxi,t) = \sum_{l=0}^{+\infty} \sum_{m=-l}^l \sum_{n=0}^{+\infty} f_{lmn}(\bx,t) \psi_{lmn}(\bx,\bxi,t),
\end{equation}
where $\psi_{lmn}(\bx,\bxi,t)$ are orthogonal basis functions with the form
\begin{displaymath}
\psi_{lmn}(\bx,\bxi,t) = [\theta(\bx,t)]^{-\frac{2n+l}{2}}
  p_{lmn} \left( \frac{\bxi - \bv(\bx,t)}{\sqrt{\theta(\bx,t)}} \right)
  \cdot \frac{1}{\mathfrak{m}[2\pi\theta(\bx,t)]^{3/2}}
    \exp \left(-\frac{|\bxi - \bv(\bx,t)|^2}{2\pi \theta(x,t)} \right),
\end{displaymath}
and $p_{lmn}$ are orthogonal polynomials satisfying
\begin{displaymath}
\int_{\mathbb{R}^3} p_{l_1 m_1 n_1}^*(\boldsymbol{c}) p_{l_2 m_2 n_2}(\boldsymbol{c})
  \cdot \frac{1}{(2\pi)^{3/2}} \exp \left( -\frac{|\boldsymbol{c}|^2}{2} \right) \,\mathrm{d} \boldsymbol{c} = \delta_{l_1 l_2}\delta_{m_1 m_2}\delta_{n_1 n_2}.
\end{displaymath}
Here $p_{l_1 m_1 n_1}^*$ means the complex conjugate of the function $p_{l_1 m_1 n_1}$. Note that the basis functions are chosen to be complex for easier formulation of the moment equations, and the basis functions satisfy $\psi^*_{lmn} = (-1)^m \psi_{l,-m,n}$, so that in order to ensure that $f(\bx,\bxi,t)$ is real, the coefficients $f_{lmn}$ must satisfy $f^*_{lmn} = (-1)^m f_{l,-m,n}$.

Among the coefficients in the expansion expansion \eqref{eq:expansion}, the coefficients $f_{001}$ and $f_{1m0}$ for $m = -1,0,1$ are always zero so that $\bv$ and $\theta$ correspond exactly to the velocity and the temperature, respectively. The first coefficient $f_{000}$ equals the density $\rho$, and other variables in the R13 equations are related to the coefficients in \eqref{eq:expansion} by
\begin{equation} \label{eq:sigam_q}
\begin{gathered}
\sigma_{11} = \sqrt{2}\re(f_{220}) - f_{200}/\sqrt{3}, \quad \sigma_{12} = -\sqrt{2}\im(f_{220}), \quad \sigma_{13} = -\sqrt{2}\re(f_{210}), \\
\sigma_{22} = -\sqrt{2}\re(f_{220}) - f_{200}/\sqrt{3}, \quad \sigma_{23} = \sqrt{2}\im(f_{210}),\quad \sigma_{33} = 2f_{200}/\sqrt{3}, \\
q_1 = \sqrt{5}\re(f_{111}), \qquad q_2 = -\sqrt{5}\im(f_{111}), \qquad q_3 = -\sqrt{5/2}f_{101}.
\end{gathered}
\end{equation}
Therefore it is sufficient to derive the equations for the above coefficients. Further derivation is based on the assumption that large temporal and spatial scales are considered, so that it makes sense to replace $t$ by $t/\epsilon$ and $\bx$ by $\bx/\epsilon$, and consider the asymptotic expansion
\begin{displaymath}
f_{lmn} = f_{lmn}^{(0)} + \epsilon f_{lmn}^{(1)} + \epsilon^2 f_{lmn}^{(2)} + \epsilon^3 f_{lmn}^{(3)} + \cdots.
\end{displaymath}
Based on the idea of Chapman-Enskog expansion, it can be derived that
\begin{itemize}
\item $f_{lmn}^{(0)} = 0$ if $(l,m,n) \neq (0,0,0)$.
\item $f_{lmn}^{(1)} = 0$ if $l \geqslant 3$ or $l = 0$.
\item For $k \geqslant 2$, $f_{lmn}^{(k)} = 0$ if $l \geqslant 2k+1$.
\item $f_{1mn}^{(1)} = \tilde{A}_{1n} \theta^{n-1} f_{1m1}^{(1)}$ and $f_{2mn}^{(1)} = \tilde{A}_{2n} \theta^n f_{2m0}^{(1)}$, where $\tilde{A}_{1n}$ and $\tilde{A}_{2n}$ are constants.
\end{itemize}
The above properties show that $f_{1mn} - \tilde{A}_{1n} \theta^{n-1} f_{1m1}$ and $f_{2mn} - \tilde{A}_{2n} \theta^n f_{2m0}$ are both $O(\epsilon^2)$ quantities. Therefore we can rewrite the asymptotic expansion of $f_{1mn}$ and $f_{2mn}$ as
\begin{equation} \label{eq:new_expansion}
f_{1mn} = \tilde{A}_{1n} \theta^{n-1} f_{1m1} + \epsilon^2 \tilde{f}_{1mn}^{(2)} + \epsilon^3 \tilde{f}_{1mn}^{(3)} + \cdots, \qquad
f_{2mn} = \tilde{A}_{2n} \theta^n f_{2m0} + \epsilon^2 \tilde{f}_{2mn}^{(2)} + \epsilon^3 \tilde{f}_{2mn}^{(3)} + \cdots.
\end{equation}
Note that $\tilde{f}_{lmn}^{(k)}$ is defined only for $l = 1,2$ and $k \geqslant 2$, and their definitions are different from $f_{lmn}^{(k)}$ by
\begin{displaymath}
\tilde{f}_{1mn}^{(k)} = f_{1mn}^{(k)} - \tilde{A}_{1n} \theta^{n-1} f_{1m1}^{(k)}, \qquad
\tilde{f}_{2mn}^{(k)} = f_{2mn}^{(k)} - \tilde{A}_{2n} \theta^n f_{2m0}^{(k)}.
\end{displaymath}
For consistency, when $l = 0$ or $l \geqslant 3$, we define $\tilde{f}_{lmn}^{(k)} = f_{lmn}^{(k)}$.

Based on the new expansions \eqref{eq:new_expansion}, we can further apply asymptotic analysis to find expressions for $\tilde{f}_{lmn}^{(2)}$ and $\tilde{f}_{lmn}^{(3)}$, which can all be represented by $\rho, \bv, \theta, f_{1m1}$, and $f_{2m0}$. The detailed derivation is highly technical and we refer the readers to \cite{Cai2020} for more details. In general, the second-order terms $\tilde{f}_{lmn}^{(2)}$ include the first-order derivatives, and the third-order terms include second-order derivatives. Finally, the R13 equations are obtained by the following steps:
\begin{itemize}
\item Approximate the distribution function by
  \begin{equation} \label{eq:ansatz}
  \begin{split}
  f(\bx,\bxi,t) &= \mathcal{M}(\bx,\bxi,t) + \sum_{m=-1}^1 \sum_{n=1}^{+\infty} \tilde{A}_{1n} f_{1m1}  \psi_{1mn}(\bx,\bxi,t)
  + \sum_{m=-2}^2 \sum_{n=0}^{+\infty} \tilde{A}_{2n} f_{2m0}  \psi_{2mn}(\bx,\bxi,t) \\
  & + \sum_{l=0}^4 \sum_{m=-l}^l \sum_{n=0}^{+\infty} \epsilon^2 \tilde{f}_{lmn}^{(2)}(\bx,t) \psi_{lmn}(\bx,\bxi,t) + \sum_{l=0}^6 \sum_{m=-l}^l \sum_{n=0}^{+\infty} \epsilon^3 \tilde{f}_{lmn}^{(3)}(\bx,t) \psi_{lmn}(\bx,\bxi,t).
  \end{split}
  \end{equation}
  In our computation, we cut off the series by discarding all coefficients with $l + 2n > 20$ so that only finite terms are included.
\item Insert this distribution function into the scaled Boltzmann equation $\partial_t f + \bxi \cdot \nabla_{\bx} f = \epsilon^{-1} \mathcal{L}[f]$, and drop the $\epsilon^3$ terms on the left-hand side.
\item Take the thirteen moments of the resulting equation by multiplying it by the following quantities and then taking the integral over $\bxi$:
\begin{displaymath}
1, \quad \xi_j, \quad (\xi_i - v_i)(\xi_j - v_j), \quad \frac{1}{2} |\bxi - \bv|^2 (\xi_j - v_j), \qquad i,j = 1,2,3.
\end{displaymath}
In the resulting 13 equations, convert the coefficients to the variables $\sigma_{ij}$ and $q_j$ according to \eqref{eq:sigam_q}.
\end{itemize}
The final R13 equations contain at most second-order derivatives, which come from $\tilde{f}_{lmn}^{(3)}$ terms in the collision part, and $\tilde{f}_{lmn}^{(2)}$ terms in the advection part.

By such procedure, a complete set of three-dimensional nonlinear R13 equations can be derived. Due to the massive calculations, all the steps are carried out using computer algebra systems. We refer the readers to the supplementary materials of \cite{Cai2020}. The resulting equations are highly involved, which need to be simplified to obtain the equations in Section \ref{sec:R13}.

\subsection{Nondimensionalization, dimension reduction, and semi-linearization of R13 equations} \label{sec:simplification}
The simplification of the equations includes nondimensionalization, dimension reduction, and semi-linearization. Note that all the steps will be carried out using Mathematica, and below we only sketch the general idea of these operations without providing the lengthy equations in the intermediate steps.

We first carry out the nondimensionalization. Let $\rho_0$ be the average density and $T_0$ be the reference temperature. The dimensionless variables are defined by
\begin{displaymath}
\hat{\rho} = \frac{\rho}{\rho_0}, \quad
\hat{\bv} = \frac{\bv}{\sqrt{k_B T_0 / \mathfrak{m}}}, \quad
\hat{\theta} = \frac{\theta}{k_B T_0 / \mathfrak{m}}, \quad
\hat{\sigma}_{ij} = \frac{\sigma_{ij}}{\rho_0 k_B T_0 / \mathfrak{m}}, \quad
\hat{q}_j = \frac{q_j}{\rho_0 (k_B T_0 / \mathfrak{m})^{3/2}},
\end{displaymath}
where $k_B$ is the Boltzmann constant. Let $L$ be the distance (with dimension) between two parallel plates. Then the dimensionless spatial and time variables are given by
\begin{displaymath}
\hat{\bx} = \bx / L, \qquad \hat{t} = t \bigg/ \frac{L}{\sqrt{k_B T_0 / \mathfrak{m}}}.
\end{displaymath}
In order to define the Knudsen number, we consider the approximation of the inverse-power-law model with the variable-hard-sphere model \cite{Bird1994}, whose basic idea is to model the gas molecules as hard spheres with diameter proportional to the inverse power of the relative velocity. For variable-hard-sphere gases, the equilibrium mean free path at density $\rho_0$ and temperature $T_0$ is
\begin{equation} \label{eq:mfp}
\lambda_0 = \frac{2}{15} (5-2\omega)(7-2\omega)
  \left( \frac{\mathfrak{m}}{2\pi k_B T_0} \right)^{1/2} \frac{\mu_0}{\rho_0},
\end{equation}
where $\omega$ is the viscosity index and $\mu_0$ is the viscosity coefficient at density $\rho_0$ and temperature $T_0$. Since the variable-hard-sphere gas is considered as an approximation of the inverse-power-law gas, we choose $\omega$ and $\mu_0$ to be the same as those for the inverse-power-law gas:
\begin{equation} \label{eq:mu_omega}
\mu_0 = \frac{5\mathfrak{m} (k_B T_0 / (\mathfrak{m} \pi))^{1/2} (2k_B T / \kappa)^{2/(\eta-1)}}{8A_2(\eta) \Gamma[4-2/(\eta-1)]}, \qquad \omega = \frac{1}{2} + \frac{2}{\eta-1},
\end{equation}
where $A_2(\eta)$ is a dimensionless constant depending on $\eta$. Some values of $A_2(\cdot)$ are given by (up to 4 significant figures)
\begin{displaymath}
A_2(5) = 0.4362, \quad A_2(7) = 0.3568, \quad A_3(10) = 0.3235, \quad A_4(17) = 0.3079, \quad A_2(+\infty) = 0.3333.
\end{displaymath}
Inserting \eqref{eq:mu_omega} into \eqref{eq:mfp}, the reference mean free path $\lambda_0$ can be calculated, and the Knudsen number is then accordingly defined by $\Kn = \lambda_0/L$. The reciprocal of the Knudsen number will appear as the coefficient of the dimensionless collision terms.

To study the steady-state flows between two parallel plates, we can first remove the time derivative in the differential equations. According to our choice of the coordinate system, the $\hat{x}_2$-axis is perpendicular to the plates, so that $\hat{v}_2 = 0$ due to mass conservation. By further assuming that the plates move only along the $\hat{x}_1$-axis, we can further reduce the dimensionality by imposing the symmetry $f(\xi_1, \xi_2, \xi_3) = f(\xi_1, \xi_2, -\xi_3)$ to the distribution function, so that all the moments that is odd in $\xi_3$ vanish. More precisely, we have
\begin{displaymath}
\hat{v}_3 = \hat{q}_3 = \hat{\sigma}_{13} = \hat{\sigma}_{23} = 0.
\end{displaymath}
Moreover, all the derivatives with respect to $\hat{x}_1$ and $\hat{x}_3$ can be taken away from the equations since the flow is homogeneous on every plane parallel to the plates. Thus, what remains is a system of eight ordinary differential equations. These equations are still fully nonlinear whose exact solution cannot be obtained.

The final equations given in Section \ref{sec:R13} are obtained by semi-linearization. As mentioned before, the semi-linearization generally follows the work \cite{Taheri2009}, where it is assumed that the flow is close to a global equilibrium state:
\begin{displaymath}
\hat{\rho} = 1 + \varepsilon \bar{\rho}, \quad
\hat{v}_1 = \varepsilon \bar{v}_1, \quad
\hat{\theta} = 1 + \varepsilon \bar{\theta}, \quad
\hat{\sigma}_{ij} = \varepsilon \bar{\sigma}_{ij}, \quad
\hat{q}_j = \varepsilon \bar{q}_j.
\end{displaymath}
Plugging these equations into the eight dimensionless R13 equations and dropping all terms of order $O(\varepsilon^2)$ or higher, we can obtain the linear R13 moment equations. Note that in the linear R13 equations, the moments $\bar{q}_1, \bar{\sigma}_{12}$ and $\bar{v}_1$ are completely decoupled from other five equations. This can be observed by removing all the underlined terms from \eqref{eq:v1}--\eqref{eq:q2}, and in the result one can see that the equations \eqref{eq:v1}, \eqref{eq:sigma12}, and \eqref{eq:q1} contain only $v_1$, $\sigma_{12}$ and $q_1$, while the remaining five equations only include the remaining five variables. In order to preserve the coupling of all quantities, we follow \cite{Taheri2009} to consider semi-linear R13 equations, where we preserve $O(\varepsilon^2)$ terms including the product of any two of the following terms or their derivatives:
\begin{displaymath}
\bar{q}_1, \quad \bar{\sigma}_{12}, \quad \frac{\mathrm{d}\bar{v}_1}{\mathrm{d}\hat{x}_2}.
\end{displaymath}
Finally, by removing all the accents, the semi-linear R13 equations in Section \ref{sec:R13} can be obtained.

\subsection{Derivation of wall boundary conditions}
The derivation of wall boundary conditions follows the same workflow as the derivation of semi-linear R13 equations. We first derive the boundary conditions for the nonlinear R13 equations. The boundary conditions are formulated based on the Maxwell boundary condition for the Boltzmann equation. As an example, we consider the boundary condition on the left solid wall, whose outer unit normal is $(0,-1,0)^T$. On this wall, the Maxwell boundary condition can be formulated as
\begin{equation} \label{eq:Maxwell_bc}
f(x_L, \bxi, t) = \chi \frac{\rho_L}{\mathfrak{m}(2\pi\theta_L)^{3/2}} \exp \left(
  -\frac{|\bxi-\bv_L|^2}{2\theta_L}
\right) + (1-\chi) f(x_L, \bxi^*, t), \quad \text{if } \xi_2 > 0,
\end{equation}
where the parameters are interpreted as follows:
\begin{itemize}
\item $\theta_L$: the temperature of the left wall;
\item $\bv_L = (v_L, 0, 0)^T$: the velocity of the left wall;
\item $\bxi^* = (\xi_1, -\xi_2, \xi_3)^T$: the reflected velocity;
\item $\chi$: the accommodation coefficient indicating the proportion of diffusive reflection;
\item $\rho_L$: the density of reflected particles defined by
  \begin{equation} \label{eq:no_flux}
  \int_{\mathbb{R}^3} \xi_2 f(x_L, \bxi, t) \,\mathrm{d}\bxi = 0,
  \end{equation}
  which means the mass flux on the boundary is zero.
\end{itemize}
Note that the boundary condition \eqref{eq:Maxwell_bc} is given only for the velocities pointing into the domain.

To derive boundary conditions for moment equations, we take moments for Maxwell's boundary condition:
\begin{equation} \label{eq:half_space_moments}
\int_{\xi_2 > 0} P(\bxi) f(x_L, \bxi, t) \,\mathrm{d}\bxi
  = \frac{\chi \rho_L}{\mathfrak{m} (2\pi \theta_L)^{3/2}} \int_{\xi_2 > 0} P(\bxi) \exp \left( -\frac{|\bxi - \bv_L|^2}{2\theta_L} \right) \mathrm{d}\bxi
  + (1-\chi) \int_{\xi_2 > 0} P(\bxi) f(x_L, \bxi^*, t) \,\mathrm{d} \bxi,
\end{equation}
where $P(\bxi)$ is a polynomial of $\bxi$. Note that the moments are taken only on the half space with $\xi_2 > 0$ since the boundary condition only applies to such velocities. Follow Grad's work \cite{Grad1949}, in order to guarantee the continuity of the boundary conditions with respect to the accommodation coefficient $\chi$, we include only odd moments, \textit{i.e.} $P(\bxi) = P(\bxi^*)$, into the boundary conditions. More precisely, we choose $P(\bxi)$ to be the following polynomials:
\begin{gather*}
\im p_{111} \left( \frac{\bxi - \bv(x_L,t)}{\sqrt{\theta(x_L,t)}} \right), \qquad
\im p_{220} \left( \frac{\bxi - \bv(x_L,t)}{\sqrt{\theta(x_L,t)}} \right), \\
\im p_{221} \left( \frac{\bxi - \bv(x_L,t)}{\sqrt{\theta(x_L,t)}} \right), \qquad
\im p_{310} \left( \frac{\bxi - \bv(x_L,t)}{\sqrt{\theta(x_L,t)}} \right), \qquad
\im p_{330} \left( \frac{\bxi - \bv(x_L,t)}{\sqrt{\theta(x_L,t)}} \right).
\end{gather*}
Similar to the derivation of the R13 equations, we insert \eqref{eq:ansatz} into the boundary condition \eqref{eq:Maxwell_bc}, drop all $O(\epsilon)^3$ terms, and then take the above moments as in \eqref{eq:half_space_moments}. Note that this is equivalent to first removing the $O(\epsilon^3)$ terms in \eqref{eq:ansatz}, and then inserting the result into \eqref{eq:half_space_moments}. The above procedure provides us five boundary conditions on the left solid wall. The five boundary conditions on the right solid wall can be similarly derived.

Note that the derivation of boundary conditions is slightly different from the approach used in \cite{Torrilhon2008, Taheri2009}. In these works, the boundary conditions are derived based on the 26-moment expansion of the distribution function:
\begin{equation} \label{eq:26m}
\begin{split}
f(\bx,\bxi,t) &= \mathcal{M}(\bx,\bxi,t) + \sum_{m=-1}^1 \tilde{A}_{11} f_{1m1}  \psi_{1m1}(\bx,\bxi,t) \\
& \quad + \sum_{m=-2}^2 \sum_{n=0}^1 \tilde{A}_{2n} f_{2m0}  \psi_{2mn}(\bx,\bxi,t) + \epsilon^2 \tilde{f}_{002}^{(2)}(\bx,t) \psi_{002}(\bx,\bxi,t) + \sum_{m=-3}^3 \epsilon^2 \tilde{f}_{3m0}^{(2)}(\bx,t) \psi_{3m0}(\bx,\bxi,t),
\end{split}
\end{equation}
while in our work, more coefficients are included. For Maxwell molecules ($\eta = 5$), these two approaches are equivalent since these 26 moments have fully covered the second-order contributions. However, for other gas molecules, if the above ansatz is used, although the general form of the boundary conditions given in \eqref{eq:bc1}--\eqref{eq:bc5} does not change, the values of the coefficients will differ slightly from the ``complete'' boundary conditions derived using \eqref{eq:ansatz} without $O(\epsilon^3)$ terms. Such a difference is given for the hard-sphere gas ($\eta = \infty$) in the appendix (see Table \ref{tab:comp}).

The boundary conditions will also be simplified through nondimensionalization, dimension reduction, and semi-linearization. The general steps are similar to Section \ref{sec:simplification} and the details are omitted.

\section{Conclusion} \label{sec:conclusion}
This work is an extension of \cite{Taheri2009} to a broader range of gas molecules. Based on the R13 equations for general inverse power law models derived in \cite{Cai2020}, we study its performance for channel flows including wall velocities, temperature differences, and body forces. The equations are semi-linearized to avoid complicated expressions, and the analytical solutions to these equations can be obtained for flows between parallel plates. By comparing these solutions with the DSMC results, it is demonstrated that the R13 equations can accurately capture a number of rarefied effects including velocity slip, temperature jump, and Knudsen layers. It has also been illustrated that the ``genuine'' R13 equations derived from general inverse power law models have better accuracy than the R13 equations for Maxwell molecules with a modified viscosity coefficient, especially for non-equilibrium quantities such as heat fluxes.

The current work focuses only on the single-species and monatomic gases. It can be foreseen that the same procedure can be carried out for gas mixtures and polyatomic gases. In these cases, we expect that the major challenges may come from the formulation of regularized moment equations. Existing works \cite{Rahimi2014,Gupta2016} on Maxwell molecules provide us good guidance in carrying out Chapman-Enskog expansions, while the details for the more involved inverse-power-law models need to be carefully worked out. The strategy presented in this paper can still be used as validations of such models.

The results of the analytical study is encouraging. It inspires us to carry out numerical study for more complicated cases including multidimensional settings and complex geometry, as will be considered in our future works.

\appendix
\section{Tables of Coefficients} \label{sec:coefficients}
In this appendix, we list the values of coefficient $\alpha^{(\eta)}_{i,j}$, $\beta^{(\eta)}_{i,j}$, $\gamma^{(\eta)}_{i,j}$ and $\delta^{(\eta)}_{i,j}$ appearing in Section \ref{sec:R13}. These values are tabulated in Tables \ref{tab:alpha}, \ref{tab:beta} and \ref{tab:gamma} for $\eta = 5,7,10,17,\infty$. All numbers are given with five significant figures. In Table \ref{tab:comp}, we compare the coefficients in the boundary conditions derived using the complete ansatz \eqref{eq:ansatz} and the simplified ansatz \eqref{eq:26m}.

\begin{table}[h]
\centering \footnotesize
\begin{displaymath}
\begin{array}{| c || c | c | c | c | c | c | c | c | c | c  | }
\hline
\eta &   \alpha^{(\eta)}_{1,1}   & \alpha^{(\eta)}_{1,2}  & \alpha^{(\eta)}_{1,3} & \alpha^{(\eta)}_{1,4} & \alpha^{(\eta)}_{1,5} & \alpha^{(\eta)}_{1,6}  & \alpha^{(\eta)}_{1,7} & \alpha^{(\eta)}_{1,8} & \alpha^{(\eta)}_{1,9} & \alpha^{(\eta)}_{1,10} \\
\hline
5  &  -1       & \frac{4}{15}  &  -\frac{4}{3}  & \frac{32}{75} & 0  &  0 &  0 & \frac{32}{75} & \frac{2}{3} & -\frac{4}{15}\\
7  &  -0.99545 &   0.23683  &  -1.3085 &  0.31951 & 0.0097118 & -0.040474 & -0.00087038& 0.35007 & 0.64398 & -0.27363\\
10 &  -0.98726 &   0.21696  &  -1.2922 &  0.25135 & 0.014075  & -0.067622 & -0.0024798 & 0.30261 & 0.63365 & -0.28014 \\
17 & -0.97660 &   0.19951   &  -1.2780  & 0.19253 & 0.016503  & -0.091995 & -0.0046493 & 0.26251 & 0.62764 & -0.28723\\
\infty & -0.95794 &   0.17684  &  -1.2596  & 0.11594 & 0.017613 & -0.12494& -0.0086566& 0.21135& 0.62411 & -0.29863\\
\hline
\eta &   \alpha^{(\eta)}_{2,1}   & \alpha^{(\eta)}_{2,2}  & \alpha^{(\eta)}_{2,3} & - & - & -  & - & - & - & - \\
\hline
5 &     -\frac{2}{5} &   -1        &  \frac{16}{15}  & - & - & -& -& -& - & -\\
7  &        -0.35525 &   -0.99719  &  1.0544  & - & - & -& -& -& - & -\\
10 &        -0.32544 &   -0.99208  &  1.0539  & - & - & -& -& -& - & -\\
17 &        -0.29926 &   -0.98534  &  1.0585  & - & - & -& -& -& - & -\\
\infty &    -0.26526 &   -0.97330  &  1.0720  & - & - & -& -& -& - & -\\
\hline
\eta &   \alpha^{(\eta)}_{3,1}   & \alpha^{(\eta)}_{3,2}  & \alpha^{(\eta)}_{3,3} & \alpha^{(\eta)}_{3,4} & \alpha^{(\eta)}_{3,5} & \alpha^{(\eta)}_{3,6} & - & - & - & - \\
\hline
5  & -\frac{8}{15} & \frac{2}{3}  &  -\frac{8}{25}  & 0 & -\frac{8}{25} & \frac{6}{5} & -& -& - & -\\
7  &  -0.47366 &   0.69154  &  -0.25070 &  0.012323 & -0.26914 & 1.1912 & -& -& - & -\\
10 &  -0.43393 &   0.70783  &  -0.20678 &  0.021515 & -0.23788 & 1.1939 & - & - & - & -\\
17 &  -0.39902 &   0.72202  &  -0.16899 &  0.030451 & -0.21165 & 1.2021 & -& -& - & -\\
\infty &-0.35368 & 0.74037  &  -0.11982  & 0.043582 & -0.17841 & 1.2214& -& -& - & -\\
\hline
\eta &   \alpha^{(\eta)}_{4,1}   & \alpha^{(\eta)}_{4,2}  & \alpha^{(\eta)}_{4,3} & \alpha^{(\eta)}_{4,4} & - & - & - & - & - & -  \\
\hline
5  &  -\frac{2}{3} &   -1  &  \frac{6}{5}  & 0 & - & -& -& -& - & -\\
7  &  -0.66280 &   -0.86755  &  1.1449 &  0.059935 & - & -& -& -& - & -\\
10 &  -0.65589 &   -0.77893  &  1.1173 &  0.099193 & - & - & - & - & - & -\\
17 &  -0.64696 &   -0.70066  &  1.0987  & 0.13375 & - & -& -& -& - & -\\
\infty &-0.63148 &   -0.59824  &  1.0823  & 0.17941 & - & -& -& -& - & -\\
\hline
\eta &   \alpha^{(\eta)}_{5,1}   & \alpha^{(\eta)}_{5,2}  & \alpha^{(\eta)}_{5,3} & \alpha^{(\eta)}_{5,4} & \alpha^{(\eta)}_{5,5} & \alpha^{(\eta)}_{5,6} & - & - & -& - \\
\hline
5  &  \frac{18}{5} &   \frac{18}{7}  &  -\frac{5}{2}  & -\frac{2}{5} & 1 &  \frac{382}{105}& -& -& - & -\\
7  &  3.3289 &   2.3144  &  -2.4923  &  -0.32671 & 0.99878 & 3.3758 & -& -& - & -\\
10 &  3.1847 &   2.1752  &  -2.4784  &  -0.27840 & 0.99651 & 3.2399 & - & - & - & -\\
17 &  3.0799 &   2.0716  &  -2.4602  & -0.23619 & 0.99349 & 3.1440& -& -& - & -\\
\infty &    2.9722 &   1.9609  &  -2.4278  & -0.18158 & 0.98799 & 3.0503 & -& -& - & -\\
\hline
\end{array}
\end{displaymath}
\caption{The value of coefficient $\alpha^{(\eta)}_{i,j}$.} \label{tab:alpha}
\end{table}

\begin{table}[h]
\centering
\begin{displaymath}
\begin{array}{| c || c | c | c | c | c | c | c | }
\hline
\eta &   \beta^{(\eta)}_{1,1}   & \beta^{(\eta)}_{1,2}  & \beta^{(\eta)}_{1,3} & \beta^{(\eta)}_{1,4} & \beta^{(\eta)}_{1,5} & - & -   \\
\hline
5  &      1.5958   & 0.39894  &  -1.0942  & -0.39894 & -0.89850  &  - & -   \\
7  & 1.5958 &   0.35845  &  -1.0824 &  -0.39894 & -0.86861 & - & - \\
10 &  1.5958 &  0.33020  &  -1.0808 &  -0.39894 & -0.85373  & - & - \\
17  & 1.5958 &   0.30445  &  -1.0835 &  -0.39894 & -0.84375 & - & - \\
\infty  & 1.5958 &   0.26956  &  -1.0930 &  -0.39894 & -0.83498 & - & - \\
\hline
\eta &   \beta^{(\eta)}_{2,1}   & \beta^{(\eta)}_{2,2}  & \beta^{(\eta)}_{2,3} & \beta^{(\eta)}_{2,4} & \beta^{(\eta)}_{2,5} & \beta^{(\eta)}_{2,6} &  \beta^{(\eta)}_{2,7 }\\
\hline
5  &      -0.20656   & 0.77460  &  0.20601  & -0.72105 & 0.035316  &  -0.30902& -0.36717   \\
7  & -0.18630 &   0.76219  &  0.20601 &  -0.71108 & 0.036855 & -0.30902& -0.36758\\
10 &  -0.17358 &  0.75764  &  0.20601 &  -0.70432 & 0.037886  & -0.30902 & -0.37068\\
17  & -0.16278 &   0.75610  &  0.20601 &  -0.69831 & 0.038819 & -0.30902 & -0.37549\\
\infty  & -0.14901 &   0.75765  &  0.20601 &  -0.69039 & 0.040109 & -0.30902& -0.38586\\
\hline
\eta &   \beta^{(\eta)}_{3,1}   & \beta^{(\eta)}_{3,2}  & \beta^{(\eta)}_{3,3} & \beta^{(\eta)}_{3,4} & \beta^{(\eta)}_{3,5} & \beta^{(\eta)}_{3,6} & -   \\
\hline
5  &      0   & 0.45356  &  0  & -0.33173 & 0.15079  &  0.080419& -   \\
7  & 0.050198 &   0.44558  &  0.023218 &  -0.32125 & 0.15079 & 0.053283 & -\\
10 &  0.083883 &  0.44219  &  0.038957 &  -0.31407 & 0.15079  & 0.035052 & -\\
17  & 0.11365 &   0.44049  &  0.053084 &  -0.30763 & 0.15079 & 0.018687 & -\\
\infty  & 0.15253 &   0.44007  &  0.071993 &  -0.29907 & 0.15079 & -0.0035168& -\\
\hline
\eta &   \beta^{(\eta)}_{4,1}   & \beta^{(\eta)}_{4,2}  & \beta^{(\eta)}_{4,3} & -& - & - & -  \\
\hline
5  &  0.15958   & 0.79788  &  -0.42554  & - & -  &  -& -   \\
7  & 0.16191 &   0.79788  &  -0.42821 &  - & - & -& -\\
10 &  0.16354 &  0.79788  &  -0.43236 &  - & -  & - & -\\
17  & 0.16504 &   0.79788  &  -0.43769 &  - & - & -& -\\
\infty  & 0.16707 &   0.79788  &  -0.44720 &  - & - & -& -\\
\hline
\eta &   \beta^{(\eta)}_{5,1}   & \beta^{(\eta)}_{5,2}  & \beta^{(\eta)}_{5,3} & \beta^{(\eta)}_{5,4} & \beta^{(\eta)}_{5,5} & - & -  \\
\hline
5  &      0.18856   & 0.23570  &  0.28209  & -0.28209 & 0.32726  &  -& -   \\
7  & 0.17007 &   0.23193  &  0.28209 &  -0.27655 & 0.32996 & -& -\\
10 &  0.15845 &  0.23054  &  0.28209 &  -0.27275 & 0.33418  & - & -\\
17  & 0.14859 &   0.23007  &  0.28209 &  -0.26934 & 0.33972 & -& -\\
\infty  & 0.13603 &   0.23054  &  0.28209 &  -0.26480 & 0.35049 & -& -\\
\hline
\end{array}
\end{displaymath}
\caption{The value of coefficient $\beta^{(\eta)}_{i,j}$.} \label{tab:beta}
\end{table}

\begin{table}[h]
\centering
\begin{displaymath}
\begin{array}{| c || c | c | c | c | c | c | c | c |}
\hline
\eta &   \gamma^{(\eta)}_{1,1}   & -  &- & - & - & - & - & -  \\
\hline
5  &      -\frac{3}{2}   & -  &  -  & - & -  &  - & - & -  \\
7   &      -1.3992   & -  &  -  & - & -  &  - & - & -  \\
10  &      -1.3381   & -  &  -  & - & -  &  - & - & -  \\
17  &      -1.2879   & -  &  -  & - & -  &  - & - & -  \\
\infty  &  -1.2270   & -  &  -  & - & -  &  - & - & -  \\
\hline
\eta &   \gamma^{(\eta)}_{2,1}   & \gamma^{(\eta)}_{2,2}  &- & - & - & - & -  & - \\
\hline
5  &      -\frac{1}{2}   & \frac{2}{5}  &  -  & - & -  &  - & -  & - \\
7   &      -0.49913   & 0.35625  &  -  & - & -  &  - & -  & - \\
10  &      -0.49757   & 0.32804  &  -  & - & -  &  - & -  & - \\
17  &      -0.49557   & 0.30372  &  -  & - & -  &  - & -  & - \\
\infty  &  -0.49211   & 0.27253  &  -  & - & -  &  - & -  & - \\
\hline
\eta &   \gamma^{(\eta)}_{3,1}   & \gamma^{(\eta)}_{3,2}  & \gamma^{(\eta)}_{3,3} &  \gamma^{(\eta)}_{3,4} &  \gamma^{(\eta)}_{3,5} &  \gamma^{(\eta)}_{3,6} & - & -  \\
\hline
5  &-\frac{84}{25}   & -\frac{6}{5}&  0 & -\frac{32}{25} & -\frac{4\sqrt{5}}{25}  &  -\frac{32}{375} & - & -  \\
7   &      -3.0778   & -1.1773  &  0.0059300  & -1.4332 & -0.36281  &  -0.065584 & - & -  \\
10  &      -2.9442   & -1.1640  &  0.0096643  & -1.6151 & -0.37392  &  -0.053318 & - & -  \\
17  &      -2.8589   & -1.1537  &  0.012712   & -1.8506 & -0.39054  &  -0.043229 & - & -  \\
\infty  &  -2.7906   & -1.1421  &  0.016240   &      -2.3157 & -0.42572  &  -0.031131 & -  & - \\
\hline
\eta &   \gamma^{(\eta)}_{4,1}   & \gamma^{(\eta)}_{4,2}  & \gamma^{(\eta)}_{4,3} &  \gamma^{(\eta)}_{4,4} &  \gamma^{(\eta)}_{4,5} &  \gamma^{(\eta)}_{4,6} &  \gamma^{(\eta)}_{4,7} &  \gamma^{(\eta)}_{4,8}     \\
\hline
5  &-\frac{1}{45}   & \frac{398}{525}  & -\frac{2}{15}  & -\frac{4}{15} & \frac{128}{375}  &  \frac{132\sqrt{5}}{875} & \frac{124}{1875} & -\frac{2}{5}   \\
7   &  -0.022122  & 0.71239  &  -0.13273  & -0.26593 & 0.38118  &  0.28469 & 0.046178  & -0.34809 \\
10  &  -0.021946  & 0.68367  &  -0.13167  & -0.26463 & 0.41961  &  0.25354 & 0.035181 & -0.31428   \\
17  &  -0.021719  & 0.65939  &  -0.13032  & -0.26297 & 0.46273  &     0.22811 & 0.026891 & -0.28480  \\
\infty&-0.021334  & 0.62893  &  -0.12800  & -0.26011 & 0.53488  &     0.19714 & 0.017863 & -0.24642  \\
\hline
\eta &   \gamma^{(\eta)}_{5,1}   & \gamma^{(\eta)}_{5,2}  & \gamma^{(\eta)}_{5,3} &  \gamma^{(\eta)}_{5,4} &  \gamma^{(\eta)}_{5,5} &  \gamma^{(\eta)}_{5,6} &  \gamma^{(\eta)}_{5,7} & -    \\
\hline
5  &      \frac{256}{75}   & \frac{8}{5}  &  0  & \frac{4384}{7225} & \frac{64\sqrt{5}}{425}  &  \frac{1408}{4875} & -\frac{1}{2}& -   \\
7   &  3.1109  & 1.5699  &  -0.014371  & 0.79172 & 0.38157  &  0.19478 & -0.49999  & - \\
10  &  2.9658  & 1.5527  &  -0.022455  & 0.95689 & 0.41450  &  0.14431 & -0.50002 & -   \\
17  &  2.8714  & 1.5394  &  -0.028517 & 1.14267 & 0.44657  &  0.10644 & -0.50000 & -  \\
\infty &2.7920 & 1.5248  &  -0.034844 & 1.4708  &    0.49540  &  0.064697& -0.50002 & -  \\
\hline
\eta &   \delta^{(\eta)}_{1}   &  \delta^{(\eta)}_{2}    &  \delta^{(\eta)}_{3}   &  - &  - &  - &  - & -    \\
\hline
5  &\frac{\sqrt{5}}{3}   & \frac{\sqrt{30}}{6}  &  \frac{\sqrt{6}}{2}  & - & -  &  - & -& -   \\
7   &  0.76807  & 0.92559  & 1.2433  & - & -  &  - & -  & - \\
10  &  0.77759  & 0.92734  & 1.2482  & - & -  &  - & - & -   \\
17 &  0.78195  & 0.92433  & 1.2474  & - & -  &  - & - & -  \\
\infty  & 0.78171 & 0.91399 &  1.2389  & - & -  &  - & - & -  \\
\hline
\end{array}
\end{displaymath}
\caption{The value of coefficient $\gamma^{(\eta)}_{i,j}$ and $\delta^{(\eta)}_{k}$.} \label{tab:gamma}
\end{table}

\begin{table}[h]
\centering
\begin{displaymath}
\begin{array}{| c || c | c | c | c | c | c | c | }
\hline
\eta = \infty &   \beta^{(\eta)}_{1,1}   & \beta^{(\eta)}_{1,2}  & \beta^{(\eta)}_{1,3} & \beta^{(\eta)}_{1,4} & \beta^{(\eta)}_{1,5} & - & -   \\
\hline
\text{BC from \eqref{eq:26m}}  & 1.5958 &   0.26956  &  -1.0930 &  -0.39894 & -0.83498 & - & - \\
\text{BC from \eqref{eq:ansatz}}  & 1.5958 &   0.28394  &  -1.0056 &  -0.39894 & -0.68790 & - & - \\
\hline
\eta = \infty &   \beta^{(\eta)}_{2,1}   & \beta^{(\eta)}_{2,2}  & \beta^{(\eta)}_{2,3} & \beta^{(\eta)}_{2,4} & \beta^{(\eta)}_{2,5} & \beta^{(\eta)}_{2,6} &  \beta^{(\eta)}_{2,7 }\\
\hline
\text{BC from \eqref{eq:26m}} & -0.14901 &   0.75765  &  0.20601 &  -0.69039 & 0.040109 & -0.30902& -0.38586\\
\text{BC from \eqref{eq:ansatz}} & -0.14901 &   0.75765  &  0.20601 &  -0.69135 & 0.041503 & -0.30902& -0.028990\\
\hline
\eta = \infty  &   \beta^{(\eta)}_{3,1}   & \beta^{(\eta)}_{3,2}  & \beta^{(\eta)}_{3,3} & \beta^{(\eta)}_{3,4} & \beta^{(\eta)}_{3,5} & \beta^{(\eta)}_{3,6} & -   \\
\hline
\text{BC from \eqref{eq:26m}}  & 0.15253 &   0.44007  &  0.071993 &  -0.29907 & 0.15079 & -0.0035168& -\\
\text{BC from \eqref{eq:ansatz}}  & 0.15253 &   0.44007  &  0.071993 &  -0.33173 & 0.15079 & 0.078659& -\\
\hline
\eta = \infty  &   \beta^{(\eta)}_{4,1}   & \beta^{(\eta)}_{4,2}  & \beta^{(\eta)}_{4,3} & -& - & - & -  \\
\hline
\text{BC from \eqref{eq:26m}} & 0.16707 &   0.79788  &  -0.44720 &  - & - & -& -\\
\text{BC from \eqref{eq:ansatz}} & 0.15958 &   0.79788  &  -0.41623 &  - & - & -& -\\
\hline
\eta = \infty  &   \beta^{(\eta)}_{5,1}   & \beta^{(\eta)}_{5,2}  & \beta^{(\eta)}_{5,3} & \beta^{(\eta)}_{5,4} & \beta^{(\eta)}_{5,5} & - & -  \\
\hline
\text{BC from \eqref{eq:26m}}  & 0.13603 &   0.23054  &  0.28209 &  -0.26480 & 0.35049 & -& -\\
\text{BC from \eqref{eq:ansatz}}  & 0.13603 &   0.23054  &  0.28209 &  -0.26583 & 0.065526 & -& -\\
\hline
\end{array}
\end{displaymath}
\caption{Comparison of coefficients $\beta^{(\eta)}_{i,j}$ for boundary conditions of the hard-sphere gas derived using the complete ansatz \eqref{eq:ansatz} and the 26-moment ansatz \eqref{eq:26m}.}
\label{tab:comp}
\end{table}

\bibliographystyle{amsplain}
\bibliography{R13}
\end{document}